\documentclass[twocolumn,floatfix,superscriptaddress,aps, prr]{revtex4-2}
\usepackage{graphicx}
\usepackage{dcolumn}
\usepackage{bm}
\usepackage{xspace}
\usepackage{amsmath}
\usepackage{xcolor}
\usepackage{hyperref}
\definecolor{link}{rgb}{0.1,0.1,0.9}
\hypersetup{colorlinks=true,linkcolor=link,citecolor=link,urlcolor=link,linktocpage}
\usepackage{makecell}

\newcolumntype{M}[1]{>{\centering\arraybackslash}m{#1}}

\newcommand{\NMO}{Ni$_{2}$Mo$_{3}$O$_{8}$\xspace}
\newcommand{\AMO}{$\mathcal{A}$$_{2}$Mo$_{3}$O$_{8}$ ($\mathcal{A}$ = Fe, Mn, Co, Ni\xspace)}

\newcommand{\AMOshort}{$\mathcal{A}$$_{2}$Mo$_{3}$O$_{8}$\xspace}
\newcommand{\NZMO}{Ni$_{2-x}$Zn$_{x}$Mo$_{3}$O$_{8}$\xspace}
\newcommand{\NMMO}{Ni$_{2-x}$Mg$_{x}$Mo$_{3}$O$_{8}$\xspace}
\newcommand{\FMO}{Fe$_{2}$Mo$_{3}$O$_{8}$\xspace}
\newcommand{\FZMO}{Fe$_{2-x}$Zn$_{x}$Mo$_{3}$O$_{8}$\xspace}

\newcommand{\MMO}{Mn$_{2}$Mo$_{3}$O$_{8}$\xspace}
\newcommand{\CMO}{Co$_{2}$Mo$_{3}$O$_{8}$\xspace}
\newcommand{\ZMO}{Zn$_{2}$Mo$_{3}$O$_{8}$\xspace}

\begin{document}

\preprint{APS/123-QED}

\title{Noncollinear magnetic order, in-plane anisotropy, and magnetoelectric coupling~\\ in the pyroelectric honeycomb antiferromagnet Ni$_{2}$Mo$_{3}$O$_{8}$}

\author{Poonam Yadav}
\affiliation{Center for Integrated Nanostructure Physics, Institute for Basic Science, Suwon 16419, Republic of Korea}
\affiliation{Sungkyunkwan University, Suwon 16419, Republic of Korea}

\author{Suheon Lee}
\affiliation{Center for Integrated Nanostructure Physics, Institute for Basic Science, Suwon 16419, Republic of Korea}
\affiliation{Sungkyunkwan University, Suwon 16419, Republic of Korea}

\author{G. L. Pascut}
\affiliation{Department of Physics and Astronomy, Rutgers University, Piscataway, New Jersey 08854, USA}
\affiliation{MANSiD Research Center and Faculty of Forestry, Applied Ecology Laboratory, Stefan Cel Mare University (USV), 13 University Road, Suceava 720229, Romania}

\author{Jaewook Kim}
\altaffiliation[Present address: ] {Korea Atomic Energy Research Institute, Daejeon, Republic of Korea 34057}
\affiliation{Department of Physics and Astronomy, Rutgers University, Piscataway, New Jersey 08854, USA}
\affiliation{Rutgers Center for Emergent Materials, Rutgers University, Piscataway, New Jersey 08854, USA}

\author{Matthias J. Gutmann}
\affiliation{ISIS Facility, Rutherford Appleton Laboratory, Chilton, Didcot, OX11 0QX, UK}

\author{Xianghan Xu}
\altaffiliation[Present address: ] {Department of Chemistry, Princeton University, Princeton, New Jersey 08544, United States}
\affiliation{Department of Physics and Astronomy, Rutgers University, Piscataway, New Jersey 08854, USA}
\affiliation{Rutgers Center for Emergent Materials, Rutgers University, Piscataway, New Jersey 08854, USA}

\author{Bin Gao}
\altaffiliation[Present address: ] {Department of Physics and Astronomy, Rice University, Houston, TX 77005, USA.}
\affiliation{Department of Physics and Astronomy, Rutgers University, Piscataway, New Jersey 08854, USA}
\affiliation{Rutgers Center for Emergent Materials, Rutgers University, Piscataway, New Jersey 08854, USA}

\author{Sang-Wook Cheong}
\affiliation{Department of Physics and Astronomy, Rutgers University, Piscataway, New Jersey 08854, USA}
\affiliation{Rutgers Center for Emergent Materials, Rutgers University, Piscataway, New Jersey 08854, USA}

\author{Valery Kiryukhin}
\affiliation{Department of Physics and Astronomy, Rutgers University, Piscataway, New Jersey 08854, USA}

\author{Sungkyun Choi}
\email{sungkyunchoi@skku.edu}
\affiliation{Center for Integrated Nanostructure Physics, Institute for Basic Science, Suwon 16419, Republic of Korea}
\affiliation{Sungkyunkwan University, Suwon 16419, Republic of Korea}
\affiliation{Department of Physics and Astronomy, Rutgers University, Piscataway, New Jersey 08854, USA}

\begin{abstract}
Ni$_{2}$Mo$_{3}$O$_{8}$ is a pyroelectric honeycomb antiferromagnet exhibiting peculiar changes in its electric polarization at magnetic transitions.
Ni$_{2}$Mo$_{3}$O$_{8}$ stands out from isostructural magnetic compounds, showing an anomalously low magnetic transition temperature and unique magnetic anisotropy.
We determine the magnetic structure of Ni$_{2}$Mo$_{3}$O$_{8}$ utilizing high-resolution powder and single-crystal neutron diffraction.
A noncollinear stripy antiferromagnetic order is found in the honeycomb planes. The magnetic space group is \textit{P$_C$na}2$_1$.
The in-plane magnetic connection is of the stripy type for both the $ab$-plane and the $c$-axis spin components. This is a simpler connection than the one proposed previously.
The ferromagnetic interlayer order of the $c$-axis spin components in our model is also distinct. The magnetic anisotropy of Ni$_{2}$Mo$_{3}$O$_{8}$ is characterized by orientation-dependent
magnetic susceptibility measurements on a single crystal, consistent with neutron diffraction analysis.
The local magnetoelectric tensor analysis using our magnetic models provides insights into its magnetoelectric coupling and polarization.
Thus, our results deliver essential information for understanding both the unusual magnetoelectric properties of Ni$_{2}$Mo$_{3}$O$_{8}$ and
the prospects for observing exotic nonreciprocal, Hall, and magnonic effects characteristic to this compound family.
\end{abstract}
\maketitle

\section{Introduction}
\label{sec:intro}
Multiferroic materials are compounds in which electric polarization and magnetic order coexist in the same phase. The magnetic and
dielectric properties of multiferroics can often be manipulated by the conjugate fields. In particular, their magnetic order may be affected by
an applied electric field, and the electric polarization by a magnetic field.
Such effects are called magnetoelectric (ME). Compounds exhibiting large ME effects are highly sought because
of fundamental and technological interest.

Pyroelectric magnets are a subset of polar multiferroics with fixed-direction polarization. Because
such a built-in polarization often has a significant magnitude, it holds the potential for large magnetically-induced changes. Many pyroelectrics
can be grown as monodomain crystals, and therefore, they do not require complex multi-field poling procedures to induce a single-domain multiferroic
state. These properties are important from both the technological and fundamental perspectives because they allow simplified manipulations,
as well as definitive studies, of the ME effects with a potentially large magnitude.

A family of polar magnetoelectric \AMO~\cite{McCarroll1957, Strobel1983} compounds with widely varying ME properties has recently attracted
significant interest. \FMO and its derivative \FZMO have been studied the most extensively so far.
They were found to exhibit such intriguing properties as a hidden ferrimagnetic
order with a hybridized band-Mott gap~\cite{Wang2015,Park2021}, tunable magnetoelectricity~\cite{Kurumaji2015}, the giant thermal Hall
effect~\cite{Ideue2017}, the optical diode effect~\cite{Yu2018}, and axion-type coupling at optical ME resonance~\cite{Kurumaji2017:FZMO}.
This compound family exhibits many other exotic properties, including diagonal ME susceptibility in
\MMO~\cite{Kurumaji2017, Stanislavchuk2020, Szaller2020} and successive electric polarization transitions and nonreciprocal light propagation
in \CMO~\cite{Tang2019, Tang2022, Reschke2022}.

\begin{figure}[t]
\includegraphics[width=\linewidth]{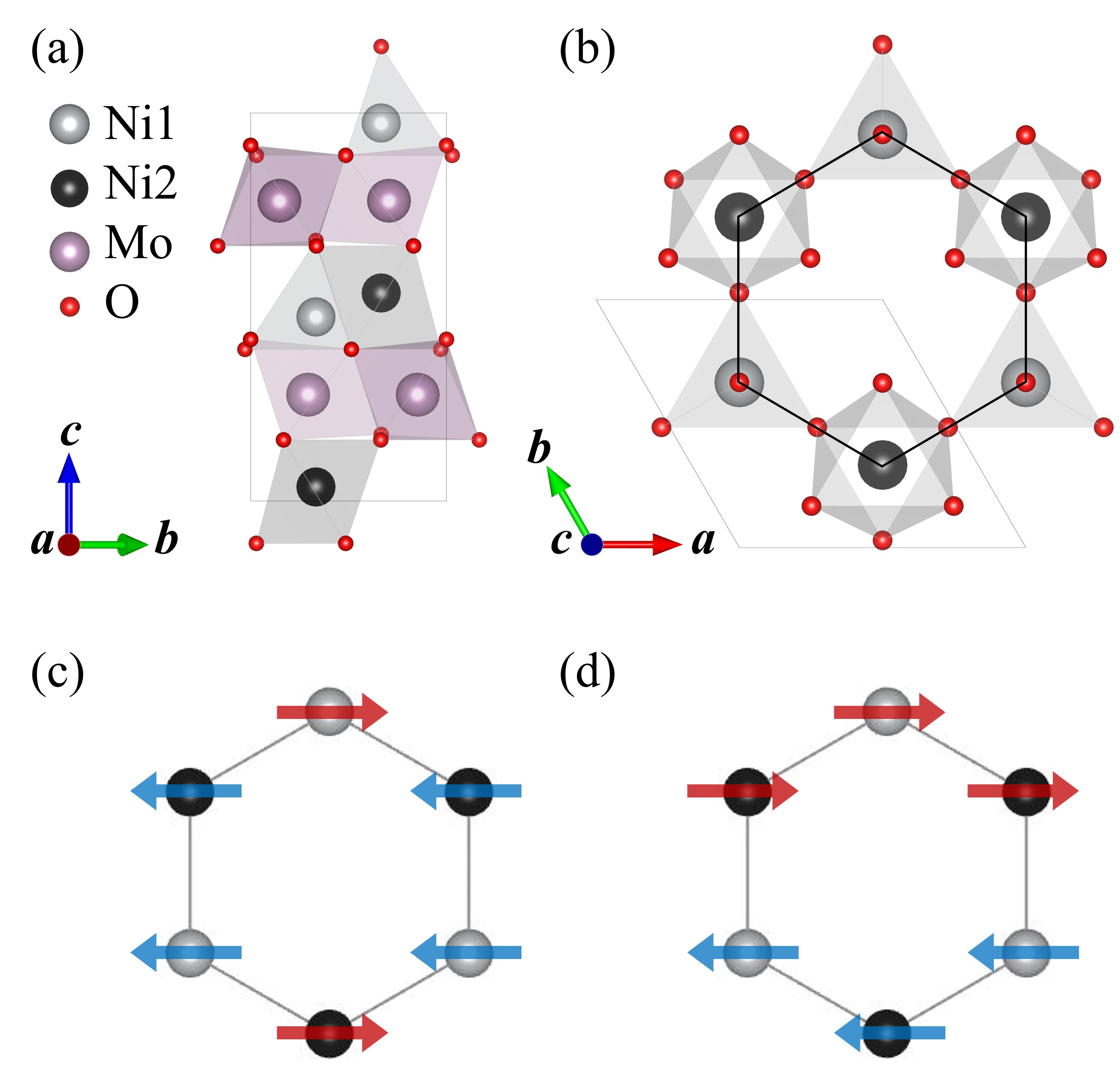}
\caption{(a) and (b) The crystal structure of \NMO and other \AMOshort compounds (0~$\leq$~z~$\leq$~1). (b) depicts the
magnetic honeycomb layer composed of the alternating NiO$_4$ tetrahedra and NiO$_6$ octahedra ($-$0.1~$\leq$~z~$\leq$~0.1).
(c) Stripy and (d) zigzag order in a generic honeycomb lattice. The easy-axis (spin) direction is arbitrary.}
\label{Fig:1}
\end{figure}

The \AMOshort structure is shown in Figs.~\ref{Fig:1}(a) and ~\ref{Fig:1} (b). This structure is noncentrosymmetric and polar.
It consists of honeycomb layers formed by alternating NiO$_4$
tetrahedra and NiO$_6$ octahedra. The formal spin of the Ni$^{2+}$ ions in the $d^8$ state is 1. These layers are separated by trimerized
nonmagnetic Mo ions. The $\mathcal{A}$=Fe, Mn, and Co compounds exhibit magnetic ordering temperatures in the range of 40 to 60~K
\cite{Wang2015,Tang2019}. Their magnetic order within the honeycomb layers is of the simple collinear N\'{e}el type, in which the nearby
spins are antiparallel and point along the $c$ axis~\cite{Bertrand1975,McAlister1983}.

\begin{figure}[t]
\includegraphics[width=\linewidth]{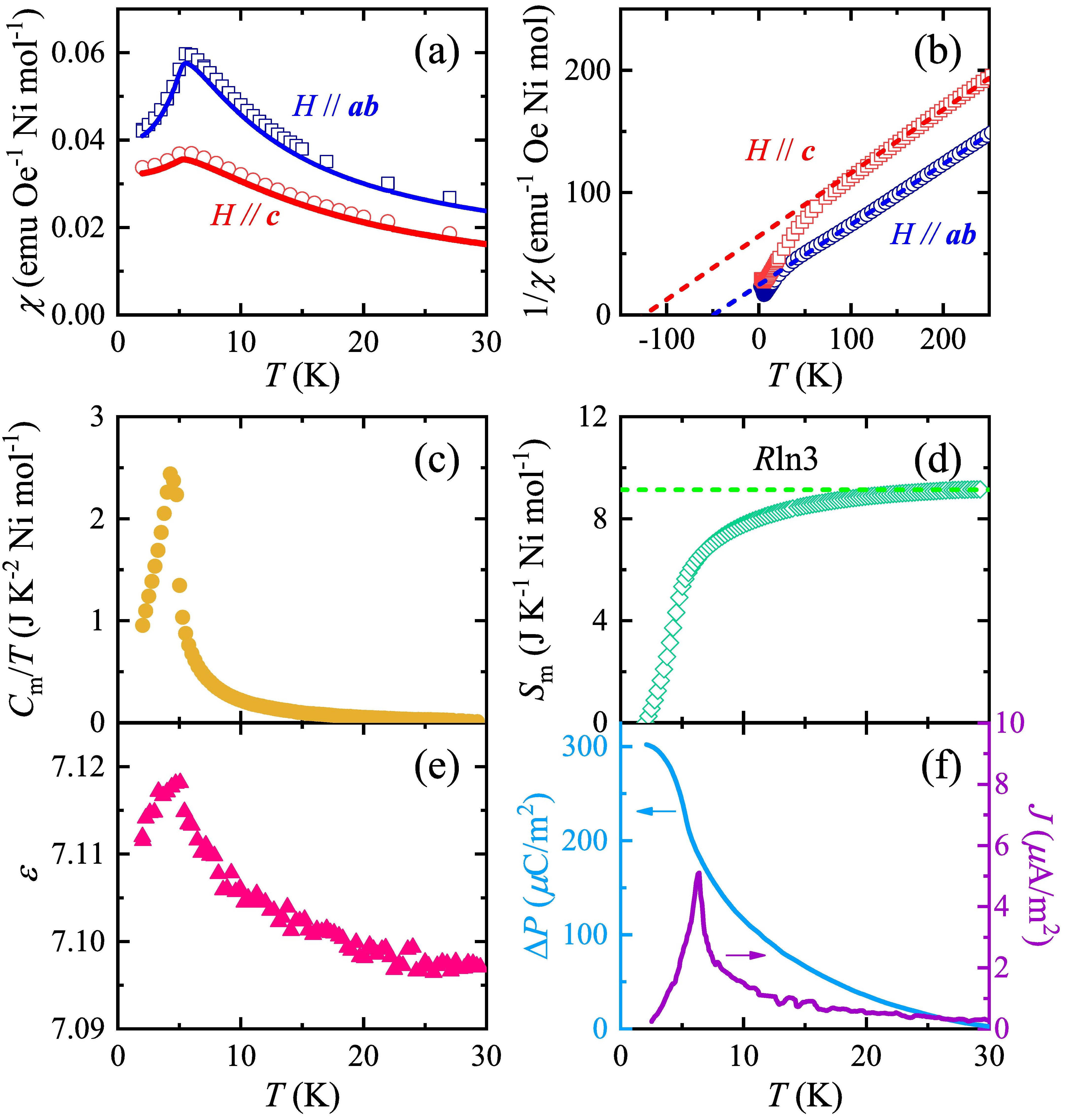}
\caption{(a) In-plane and $c$-axis magnetic susceptibility versus temperature. The open symbols and solid lines are
the data from different crystals. The single crystal corresponding to the solid-line data is used for the orientation-dependent susceptibility
measurements shown in Fig.~\ref{Fig:10}. The solid-line data for $H$~$\vert\vert$~$ab$ is taken for the [$\overline{2}$ $\overline{1}$ 0] field direction,
corresponding to direction 1 in Fig.~\ref{Fig:10}. (b) Inverse magnetic susceptibility.
The dashed lines are fits to the Curie--Weiss law.
(c) Magnetic specific heat divided by the temperature.
(d) Magnetic entropy extracted from the data shown in (c). (e) Dielectric constant along the $c$ axis as a function of temperature.
(f) Temperature dependence of the polarization change $\Delta P$ [$P(T)$ – $P(T = 30~K)$] and the pyroelectric current density $J$.}
\label{Fig:2}
\end{figure}

On the other hand, the magnetic properties of \NMO are unique among the compounds of the \AMOshort family. Its magnetic ordering temperature
($\sim$5.5 K) is anomalously low, and the magnetic anisotropy is of the easy-plane type, with magnetic moments lying predominantly in the $ab$
plane \cite{Tang2021}. Recent powder diffraction measurements suggest that the magnetic order is noncollinear \cite{Morey2019}.
There are several patterns of magnetic order typical of the honeycomb lattice. In addition to the N\'{e}el order described above, they are the stripy and zigzag orders. Both orders consist of ferromagnetic spin chains antiferromagnetically coupled to each other. The chains are of
the linear and zigzag types in the stripy and zigzag order, respectively [see Figs.~\ref{Fig:1}(c) and \ref{Fig:1}(d)]. The magnetic structure of \NMO proposed in Ref.\ [\onlinecite{Morey2019}] is very complex. The in-plane magnetic connection is of the stripy type for the $ab$-plane
components of the spins, while the $c$-axis spin components exhibit the zigzag-type connection. Both the reduced magnetic transition temperature and the complex noncollinear structure indicate significant magnetic frustration.

Frustrated honeycomb magnetic lattices are of significant interest. For example, topological magnons have been predicted in the stripy and zigzag states~\cite{Owerre2016}. In the \AMOshort compounds, the potentially exotic magnetism is coupled to the electric polarization. In the specific case of \NMO, a rich and
complex magnetoelectric behavior was observed~\cite{Tang2021}. The polarization dependence on the applied magnetic field exhibited linear, parabolic,
and plateau-like features, depending on the field direction and its magnitude. These unusual phenomena were discussed
using symmetry-based local ME tensor analysis. It was proposed that both the spin current and $p$-$d$ hybridization mechanisms
could contribute to the electric polarization and the ME response. The magnetic structure is the necessary starting point
for understanding the mechanism of the ME coupling and the possible observation of the topological magnons and exotic magnetic states in \NMO,
as well as for the analysis of the potential nonreciprocity and exotic Hall effects, as found in the other compounds in the \AMOshort family.

In this work, we revise the magnetic structure of \NMO using both powder and single-crystal neutron diffraction.
We find that the in-plane magnetic connection is of the stripy type for both the $ab$-plane and the $c$-axis spin components.
This connection is referred to as stripy/stripy hereafter, in contrast to the stripy/zigzag connection found in the previous powder diffraction studies \cite{Morey2019}.
The revised magnetic structure presents a significantly simpler magnetic connection than the one proposed previously.
The ferromagnetic interlayer order of the $c$-axis spin components in our model is also distinct.
In our study, the referential magnetic orders converge in our models for both the powder and single-crystal data.
We further investigate orientation-dependent magnetic susceptibility in a single crystal by quantifying the magnetic anisotropy from the Curie--Weiss fits.
The results are consistent with the magnetic order determined by our neutron diffraction study.
The magnetic structure and anisotropy found in our experiments provide key information for understanding the magnetic and ME properties of \NMO,
and should help us understand the uniqueness of \NMO in the \AMOshort compound family and the prospects for the observation of exotic nonreciprocal, Hall,
and magnonic effects characteristic to this compound family.

This paper is organized as follows. Section~\ref{sec:exp} describes the experimental details. The magnetic, thermodynamic, and electric properties of \NMO are described in Sec.~\ref{sec:Chi:Cp:P}. Single-crystal x-ray diffraction results are presented in Sec.~\ref{sec:sXRD}. The powder neutron diffraction and single-crystal diffraction results are presented in Secs.~\ref{sec:NPD} and~\ref{sec:sND}, respectively. The orientation-dependent susceptibility measurements are presented in
Sec.~\ref{sec:ODChi}. The implications of our results for the magnetoelectric coupling and polarization are explained in Sec.~\ref{sec:ME}.
A discussion is given in Sec.~\ref{sec:discussion}, followed by conclusions in Sec.~\ref{sec:conclusions}. The Appendixes include additional information.

\begin{figure} [t]
\includegraphics[width=\linewidth]{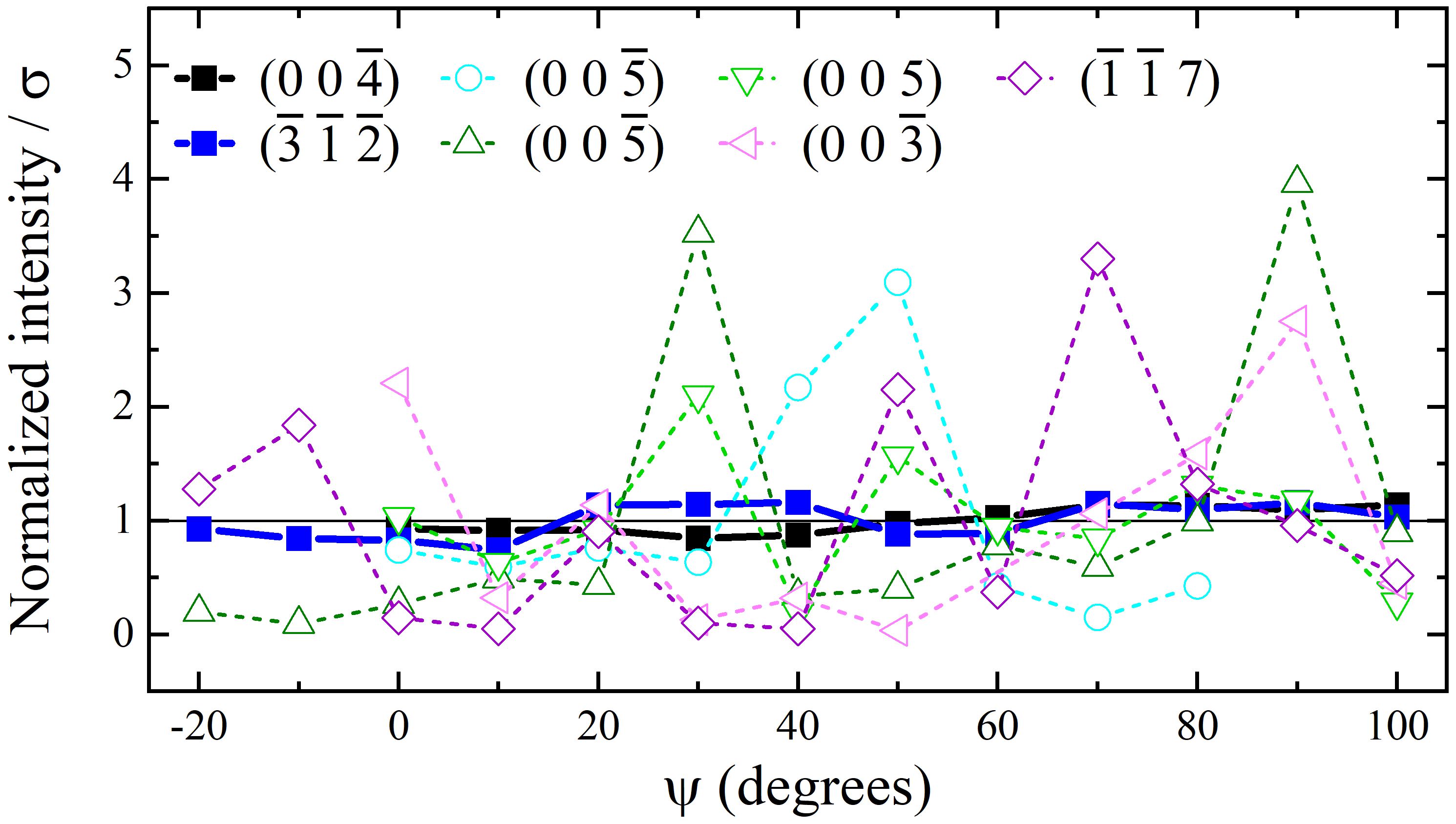}
\caption{Azimuthal x-ray scans for seven selected nuclear Bragg peaks at room temperature. The intensities are normalized as discussed in
the text. Open and solid symbols show the forbidden and allowed reflections, respectively. The (00$\overline{5}$) peak was measured using
two different scattering geometry settings to emphasize its irregular variation with $\Psi$.
}
\label{Fig:3}
\end{figure}

\section{Experimental details}
\label{sec:exp}
Polycrystalline \NMO was synthesized using a solid-state reaction by adding a small amount of ZnO powder in an evacuated quartz tube to initiate the reaction.
We did not observe any impurity peaks related to Zn ions in the powder neutron diffraction data. Single crystals were grown using a chemical vapor transport technique.
Results from powder and single-crystal neutron diffraction analysis were consistent, verifying the quality of our samples (see Appendix~\ref{app:growth} for the details of sample growth).

The temperature dependences of the specific heat, dielectric constant, and pyroelectric currents of single crystals are measured using a physical property measurement system (PPMS; Quantum Design). Magnetic susceptibility was measured using either a superconducting quantum interference device (magnetic property measurement system, Quantum Design) or a vibrating sample magnetometer in a PPMS. For the orientation-dependent susceptibility measurements, the crystallographic axes of single crystals were predetermined via Laue x-ray diffraction
(see Appendix~\ref{app:growth} for the details of measurement methods).

Time-of-flight neutron diffraction measurements of polycrystalline and single-crystal samples were conducted on the WISH (Wide angle In a Single Histogram)~\cite{Chapon2011} and SXD (Single Crystal Diffractometer)~\cite{Keen2006} instruments, respectively, at ISIS, United Kingdom. Nuclear and magnetic refinements were performed using the JANA2006 package~\cite{Petricek2014}. An
absorption correction for the single-crystal data was employed analytically using a multifaceted crystal model~\cite{Clark1993} implemented in the SXD2001 software~\cite{Gutmann:SXD2001,Keen2006}. The observed reflections with intensities
$I > 3.0 \times \sigma(I)$ were used in the refinements for both the powder and single-crystal data; $I$ and $\sigma$ represent the intensity and standard deviation, respectively. Symmetry analysis is performed to determine the candidate magnetic space groups based on
the observed magnetic wave vector $q$~=~(1/2, 0, 0) in the parent hexagonal notation using the Bilbao Crystallographic Server~\cite{Aroyo2011} (see the detailed information in Appendix~\ref{app:growth}).

\begin{figure} [t]
\includegraphics[width=\linewidth]{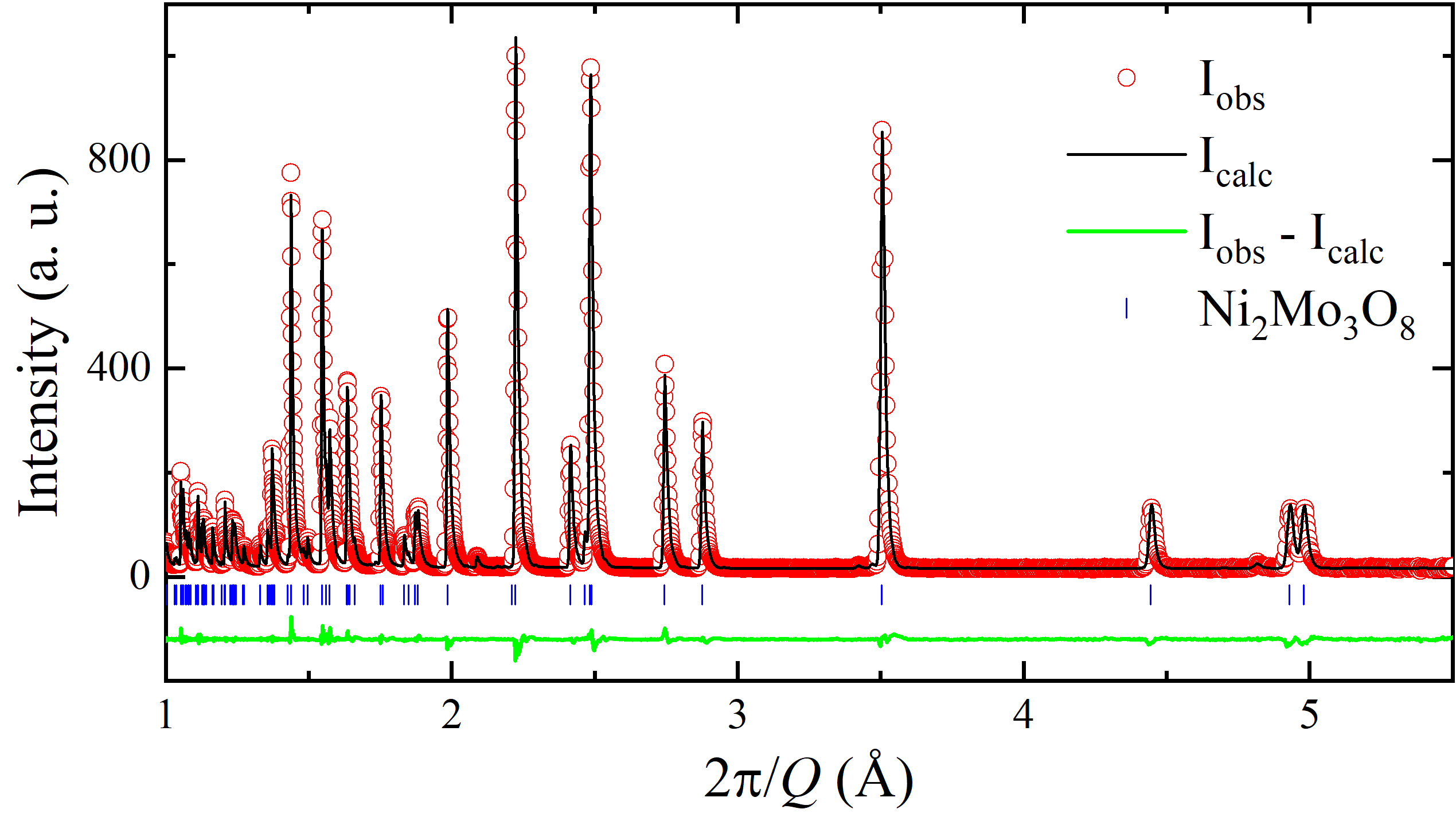}
\caption{Rietveld refinement results for 5.5~K, just above the magnetic ordering temperature.
}
\label{Fig:4}
\end{figure}

\section{Magnetic, thermodynamic, and electric properties}
\label{sec:Chi:Cp:P}
Figure~\ref{Fig:2} presents the magnetic, thermodynamic, and electric properties of \NMO single crystals. Figure~\ref{Fig:2}(a) shows the magnetic susceptibility $\chi(T)$ for $H$~$\vert\vert$~$ab$ and $H$~$\vert\vert$~$c$ measured in an applied magnetic field of 0.1 T.
The large difference in the effective moments indicates significant magnetic anisotropy between the $ab$ plane and $c$ axis, consistent with previous reports~\cite{Morey2019,Tang2021}. $\chi(T)$ exhibits a sharp maximum at $T_\textrm{N}=5.5$~K, signifying the antiferromagnetic long-range ordering. $\chi(T)$ is well reproduced by the Curie--Weiss law with
$\Theta_\textrm{CW}=-124.22$~K and $\mu_\textrm{eff}=3.93$~$\mu_\textrm{B}$ for $H$~$\vert\vert$~$c$, and $\Theta_\textrm{CW}=-49.59$~K and
$\mu_\textrm{eff}=4.02$~$\mu_\textrm{B}$ for $H$~$\vert\vert$~$ab$. The obtained $\Theta_\textrm{CW}$ values are consistent with the previous single-crystal results~\cite{Tang2021}. The large difference between these values again indicates the substantial magnetic anisotropy of \NMO.

Figure~\ref{Fig:2}(c) shows the temperature dependence of the magnetic specific heat divided by the temperature, $C_\textrm{m}/T$. The magnetic specific heat is obtained by subtracting the lattice contribution from the nonmagnetic counterpart, \ZMO.
 $C_\textrm{m}/T$ exhibits a gradual increase followed by a maximum at around $T_\textrm{N}$ as the temperature is lowered.
The magnetic entropy $S_\textrm{m}$ was calculated by integrating $C_\textrm{m}/T$ with respect to temperature. Figure~\ref{Fig:2}(d) shows
that $S_\textrm{m}$ reaches the theoretically expected value $R\textrm{ln}(2S+1)=R\textrm{ln}3$ at 30~K. $S_\textrm{m}$ decreases quite
gradually and releases 53~\% of its value through $T_\textrm{N}$ upon cooling. This indicates significant frustration in the magnetic system.
The dielectric constant $\epsilon$, electric polarization change $\Delta P$ ($P(T)$ – $P(T = 30~K)$), and pyroelectric current density $J$,
shown in Figs.~\ref{Fig:2}(e) and \ref{Fig:2}(f), present sharp maxima (or a kink) at $T_\textrm{N}$. The observed anomalies
in the magnetic, thermodynamic, and electric results corroborate the increase in the polarization associated with the onset
of the magnetic order, confirming strong ME coupling in \NMO.

\begin{figure} [t]
\includegraphics[width=\linewidth]{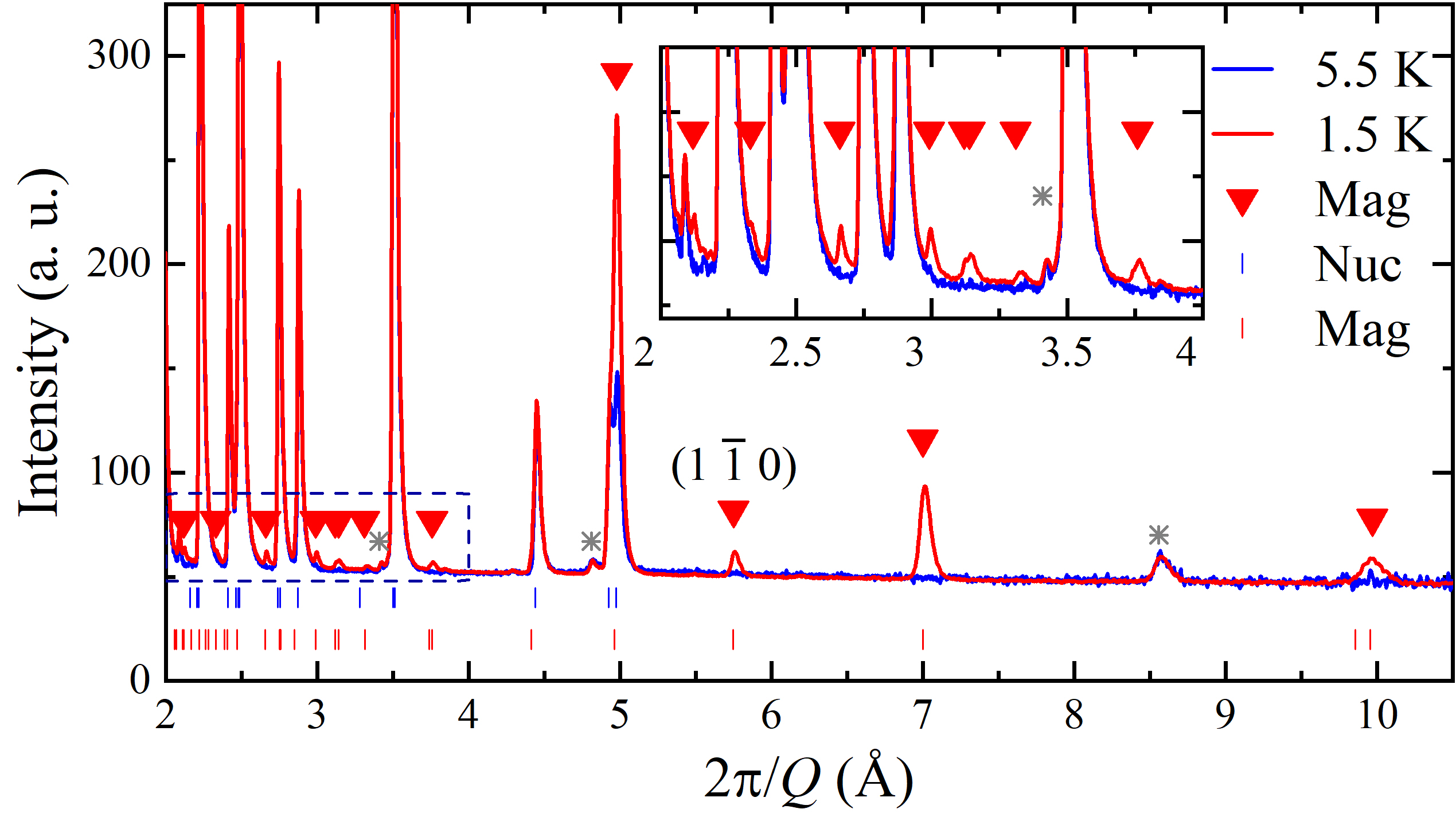}
\caption{Comparison of neutron powder diffraction data at 5.5 and 1.5 K. Red triangles point to the magnetic Bragg peaks.
Asterisk ($\ast$) mark the background peaks observed at both temperatures (e.g., possibly those from the cryostat).
The two data sets were aligned at the high $d$ spacing. Nuc (Mag) refers to the nuclear (magnetic) Bragg peaks.
}
\label{Fig:5}
\end{figure}

\begin{figure} [t]
\includegraphics[width=\linewidth]{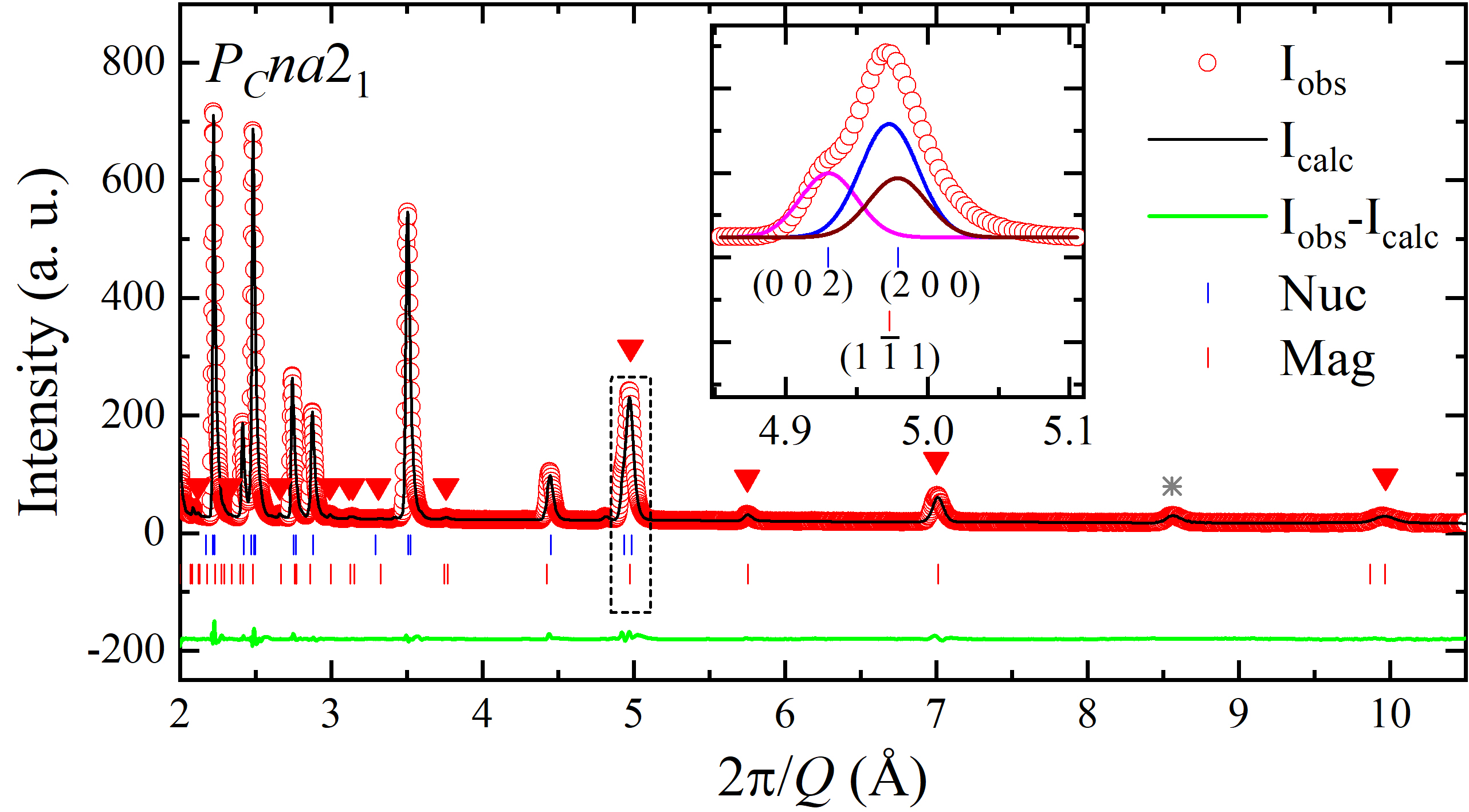}
\caption{Magnetic structural refinement fits for model 1 in Table~\ref{Table:I}. The magnetic space group is \textit{P$_C$na}2$_1$.
The inset shows the resolved (002), (1$\overline{1}$1), and (200) peaks. The solid Gaussian curves are
guides to the eye. The asterisk ($\ast$) marks the background peak; others are not marked for simplicity (see Fig.~\ref{Fig:5} for the rest).
}
\label{Fig:6}
\end{figure}

\begin{table*}
\caption{Magnetic refinement results of the powder neutron diffraction analysis for the magnetic space group \textit{P$_C$na}2$_1$ at 1.5~K.
Our best results are listed as models 1 to 4. We also show our fit results using the published magnetic structures~\cite{Morey2019} (solutions 1 and 2).
Note that solutions 1\# and 2\# are manually modified structures from solutions 1 and 2 to fit our data. GOF means the goodness of fit. Definitions of all reliability factors presented in this paper are given in Table~\ref{DefinitionR}.
}
\vspace{0.2cm}
\label{Table:I}
\setlength\extrarowheight{4pt}
\setlength{\tabcolsep}{2.8pt}
\begin{tabular}{cccccccccccc}
\hline\hline
Condition                                     	& Model	& Atom	& $M_{a}$    & $M_{b}$   & $M_{c}$    &  M          &  GOF   &Rp(All) & wRp(All)  & Rp(Mag) & wRp(Mag) \\\hline
$M_{T}$(Ni1)$>$ $M_{O}$(Ni2)    	& 1           & Ni1   & -2.040(8)  & -1.020(2) & -0.708(11) & 1.904(13) &  5.53   & 3.72    & 3.64        &1.77   & 2.77  \\
(stripy/stripy)                                &              & Ni2   & 1.181(10)  & 0.590(2)  & -0.179(9)  & 1.038(14)                                 & 		\\[0.3cm]

                                                   & 2      & Ni1   & -2.040(8)  & -1.020(2) & -0.100(9)  & 1.769(12)        & 5.52   & 3.72    & 3.64         &1.71   & 2.76  \\
                                                   &        & Ni2   & 1.181(11)  & 0.591(3)  & -0.793(10) & 1.295(15)                                  &		\\\hline

$M_{T}$(Ni1)$<$ $M_{O}$(Ni2)    & 3     & Ni1   & -1.150(10) & -0.575(2) & -0.174(9)  & 1.011(14)       & 5.55    & 3.73    & 3.66        & 1.74  & 2.71  \\
(stripy/stripy)                               &        & Ni2   & 2.047(7)   & 1.024(2)  & -0.708(11) & 1.909(13)                                 &		\\[0.3cm]

                                                   & 4     & Ni1   & -1.158(10) & -0.579(3) & -0.783(10) & 1.272(15)      & 5.56    & 3.74    & 3.66        & 1.72  & 2.78  \\
                                                   &        & Ni2   & 2.043(8)   & 1.021(2)  & -0.106(10) & 1.772(12)                                 & 		\\\hline

Solution 1 (stripy/zigzag)~\cite{{Morey2019}} &        & Ni1   & -1.988     & -0.994    & 0.140      & 1.727       & 6.35   & 4.08  & 4.19   & 2.43  & 3.61  \\
											    &        & Ni2   & 0.946      & 0.473     & -1.172     & 1.430                                     & 		\\[0.3cm]

Solution 2 (stripy/zigzag)~\cite{{Morey2019}} &        & Ni1   & -1.910     & -0.955    & -1.118     & 1.996       & 6.34   & 4.09   & 4.18  & 2.45 & 3.51	  \\
											    &        & Ni2   & 1.028      & 0.514     & 0.042      & 0.891                                     &		\\[0.3cm]	

Solution 1\# (stripy/zigzag)       &        & Ni1   & -2.040     & -1.020    & 0.120      & 1.771       & 5.94   & 3.88   & 3.92      & 1.92    & 2.89  \\
                                                &        & Ni2   & 0.980      & 0.490     & -0.900     & 1.237                                     & 		\\[0.3cm]

Solution 2\# (stripy/zigzag)       &        & Ni1   & -2.040     & -1.020    & -0.900     & 1.983       & 5.62   & 3.76   & 3.71      & 2.08   & 3.05  \\
                                                &        & Ni2   & 1.181      & 0.591     & 0.050      & 1.024                                     & 		\\
\hline\hline
\end{tabular}
\end{table*}

\section{Single-crystal x-ray Diffraction}
\label{sec:sXRD}
Single-crystal X-ray diffraction was performed to determine the crystal structure at room temperature. The crystal exhibited high quality;
more than 98~\% of the detected Bragg reflections originated from a single hexagonal domain (1124 out of 1146 peaks). In our measurements, we have
observed numerous weak reflections that could not be indexed within the published space group of \NMO (as well as other \AMOshort compounds),
\textit{P}6$_3$\textit{mc} (No.~186). One possibility is that the symmetry of \NMO at room temperature is lower.
Trigonal \textit{P}3\textit{m} (No.~156) or \textit{P}3 (No.~143) space groups, for example, would accommodate all the observed reflections.
However, before making such conclusions, one must confirm that the
forbidden reflections do not originate from multiple scattering.

For this purpose, we performed so-called azimuthal scans for a set of
the forbidden and allowed peaks. The sample is rotated around the scattering vector $\bm{Q}$ in the azimuthal scan. The direction and the magnitude
of $\bm{Q}$ are fixed, and the Bragg condition is therefore maintained. In such scans, the intensities of the forbidden peaks should show extreme
variations because the conditions for the multiple scattering are broken and restored as the azimuthal angle changes. On the other hand,
the intensity of the allowed peaks shows little variation (neglecting the absorption effects).

Figure~\ref{Fig:3} shows the evolution of the normalized intensities $I_{i}$ in a wide range of azimuthal angles $\Psi$$_{i}$ for seven
representative reflections. To minimize the extrinsic effects (such as geometry-dependent x-ray absorption), the reflections are chosen to be
nearby in momentum space. This set includes both types of forbidden reflections in the \textit{P}6$_3$\textit{mc} space group,
($H$ $H$ odd) and (0 0 odd), as well as allowed reflections for a cross-check,
based on the reflection conditions for the general Wyckoff site of \textit{P}6$_3$\textit{mc}.
Each raw intensity $I_{i}$ is first divided by the corresponding sigma $\sigma$$_{i}$ to obtain statistically well-defined parameters.
The obtained $I_{i}$/$\sigma$$_{i}$ values are normalized by the average intensity $I_{av}$=$(1/n)\sum_{i=1}^{n} I_{i}$.

Figure~\ref{Fig:3} clearly demonstrates that the intensities of the forbidden reflections vary widely,
whereas the allowed reflections show little variation with the azimuthal angle.
We therefore conclude that the forbidden reflections originate from multiple scattering~\cite{Chang2004} and that the space group of \NMO at room temperature is indeed \textit{P}6$_3$\textit{mc}, a polar group.
We refined the crystallographic structure of \NMO using this space group and found results
very similar to those obtained by neutron diffraction at 5.5 K, described in detail below.

\section{Powder Neutron Diffraction}
\label{sec:NPD}
Neutron diffraction measurements were performed to determine the magnetic order of \NMO. We start with powder diffraction. The bulk
magnetic characteristics of our high-quality polycrystalline samples can be found in the Appendix~\ref{app:pChi} (see Fig.~\ref{Fig:B1}). The nuclear structure was determined first.
The data were collected at 5.5 K, just above the magnetic ordering temperature. The corresponding Rietveld refinement is shown in
Fig.~\ref{Fig:4}. It reveals the same noncentrosymmetric structure (\textit{P}6$_3$\textit{mc}) found at room temperature in
our x-ray measurements, as well as in the literature~\cite{Morey2019, McCarroll1957}, and shown in Figs.~\ref{Fig:1}(a) and ~\ref{Fig:1}(b).
The refined lattice parameters and atomic positions are shown in Table~\ref{Table:C1} in Appendix~\ref{app:pND}.
This result is consistent with a smooth change in the bulk characteristics above the magnetic transition [Fig.~\ref{Fig:2}(f)].
Tiny amounts of impurity phases of NiO, MoO$_{2}$, and NiMoO$_4$ were detected and fitted with Le Bail fits (see Fig.~\ref{Fig:C1} in Appendix~\ref{app:pND}). The vertical bars corresponding to these phases are not shown in Fig.~\ref{Fig:4}
because the mass fractions of these phases were very small (less than 2 \%).

\begin{figure}
\includegraphics[width=\linewidth]{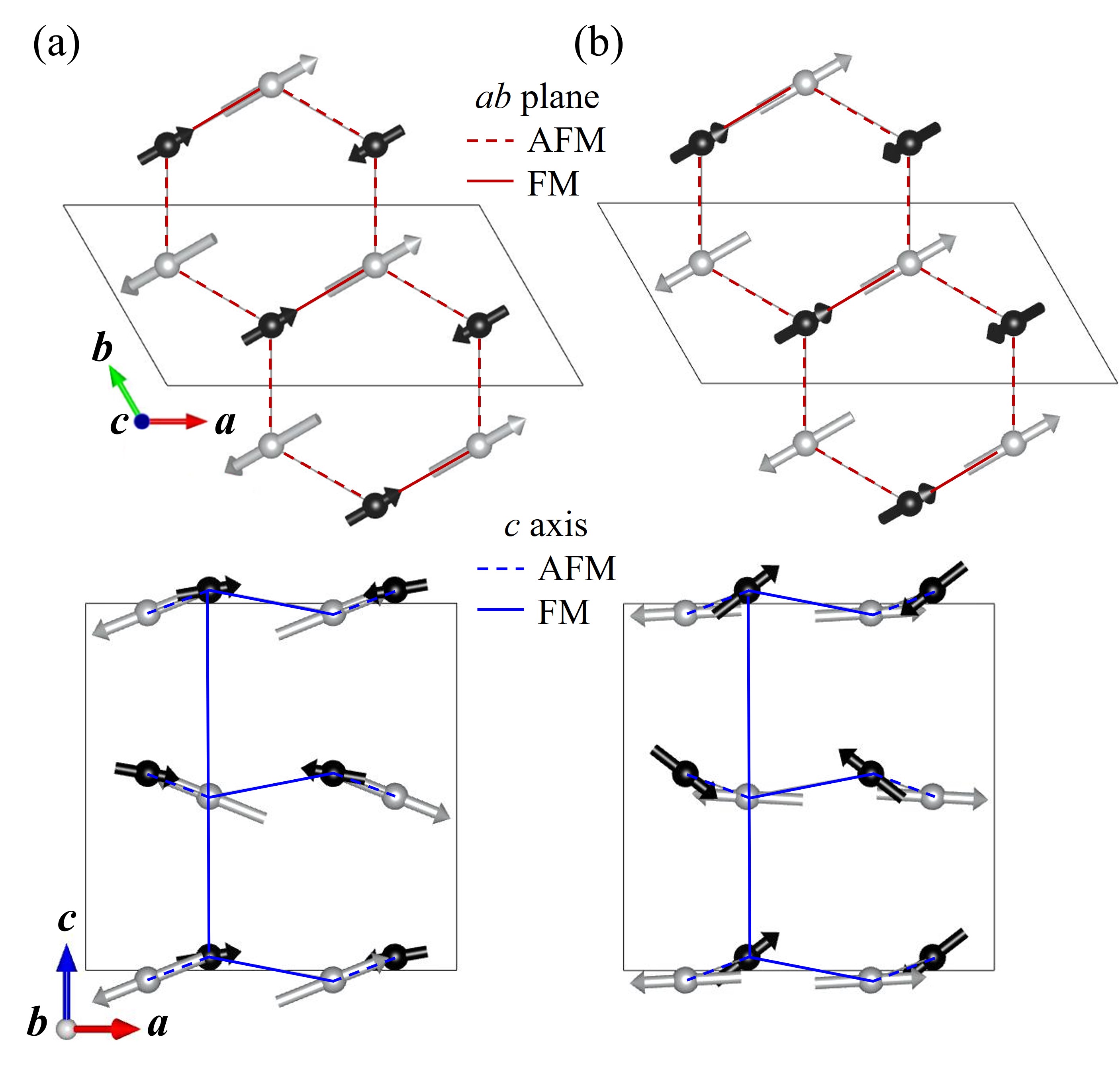}
\caption{Refined magnetic structures of (a) model 1 and (b) model 2 at 1.5~K (Table~\ref{Table:I}, $-$0.1~$\leq$ z $\leq$~0.1).
Black (gray) symbols indicate the octahedral (tetrahedral) Ni sites.
The parallelogram-shaped black solid lines delineate the magnetic unit cell in the nonstandard setting, 2$a$$\times$$a$$\times$$c$,
used in our work. Here $a$ and $c$ represent the crystallographic axes of the high-temperature hexagonal unit cell.
Red and blue lines represent the connection of the $ab$-plane and $c$-axis spin components, respectively.
}
\label{Fig:7}
\end{figure}

For the magnetic structural refinement, the data were collected at 1.5~K. Figure~\ref{Fig:5} shows both the 1.5 K and 5.5 K data for
comparison. Several new peaks appear at 1.5 K, indicating the onset of the long-range magnetic order. These peaks are listed in
Table~\ref{Table:C6}. The magnetic order is indexed with propagation vector $q$ = (1/2, 0, 0).
That is, the high-temperature unit cell is doubled along the $a$ axis, becoming orthorhombic.
In this work, we use a nonstandard magnetic unit cell for the analysis of the magnetic structure for simplicity [shown as a black parallelogram in Fig.~\ref{Fig:7}(a) below].
Group analysis performed using the parent structure with the $q$ = (1/2, 0, 0) magnetic propagation vector yields
four possible maximal magnetic space groups:
\textit{P$_C$mn}2$_1$, \textit{P$_C$na}2$_1$, \textit{P$_C$ca}2$_1$ and \textit{P$_C$mc}2$_1$ (see Fig.~\ref{Fig:C2}
and Table~\ref{Table:C2} in Appendix~\ref{app:pND}).

Among the four candidate groups, \textit{P$_C$mn}2$_1$ and \textit{P$_C$mc}2$_1$ can be excluded because the (1 $\overline{1}$ 0) magnetic
peaks are clearly observed at 1.5 K (see Fig.~\ref{Fig:5}). Indeed, these groups allow only the magnetic moments along the $b$ direction
of the parent unit cell, as shown in Table~\ref{Table:C2} in Appendix~\ref{app:pND}. The (1 $\overline{1}$ 0) wave vector is parallel to the
$b$ axis, and therefore, no magnetic neutron diffraction signal is allowed~\cite{Squires2012} at this \textbf{\textit{Q}}. Magnetic refinements using the
\textit{P$_C$ca}2$_1$ group yielded poor results, as shown in detail in Appendix~\ref{app:pND}.

Magnetic refinements using the \textit{P$_C$na}2$_1$ (No.~33.154) magnetic space group, on the other hand, produced very good results. We therefore conclude that \textit{P$_C$na}2$_1$ is the correct magnetic group for \NMO. We tested various magnetic starting structures, distinguished
by the relative sizes of the Ni$^{2+}$ moments on the tetrahedral and octahedral sites, the dominant in- or out-of-plane spin component, and the magnetic connection type (stripy/stripy or stripy/zigzag). The refinements produced four magnetic structures of similar fit quality.
They are listed as models 1 through 4 in Table~\ref{Table:I}. Table~\ref{Table:I} contains the magnetic connection type, the refined magnetic moment vectors,
and the standard reliability parameters. The obtained quality of the fits is illustrated in Fig.~\ref{Fig:6},
which shows the refinement result for model 1. The inset demonstrates the high resolution of our measurements: the magnetic (1 $\overline{1}$ 1)
peak is clearly resolved from the nearby (2 0 0) and (0 0 2) nuclear peaks.

The refined magnetic structures of models 1 and 2 are shown in Fig.~\ref{Fig:7}. Both structures are noncollinear and of the fully stripy
type; that is, they exhibit the stripy/stripy connection in the honeycomb planes. The $ab$ spin components are quite similar in the two models.
The main difference between models 1 and 2 is in the relative values of the $c$-axis spin components on the tetrahedral and octahedral sites (see the bottom panels in Fig.~\ref{Fig:7}). Models 3 and 4 exhibit the same features, except for the ratio of the tetrahedral to the octahedral Ni$^{2+}$ magnetic moment values.
This ratio is larger than 1 for models 1 and 2; it is smaller than 1 for models 3 and 4.
In other words, the magnitudes of the octahedral and tetrahedral moments
are interchanged as one goes from models 1 and 2 to models 3 and 4.

In addition to $T$~=~1.5~K, neutron diffraction data were collected at other temperatures and analyzed. The magnetic peaks smoothly disappear at the magnetic transition temperature, as expected. The refined values of the Ni$^{2+}$ moments at the two different sites exhibit smooth monotonic curves and go to zero at the transition temperature. These data can be found in Figs.~\ref{Fig:C7} and~\ref{Fig:C8} in Appendix~\ref{app:pND}.

Importantly, all the refinements using the powder data converge to the stripy/stripy structure. This is qualitatively different from previously published results, which presented a complex stripy/zigzag connection in the honeycomb planes~\cite{Morey2019}. We therefore undertook a thorough comparative analysis of our models and the published structures. We could not stabilize the stripy/zigzag solutions~\cite{Morey2019} in the refinement because they always converged to the stripy/stripy structures. We therefore fixed the magnetic connection and the corresponding moment components of solutions 1 and 2~\cite{Morey2019} in the refinements. The corresponding refinement results are much worse than those of our stripy/stripy models in Table~\ref{Table:I}. To find better solutions within the stripy/zigzag structure, we manually tweaked the magnetic moments and found two better candidates, labelled as Solutions 1\# and 2\# in Table~\ref{Table:I}. As one can see from all the reliability factors, these fits are again consistently worse than those of models 1 to 4. This can be seen, in particular, from RFw(Mag), which is known as a good metric to compare how well different models match the same set of experimental data~\cite{Toby2019}.

\begin{figure} [t]
\includegraphics[width=\linewidth]{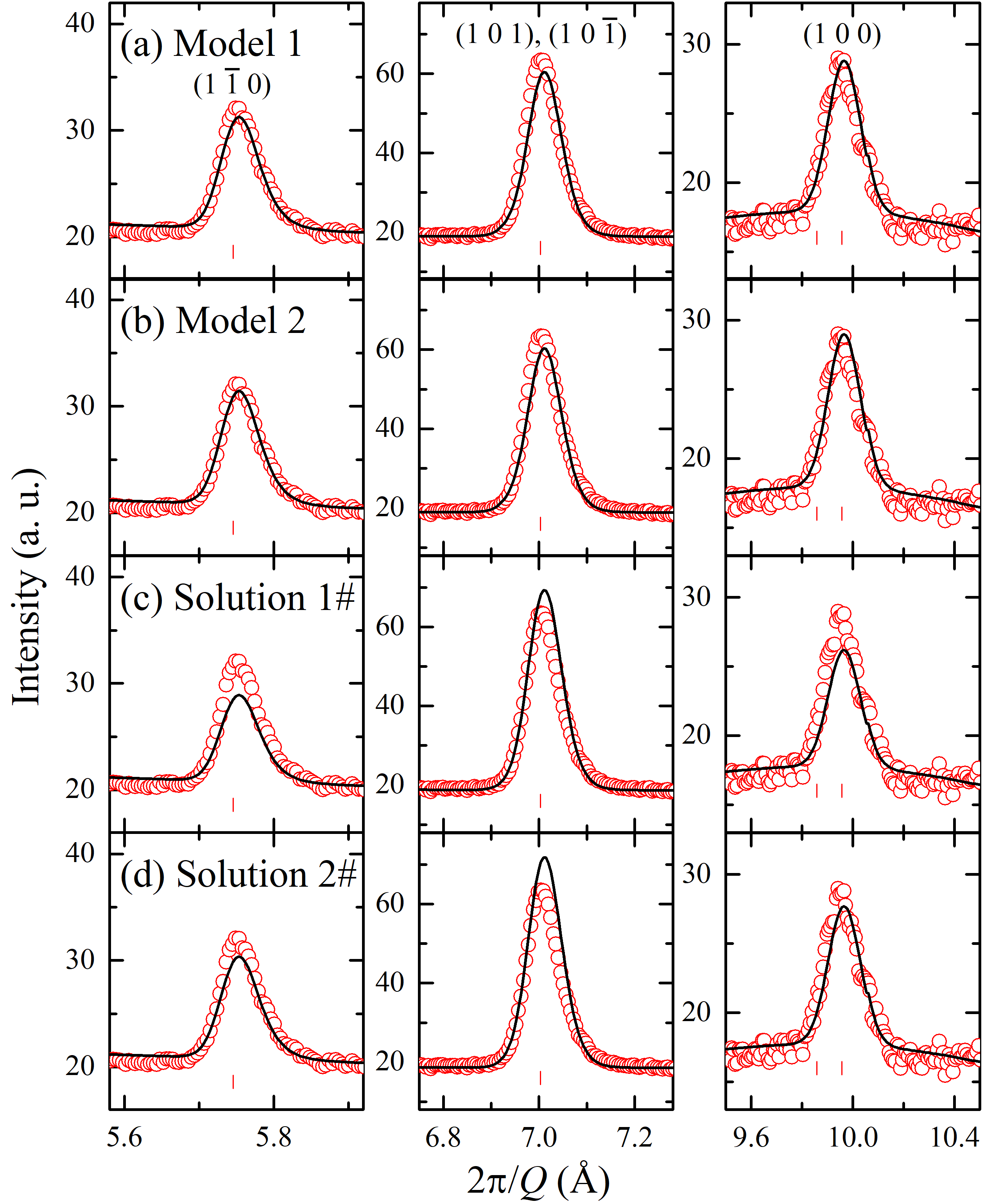}
\caption{Comparison of selected magnetic refinement results at 1.5~K using (a) model 1, (b) model 2, (c) solution 1\#, and (d) solution 2\#.
Note that models 1 and 2 are stripy/stripy and solutions 1\# and 2\# are stripy/zigzag. All plots are based on results in Table~\ref{Table:I} for consistent comparison.
}
\label{Fig:8}
\end{figure}

To illustrate the better fit quality of the stripy/stripy models directly, we compare the obtained fits in the selected ranges of the $d$ spacing (2$\pi$/$Q$) in Fig.~\ref{Fig:8}. The peaks shown in Fig.~\ref{Fig:8} are the strongest magnetic peaks representing the most reliable measurements and having a significant effect on the refinements (see the full range data in Fig.~\ref{Fig:C6}). We emphasize that all the stripy/stripy models in Table~\ref{Table:I} exhibit better fits than those for all the stripy/zigzag solutions. The observed differences between the RF(Mag) and RFw(Mag) of the stripy/stripy and stripy/zigzag structures lie in the 0.11 $-$ 0.8 range. Such differences are accepted as statistically significant in the community~\cite{Ohoyama2005, Balz2017, Duijn2017, Korshunov2020, Willwater2021}. Based on the combined results of the powder diffraction data refinement, we conclude that the magnetic structure of \NMO is of the stripy/stripy type. It is, however, difficult to determine which structures are better among the four models from the powder diffraction data alone.

\begin{table*}
\caption{Magnetic refinement results from single-crystal neutron diffraction analysis at 1.5~K.
The magnetic space group is \textit{P$_C$na}2$_1$. The reliability parameters, R(Nuc) = 4.32 and wR(Nuc) = 5.10, are the same for all magnetic structures presented.
}
\vspace{0.2cm}
\label{Table:II}
\setlength\extrarowheight{4pt}
\setlength{\tabcolsep}{3.pt}
\begin{tabular}{cccccccccccc}
\hline\hline
	                  Condition                    & Model & Atom &  $M_{a}$   &  $M_{b}$   &  $M_{c}$   &     M     & GOF  & R(All) & wR(All) & R(Mag) & wR(Mag)\\ \hline
	        $M_{T}$(Ni1) $>$ $M_{O}$(Ni2)          &   1   & Ni1  & -2.410(47) & -1.205(12) & -0.856(60) & 2.256(77) & 2.75 &  4.60  & 5.58 &  28.53 & 21.19  \\
	               (stripy/stripy)                 &       & Ni2  & 1.199(59)  & 0.599(15)  & -0.167(63) & 1.052(88) &      &        & 		&        &        \\[0.3cm]
	                                               &   2   & Ni1  & -2.406(47) & -1.203(12) & -0.229(74) & 2.096(89) & 2.75 &  4.60  & 5.60 &  28.81 & 21.51  \\
	                                               &       & Ni2  & 1.223(61)  & 0.611(15)  & -0.781(67) & 1.315(92) &      &        & 		&        &        \\[0.3cm]\hline
	        $M_{T}$(Ni1) $<$ $M_{O}$(Ni2)          &   3   & Ni1  & 1.170(57)  & 0.585(14)  & -0.115(55) & 1.020(81) & 2.75 &  4.61  & 5.59 &  29.30 & 21.41  \\
	               (stripy/stripy)                 &       & Ni2  & -2.451(45) & -1.226(11) & -0.897(54) & 2.304(71) &      &        & 		&        &        \\[0.3cm]
	                                               &   4   & Ni1  & 1.203(59)  & 0.601(15)  & -0.817(64) & 1.324(88) & 2.76 &  4.62  & 5.62 &  29.91 & 21.97  \\
	                                               &       & Ni2  & -2.469(45) & -1.234(11) & -0.182(68) & 2.146(83) &      &        & 		&        &        \\[0.3cm]\hline
	Solution 1 (stripy/zigzag)~\cite{{Morey2019}}&       & Ni1  & -1.988     & -0.994     & 0.140      & 1.727     & 2.87 &  4.68  & 5.85 &  34.90 & 26.46  \\
	                                             &       & Ni2  & 0.946      & 0.473      & -1.172     & 1.430     &      &        &	 &        &        \\[0.3cm]
	Solution 2 (stripy/zigzag)~\cite{{Morey2019}}&       & Ni1  & -1.910     & -0.955     & -1.118     & 1.996     & 2.87 &  4.68  & 5.84 &  35.35 & 26.16  \\
	                                             &       & Ni2  & 1.028      & 0.514      & 0.042      & 0.891     &      &        & 	  &        &        \\[0.3cm]
	Solution 1\# (stripy/zigzag)  				&       & Ni1  &   -2.406   &   -1.203   &   0.100    &   2.086   & 2.77 &  4.61  & 5.65 &  29.04 & 22.50  \\
	                                            &       & Ni2  &   1.224    &   0.612    &   -1.050   &   1.492   &      &        & 		&        &        \\[0.3cm]
	Solution 2\# (stripy/zigzag)  				&       & Ni1  &   -2.410   &   -1.205   &   -1.100   &   2.359   & 2.76 &  4.60  & 5.62 &  28.72 & 21.96  \\
	                                            &       & Ni2  &   1.199    &   0.599    &   0.060    &   1.040   &      &        & &        &        \\
\hline\hline
\end{tabular}
\end{table*}

\section{Single-crystal neutron diffraction}
\label{sec:sND}

\begin{figure}
\includegraphics[width=\linewidth]{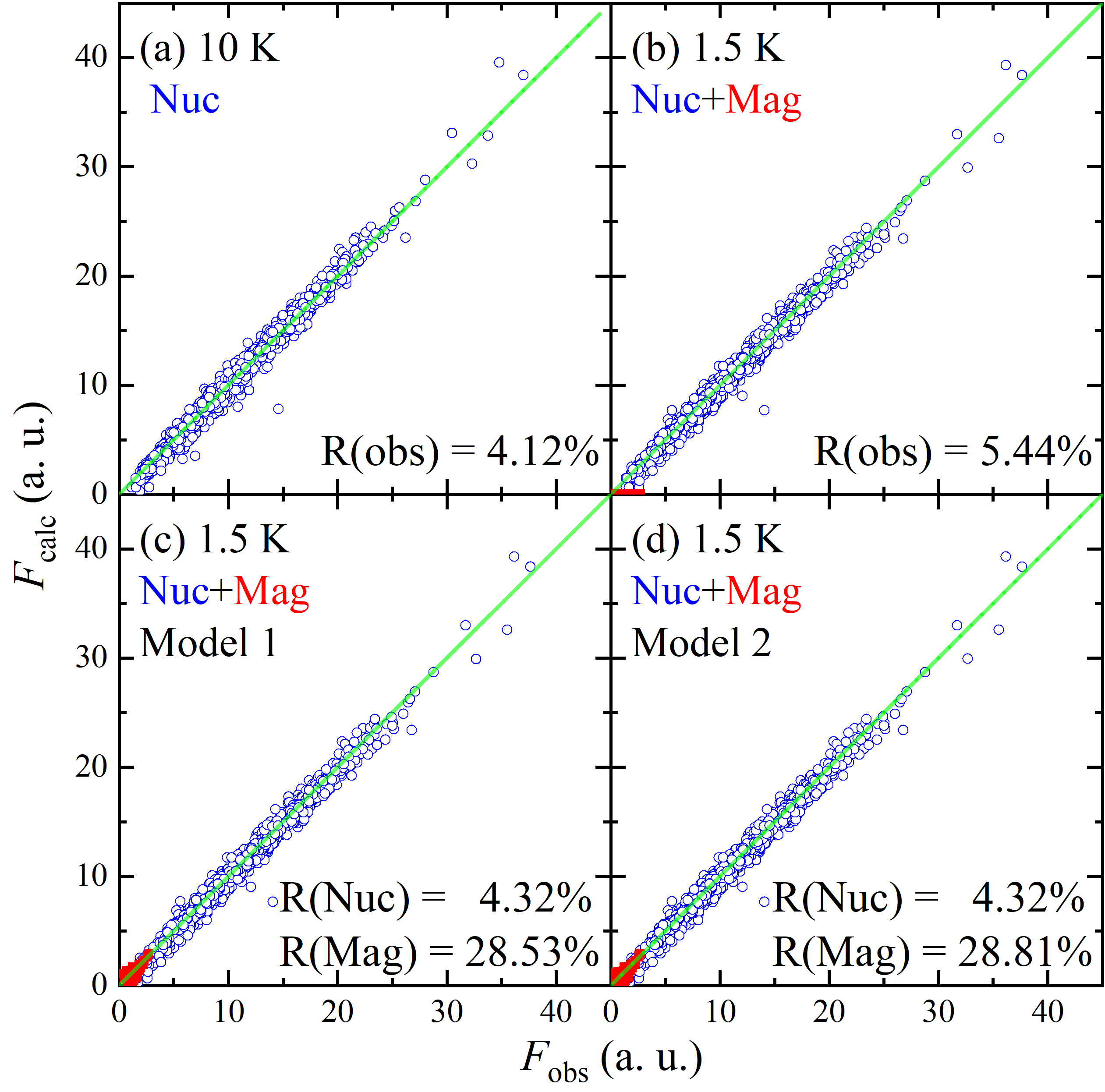}
\caption{Magnetic refinement results for single-crystal neutron diffraction data. $F_{\rm calc}$ and $F_{\rm obs}$ are the calculated
and observed structure factors, respectively. (a) Structural refinement at 10 K in the high-temperature space group. (b) Structural refinement
at 1.5~K in the extended unit cell obtained as described in the text. (c) and (d) Magnetic refinements using models 1 and 2, respectively (see
Table~\ref{Table:II}). Blue and red symbols represent nuclear and magnetic peaks, respectively.
}
\label{Fig:9}
\end{figure}

We performed single-crystal neutron diffraction measurements to reduce the ambiguity left by the powder neutron diffraction analysis. This technique exhibits several advantages, such as good sensitivity to thermal parameters and nonoverlapped Bragg peaks, and is therefore often sensitive to the features inaccessible to powder diffraction. Following the same general approach,
we first collected the data at 10~K and did structural refinements; 2341 nuclear Bragg peaks were fitted [see Fig.~\ref{Fig:9}(a)].
The obtained structural parameters are consistent with the structure at 5.5~K determined using
powder neutron diffraction; compare Tables~\ref{Table:C1} and~\ref{Table:D1} in Appendixes~\ref{app:pND} and~\ref{app:sND}.

At 1.5~K, 229 magnetic Bragg peaks at the wave vector $q$~=~(1/2, 0, 0) in the high-temperature parent cell were found. In the following
refinements, we used the fixed nuclear structural parameters determined at 10 K, as well as the same set of nuclear
peaks for a systematic analysis. Structural refinement was done first using the extended (magnetic) unit cell
2$a$$\times$$a$$\times$$c$. This was done to obtain the accurate scale factor and extinction
parameters for the following magnetic refinements. Figure~\ref{Fig:9}(b) shows
that these parameters describe the nuclear peaks of the 1.5 K data very well, justifying this approach. We then employed models 1 to 4, described in the previous section, to refine the magnetic structure. Three orientational magnetic domains are possible in a single crystal. The
populations of these domains were refined and found to be nearly identical; see Appendix~\ref{app:sND} for the details.

The obtained parameters are listed in Table~\ref{Table:II} for all the models. Reliability parameters, such as GOF, R(All), and wR(All), are very similar in Table~\ref{Table:II} because there is a much larger number of the nuclear Bragg peaks (2341) than magnetic Bragg peaks (229) in magnetic refinements. Thus, R(Mag) and wR(Mag) need to be compared to determine a better magnetic structure in Table~\ref{Table:II}.
We note that R(Nuc) and wR(Nuc) values are all the same in Table~\ref{Table:II}, which ensures reliable and fully-controlled structural calculations during the magnetic refinements in our analysis.

While all these fits are comparable, the magnetic reliability factors, R(Mag) and wR(Mag), of model 1 are better than those of models 2$-$4. The latter models are therefore less favoured by our single-crystal neutron diﬀraction analysis. We note that the differences in the magnetic reliability factors of models 1 and 2 are marginal. The reﬁnement results for models 1 and 2 are shown in Figs.~\ref{Fig:9}(c) and~\ref{Fig:9}(d) for comparison.

We have also tested the previously published magnetic structure~\cite{Morey2019} using our experimental data.
As in the powder diffraction analysis, the published stripy/zigzag initial structures converged to the stripy/stripy order.
Specifically, solution 1 from Ref.~[\onlinecite{Morey2019}] converged to model 2, and solution 2 converged to model 1.
To estimate the goodness of fit differences, we applied the same procedure as in our powder diffraction data analysis.
The magnetic connection and the corresponding moment components in the refinements were fixed.
The results are shown in Table~\ref{Table:II}. As in the powder diffraction measurements, stripy/stripy models fit the experimental data best.

The refined magnetic structures of models 1 and 2 in Table~\ref{Table:II} show very little difference from the corresponding models obtained
in the powder diffraction experiments and shown in Fig.~\ref{Fig:7}. Considering the common features of models 1 and 2,
we conclude that the magnetic structure of \NMO exhibits the stripy/stripy magnetic connection in the honeycomb planes,
and the magnetic moment of tetrahedral Ni$^{2+}$ is larger than that of the octahedral Ni$^{2+}$ site.

\begin{figure} [t]
\includegraphics[width=\linewidth]{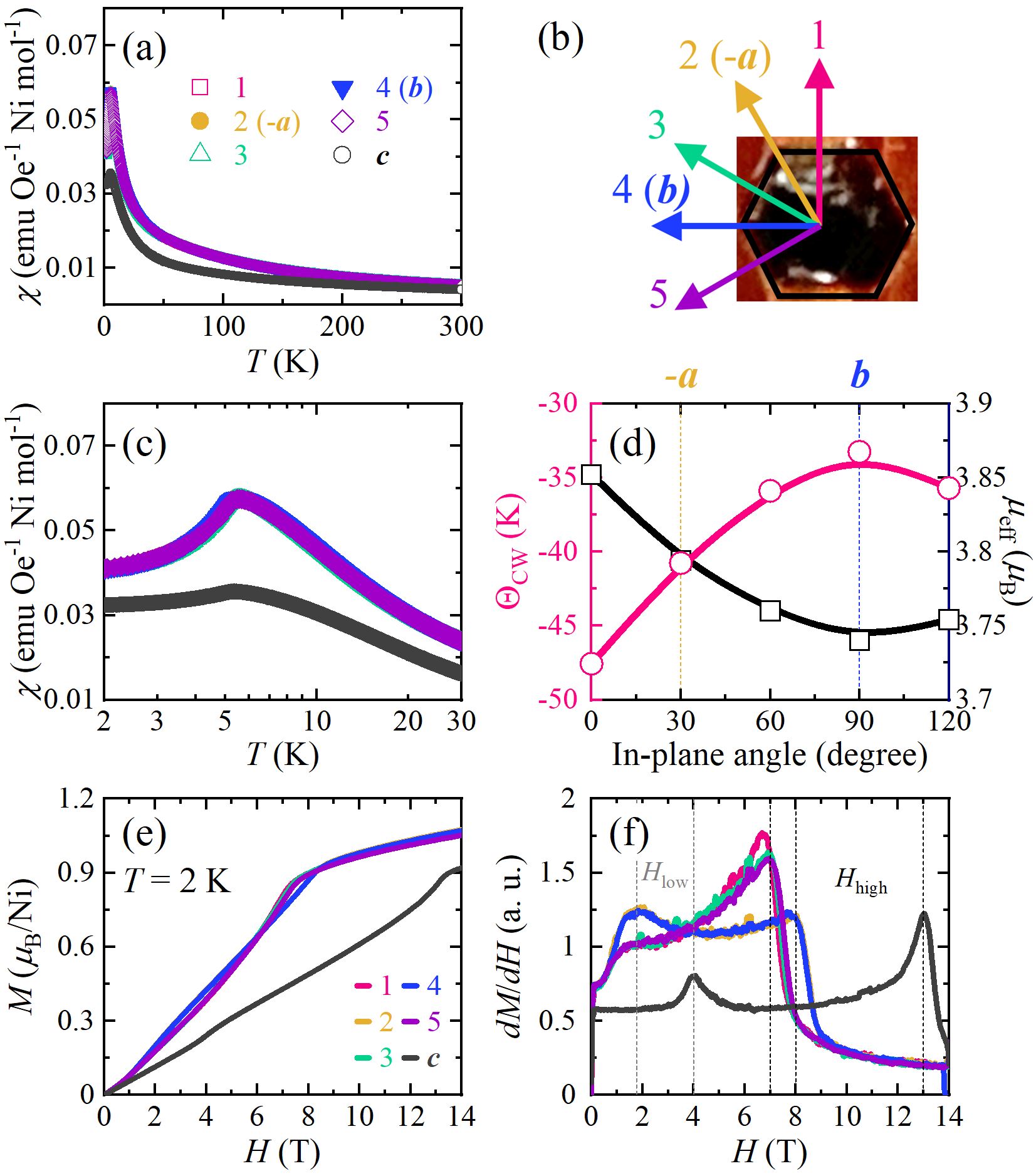}
\caption{Orientation-dependent magnetic susceptibility.
(a) Magnetic susceptibility versus temperature for the different directions of applied magnetic field (0.1 T). (b) The field directions with respect
to the crystal. The corresponding crystallographic axes are described in the text. (c) The low-temperature region of the data
shown in (a). In (a) and (c), the data variation is smaller than the symbol size for the in-plane fields.
(d) Angular dependence of the
Curie--Weiss temperature $\Theta_\textrm{CW}$ (pink circles) and the effective magnetic moment $\mu_\textrm{eff}$ (black squares) extracted from (a).
Dashed vertical lines denote the crystallographic directions $a$ and $b$. (e) Isothermal magnetization at 2~K for various field
directions. (f) The first derivative of the isothermal magnetization at 2~K. Gray and black dashed vertical lines indicate the metamagnetic
transition fields, $H_\textrm{low}$ and $H_\textrm{high}$, respectively.
}
\label{Fig:10}
\end{figure}

\section{Orientation-dependent magnetic susceptibility}
\label{sec:ODChi}
To characterize the magnetic anisotropy of \NMO in the context of its magnetic structure and the observed metamagnetic transitions, we performed
orientation-dependent magnetic susceptibility measurements on a single crystal.
Figure~\ref{Fig:10}(a) shows the magnetic susceptibility for various directions of the applied magnetic field. These directions are described
in Fig.~\ref{Fig:10}(b). Figure~\ref{Fig:10}(c) shows the low-temperature part of the data. The largest effect is the clear easy-plane
magnetic anisotropy, which is consistent with the refined magnetic structure. The $ab$-plane anisotropy is much more subtle but still
observable. Specifically, the Curie--Weiss fits using the higher-temperature data reveal a systematic modulation of the Curie--Weiss
temperature $\Theta_\textrm{CW}$ and effective magnetic moment $\mu_\textrm{eff}$. Figure~\ref{Fig:10}(d) shows that
$\Theta_\textrm{CW}$ reaches its maximum value for $H$~$\vert\vert$~$b$, while the minimum is observed for $H$~$\perp$~$b$.
$\mu_\textrm{eff}$ exhibits the opposite behavior. These systematic tendencies reflect the local magnetic anisotropies and, possibly,
those of the exchange interactions. They appear to be consistent with the direction of the magnetic moments at low temperatures; for instance, the in-plane
components are perpendicular to the $b$ axis. The full analysis should consider the magnetic domain populations and is therefore the subject of future work.

Another systematic trend for the in-plane magnetic anisotropy is found in magnetic field-dependent magnetization. Figure~\ref{Fig:10}(e)
presents the isothermal magnetization $M(H)$ at 2~K for various directions of the external magnetic field. The corresponding differential
magnetization $dM/dH$ is shown in Fig.~\ref{Fig:10}(f). Based on the direction
of $H$, the in-plane data can be classified into two categories. One set of $M(H)$ is for $H$ parallel to the hexagonal axes,
directions 2 ($-$ $a$) and 4 ($b$). The second set includes the directions 1, 3,
and 5, corresponding to [$\overline{2}$ $\overline{1}$ 0], [$\overline{1}$ 1 0], and [1 2 0], respectively.
This difference between the two sets is most clearly observed in the differential magnetization of  Fig.~\ref{Fig:10}(f). The two observed peaks correspond to the metamagnetic transitions first discussed in Refs.\ \cite{Morey2019,Tang2021}.
A systematic broadening and shift of the metamagnetic transition fields, $H_\textrm{low}$ and
$H_\textrm{high}$ [vertically dashed lines in Fig.~\ref{Fig:10}(f)] are found. $H_\textrm{high}$, for example, is larger by approximately
1 T for $H$ // $a$ and $H$ // $b$ (directions 2 and 4) than for the other measured directions. It also appears that the higher-field
transition is broader for directions 2 and 4.

These observations can be qualitatively understood using the refined magnetic structure. The octahedral moments were proposed to be involved
in the low-field transition, while the tetrahedral moment motion was presumed to be dominant in the high-field metamagnetic changes~\cite{Tang2021}.
It is well known that the spin-flop transition becomes sharper when the field direction is closer to the easy-axis direction~\cite{Jacobs1961}.
The broader $H_\textrm{high}$ for $H$ // $a$ and $H$ // $b$ (directions 2 and 4) is then understood in terms of the flop of the
magnetic moments perpendicular to the $b$ axis, as determined in our measurements. This result is valid even if magnetic domains are present
because the principal hexagonal axes are perpendicular to the moment direction. The opposite trend observed for $H_\textrm{low}$ indicates
a different mechanism, such as the establishment of a collinear structure along the $c$ axis by the octahedral moments, as proposed in Ref. \cite{Tang2021}. While the described features are understood within our magnetic models qualitatively, detailed modeling will need measurements of the magnetic exchange parameters. This will require inelastic neutron experiments.

\section{Implications on the magnetoelectric coupling and polarization}
\label{sec:ME}

\begin{figure} [t]
\includegraphics[width=\linewidth]{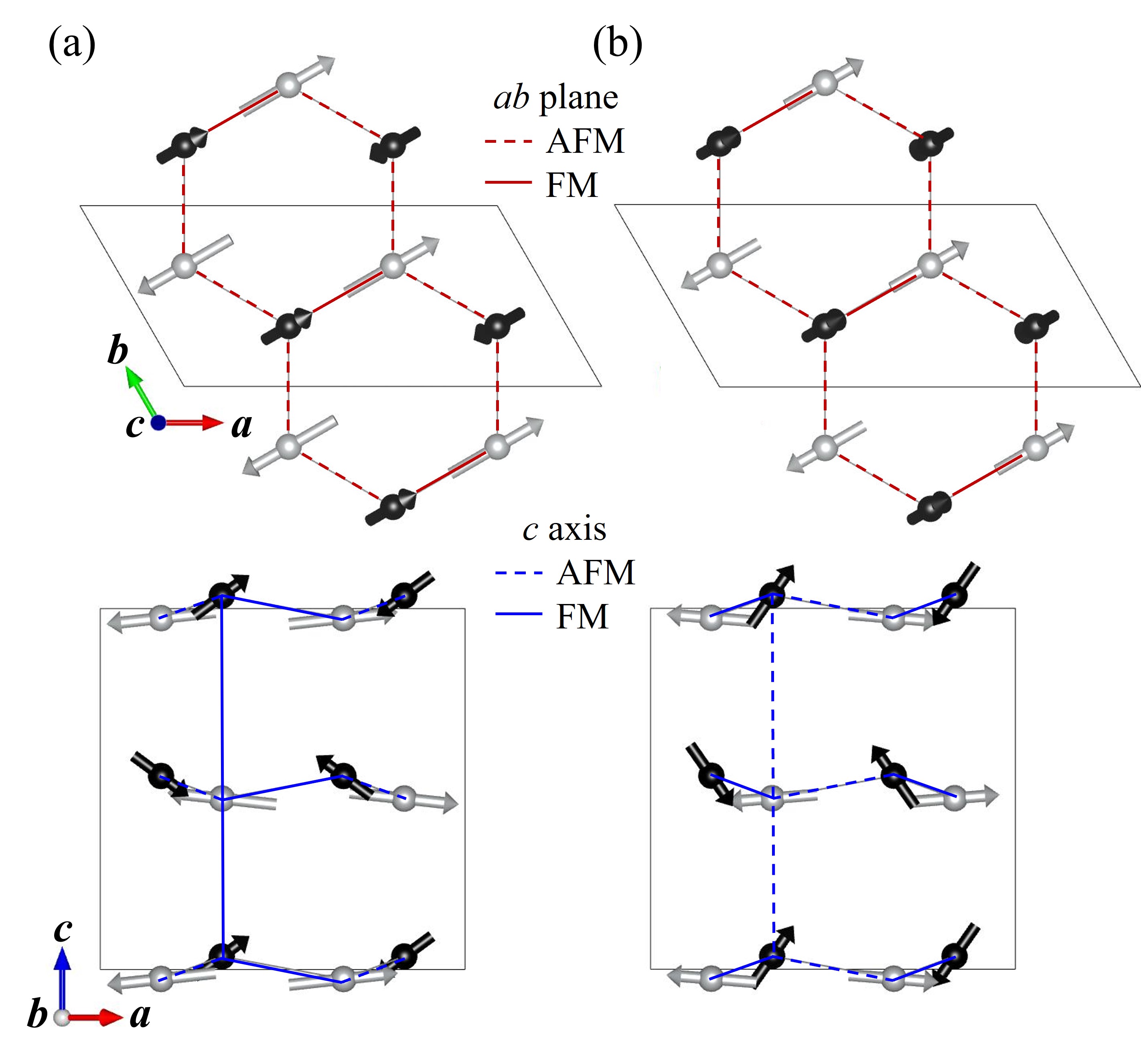}
\caption{Comparison of (a) the stripy/stripy and (b) the stripy/zigzag magnetic structures of \NMO.
The stripy/stripy structure is shown using the refined model 2 for the single crystal obtained in our experiments.
The stripy/zigzag structure is represented by the corresponding solution 1 directly from Ref.\ \cite{Morey2019}.
Red and blue lines represent the connection of the $ab$-plane and $c$-axis spin components, respectively.
}
\label{Fig:11}
\end{figure}

All the refinements of our neutron diffraction data, powder and single crystal, produce the stripy/stripy magnetic connection in the honeycomb
planes. That is, all the components of the Ni$^{2+}$ spins exhibit the stripy pattern.
Previous experiments~\cite{Morey2019} claimed the stripy/zigzag connection, in which the $c$-axis spin components order in the zigzag structure in
the honeycomb planes.
In addition, the interlayer connection of the $c$-axis spin components in our model is ferromagnetic, while it is antiferromagnetic
in the published solutions.
The two magnetic structures
are compared in Fig.~\ref{Fig:11}, using model 2 from our experiments, and the corresponding solution 1 from Ref.~\cite{Morey2019}.
As discussed above, all the starting structures of the stripy/zigzag type converged to the stripy/stripy ones for both single-crystal and powder data. We are therefore confident that the stripy/stripy structure is realized in \NMO.

We now discuss the implication of our results for the magnetoelectric coupling. \NMO exhibits significant changes in the electric polarization
with the onset of the magnetic order [Fig.~\ref{Fig:2}(f)]. The polarization also changes in the applied magnetic field,
especially at the metamagnetic transitions~\cite{Tang2021}.
This indicates a strong coupling between the magnetic and electric order parameters.

Several mechanisms of such coupling are known. They include the spin-current mechanism that utilizes the Dzyaloshinskii-Moriya (DM) interaction
and works in noncollinear systems~\cite{Katsura2005}, the exchange striction that works best in the collinear systems~\cite{Sergienko2006},
and the $p$-$d$ hybridization between spin and ligand ions mediated by the spin-orbit coupling~\cite{Arima2007} which works for various spin
orders. The type of the magnetic order clearly plays a key role in these mechanisms.
The ME properties of \NMO have been recently examined in the framework of these models~\cite{Tang2021} using a general symmetry-based ME tensor
approach~\cite{Xiang2011, Chai2018}. The stripy/zigzag magnetic order was assumed in those studies. This analysis needs to be revised
based on the stripy/stripy magnetic order found in our experiments.

The magnitudes of the terms in the local ME tensor are set by the specific spin arrangement, spin magnitudes, and type of the ME coupling.
The global polarization is the summation of the local polarizations generated in the bonds between the Ni ions and in the single-ion~\cite{Tang2021}
\begin{equation}
p=\sum_{<i,j>} p_{ij}=\sum_{<i,j>} P_{ij}^{\alpha \beta \gamma} S_{i}^{\alpha} S_{j}^{\beta},
\label{ME1}
\end{equation}
where $P_{ij}^{\alpha \beta \gamma}$ are the local ME tensor and
$(\alpha, \beta, \gamma)$ correspond to the Cartesian orthorhombic coordinate components, which are (${\bm x}$, ${\bm y}$, ${\bm z}$) in the magnetic phase;
they are defined as ${\bm x}=2{\bm a}+{\bm b}$, ${\bm y}={\bm b}$, ${\bm z}={\bm c}$, respectively. $S_{i}^{\alpha}$ ($S_{j}^{\beta}$) is the
$\alpha$ ($\beta$) spin component for the $i$ ($j$) site. $<$$i,j$$>$ runs over the Ni--Ni bonds generating the polarization. For $i=j$,
single-spin terms are generated in Eq.~(\ref{ME1}), which therefore combines the two-spin and single-spin ME tensors.

The allowed terms in the $P_{ij}^{\alpha \beta \gamma}$ tensor of \NMO~\cite{Tang2021} are restricted by the point group symmetry ($mm2$)~\cite{Chai2018}
of the experimentally determined magnetic space group (\textit{P$_C$na}2$_1$) and is given by
\begin{equation}
P_{ij}^{\alpha \beta \gamma}=\left[\begin{array}{lll}
(0,0,P_{ij}^{xx(z)})     ~& (0,0,0)                     ~&  (P_{ij}^{xz(x)},0,0) \\
(0,0,0)                      ~&  (0,0,P_{ij}^{yy(z)})   ~&  (0,P_{ij}^{yz(y)},0) \\
(P_{ij}^{zx(x)},0,0)     ~&  (0,P_{ij}^{zy(y)},0)   ~&  (0,0,P_{ij}^{zz(z)}) \\
\end{array}\right],
\label{ME2} % \tag{S3}
\end{equation}
where the coordinates inside the parenthesis give the direction of the local polarization and the other two coordinates refer to the spin components.
As there are no magnetic moment components along the $\bm y$ axis in the magnetic order of \NMO, Eq.~(\ref{ME2}) is
simplified to
\begin{equation}
P_{ij}^{\alpha \beta \gamma}=\left[\begin{array}{lll}
(0,0,P_{ij}^{xx(z)})     ~& (0,0,0)                     ~&  (P_{ij}^{xz(x)},0,0) \\
(0,0,0)                      ~&  (0,0,0)   ~&  (0,0,0) \\
(P_{ij}^{zx(x)},0,0)     ~&  (0,0,0)   ~&  (0,0,P_{ij}^{zz(z)}) \\
\end{array}\right],
\label{ME3} % \tag{S3}
\end{equation}
which consists of two diagonal and two off-diagonal components. Accordingly, $p_{ij}$ in Eq.~(\ref{ME1}) is expressed as
\begin{equation}
p_{ij} = \\
                                               P_{ij}^{xx(z)} S_{i}^{x} S_{j}^{x} +
                                               P_{ij}^{xz(x)} S_{i}^{x} S_{j}^{z} +
                                               P_{ij}^{zx(x)} S_{i}^{z} S_{j}^{x} +
                                               P_{ij}^{zz(z)} S_{i}^{z} S_{j}^{z}.
\label{ME4}
\end{equation}
Here $p_{ij}$ gives the polarization of the corresponding Ni--Ni bond for the two spins ($i \neq j$), or the single spin ($i=j$).
Since the magnetic space group is the same as the one analyzed in Ref.~\cite{Tang2021},
the conclusions given there also apply to our structure. Both the spin current (from the noncollinear spins along the ${\bm x}$
axis in the ${\bm {xz}}$ plane) and the $p$-$d$ hybridization (the single-ion terms) mechanisms may contribute to the observed
ME effect in \NMO. While the same two mechanisms are allowed for the previously published and revised magnetic structures,
the signs of the components in Eq.~(\ref{ME4}) differ for the two structures. In Eq.~(\ref{ME4}), all the signs except the one for the first term are the opposite. This can be seen in Fig.~\ref{Fig:11}, which shows the connections between the $ab$-plane and $c$-axis spin components for these structures. The sign of the fourth term involving $P_{ij}^{zz(z)}$ is important for understanding the experimentally measured polarization along the $c$ axis.

% The sign of the fourth term involving $P_{ij}^{zz(z)}$ could be of particular importance in interpreting the experimentally measured polarization along the $c$ axis. It is because the sign of the $S_{i}^{z} S_{j}^{z}$ term in Eq.~(\ref{ME4}) is opposite in the two structures. This is clearly compared in the lower panel of Figs.~\ref{Fig:11}(a, b). The connection of the $c$-axis moments is opposite in the two structures in both the $ab$ plane and along the $c$ axis.

For the full analysis of the microscopic interactions underlying the ME effect, one needs to know all the terms in the local ME tensor. So far, experiments have been done only for the $c$-axis polarization direction. Importantly, the signs of the ME tensor terms should be determined correctly. This involves the determination of the absolute direction of the polar $c$ axis. This has not been done so far, and therefore, the published changes of the $c$-axis polarization,
in fact, give only the absolute values. A detailed study of the ME tensor in \NMO, including the measurement of $P_{\bm {x}}$ in Eq.~(\ref{ME4}),
is highly desired for a better understanding of the microscopic interactions underlying the ME effect. Combined with the magnetic structure determined here,
it should lead to a more comprehensive understanding of the microscopic origin of the peculiar ME properties of \NMO.

\section{Discussion}
\label{sec:discussion}
Our measurements show that the magnetic structure of \NMO is noncollinear, of the stripy/stripy type in the honeycomb planes, the ferromagnetic along the $c$ axis,
and that the magnetic moment M$_T$ of the tetrahedral Ni$^{2+}$ site is larger than the moment M$_O$ on the octahedral site.
The numerical values of the ratio, M$_T$/M$_O$, vary in a certain range, reflecting the systematic errors of the methods used,
such as electron spin resonance (ESR) measurements~\cite{Morey2019} and powder neutron diffraction~\cite{Morey2019}.
Specifically, M$_T$/M$_O$ values determined in our neutron experiments for model 1 are 1.83 and 2.14 for the powder and single-crystal experiments, respectively.
The corresponding values for model 2 are 1.37 and 1.59. On the other hand, the ESR measurements give M$_T$/M$_O$=1.78,
and powder neutron diffraction measurements from the literature give 1.21 and 2.24 for solutions 1 and 2, respectively. A comparison of the ESR and the neutron results does not therefore favor either of the two models.
We note that the ratios obtained from our neutron diffraction analysis are at least consistent with those from powder neutron diffraction refinements of the reference
(see Table~\ref{Table:D5} for a detailed comparison).

The observed dominant easy-plane magnetism can be explained in the framework of crystal field theory.
Reference~\cite{Li2022} presented a theoretical model that reproduces the dominant in-plane directions of the magnetic moments.
The calculated single-ion anisotropies are different for the tetrahedral and octahedral Ni sites.
The observed difference in the corresponding magnetic moment values also obviously results from the different Ni environments.
A more detailed analysis requires experimental measurements of Ni crystal field levels by inelastic neutron or Raman spectroscopy.

The noncollinear magnetic order found in \NMO is rather complex, and the spin canting angles out of the $ab$ plane are large. A recent
theoretical work~\cite{Li2022} introduced the spin Hamiltonian containing, in addition to the Heisenberg terms, bilinear interactions, single-ion anisotropy,
and the DM interaction. It was proposed that the DM interaction is responsible for the spin canting.
However, given the large observed values of the spin canting angles, anisotropic magnetic interactions
and local anisotropies may also need to be considered to explain the observed noncollinear order.
Inelastic neutron scattering measurements on single crystals will eventually be required to determine the spin Hamiltonian experimentally.

Finally, \NMO is a member of the compound family in which exotic phenomena as topological magnons, nonreciprocity,
and unusual Hall effects are either observed or expected. Theoretical analysis of the revised magnetic structure of \NMO is needed to establish
whether they should be sought in this compound. The \AMOshort compound family provides many intriguing opportunities for further study.
The presence of two distinct magnetic sites is especially useful because it makes selective doping possible, creating various sublattices in the honeycomb structural motif.
Nonmagnetic Zn doping is already known to create axion-type coupling, magnetic transitions, and diagonal magnetoelectricity in \FZMO~\cite{Kurumaji2017:FZMO, Ideue2017, Yu2018}.
However, studies of doping effects in other compounds of this family are scarce. \NMO is a unique member of this series, showing a rather distinct magnetism.
This calls for examining the doping effects in this compound, for instance, in \NMMO and \NZMO.

\section{Conclusions}
\label{sec:conclusions}
The magnetic structure of the pyroelectric honeycomb antiferromagnet \NMO was determined using combined powder and single-crystal neutron diffraction.
The structure is noncollinear. The magnetic moment of Ni$^{2+}$ on the tetrahedral site is significantly larger than the moment on the octahedral site.
The magnetic order in the honeycomb planes is of the stripy type for all the spin components, which is different from the previously proposed stripy/zigzag connection.
The ferromagnetic interlayer order of the c-axis spin components in our model is also distinct.
In our study, the referential magnetic orders converge into our models for both the powder and the single-crystal data.
In single crystals, we found a subtle, but clearly detectable, magnetic anisotropy in the honeycomb planes, using orientation-dependent magnetic susceptibility measurements.
It manifests through the systematic modulation of the Curie--Weiss temperature and the effective magnetic moment and through variation of the magnetic fields at which spin-flop transitions occur,
consistent with the neutron diffraction analysis. Using our magnetic model, we revised the analysis of the magnetoelectric tensor in \NMO. Our results provide key input for future studies of the Ni-based compounds of the \AMOshort family, in which topological excitations, nonreciprocity, unusual magnetoelectricity, and other exotic effects are under investigation.

\section*{Acknowledgments}
We thank P. Manuel and F. Orlandi for helping with powder and single-crystal neutron diffraction experiments on WISH.
This work was supported by the Institute for Basic Science (IBS-R011-Y3) and Advanced Facility Center for Quantum Technology at Sungkyunkwan University.
Part of this study has been performed using facilities at IBS Center for Correlated Electron Systems, Seoul National University.
V.K. was supported by the NSF, Grant No. DMR-2103625. The work at Rutgers University was supported by the DOE under Grant No. DOE: DE-FG02-07ER46382.
G.L.P.’s work were supported by a grant of the Romanian Ministry of Education and Research, CNCS UEFISCDI, Project No. PN-III-P1-1.1-TE-2019-1767, within PNCDI III.

\bibliography{Manuscript}

\appendix

\section{Growths and characterizations}
\label{app:growth}
Polycrystalline \NMO is synthesized using a solid-state reaction. High-purity NiO:ZnO:Mo:MoO$_{3}$ powders (1.9:0.1:1:2) were mixed and pelletized. Five percent nonmagnetic ZnO is required to initiate the reaction; pure \NMO polycrystalline samples cannot be grown without it in the conventional solid-state reaction. The pellet is prepared and sealed in a vacuum quartz tube and sintered at 900~$^{\circ}$C, 1000~$^{\circ}$C, and 1100~$^{\circ}$C for 5, 5, and 10 h, respectively, with several intermediate grindings to improve the powder quality.

Single crystals of \NMO are grown using a chemical vapor transport method. Three grams of NiO:Mo:MoO$_{3}$ = 2:1:2 mixture powder are placed in a vacuum quartz tube (diameter = 15 mm, length = 200 mm). Then, 0.1 g TeCl$_{4}$ is added as a transport agent. The quartz tube is placed in a tube furnace with a hot zone at 1000~$^{\circ}$C and a cold zone at 850~$^{\circ}$C for three weeks. Single crystals with masses of approximately 20 mg are observed in the cold zone.

Magnetic susceptibility measurements are performed using superconducting quantum interference device magnetometry between 300 or 350~K and the base temperature (2 or 3 K) under 0.1~T applied magnetic field with a Quantum Design MPMS-XL7 and PPMS-14 [Figs.~\ref{Fig:2}(a) and \ref{Fig:2}(b)]; zero-field-cooled and field-cooled measurements are performed when necessary. The temperature dependence of the dielectric constant, electric polarization, specific heat, and magnetic susceptibility measurement of single crystals are measured using a PPMS-9 [Figs.~\ref{Fig:2}(c)--\ref{Fig:2}(f)], along with the magnetic susceptibility measurement of polycrystalline powder (Fig.~\ref{Fig:B1}). Orientation-dependent magnetic susceptibility measurements (Fig.~\ref{Fig:10}) are performed in the temperature range of 2 - 300 K under an applied magnetic field of 0.1 T using a PPMS-14 equipped with a vibrating sample magnetometer. Magnetization data are collected at 2 K up to 14 T by rotating the field direction by 30$^{\circ}$ within the $ab$ plane, as well as along the $c$ axis.

A time-of-flight neutron diffraction measurement on polycrystalline samples is conducted on WISH at ISIS, United Kingdom, to determine nuclear and magnetic structures and examine a magnetic transition. Multibank detectors are utilized at different scattering angles; 2.54 g of the powder sample are packed in a cylindrical vanadium can (diameter = 6 mm, height = 26 mm). The entire vanadium cell is placed in a standard cryostat. The data are collected from 1.5~ to 10~K with a step of 0.5~K going through an antiferromagnetic transition temperature at around 5~K, followed by a measurement at 100~K. Longer measurements at 1.5 and 10 K are conducted for 1 h to obtain high-quality data for detailed refinements.

Single-crystal neutron diffraction experiments are performed on SXD at ISIS, United Kingdom, where the time-of-flight Laue technique is used to access large three-dimensional volumes of reciprocal space in a single measurement. The data are collected by rotating mounted single crystals. We measured the data at two rotation angles at 10 K (the paramagnetic state) and two identical rotation angles at 1.5 K (a magnetically ordered phase) and added three more rotation angles at 1.5 K to increase the statistics. One rotation of the data is collected for 6.5 h.

Nuclear and magnetic refinements are performed using the JANA2006 package~\cite{Petricek2014}. As extinction correction is key for reliable refinements using single-crystal neutron diffraction data, we test all isotropic extinction models implemented in the JANA2006 and use the best model for the refinements; that is, the Becker \& Coppens model (a mixed Lorentzian model)~\cite{Petricek2014,Becker1974} was used for structural refinements (unless other- wise specified in this paper) at the paramagnetic temperature, which was consistently used for magnetic refinements at lower temperatures. We confirmed a single structural domain for measured single crystals and three nearly equally populated magnetic domains in the magnetic phase.

\section{Powder magnetic susceptibility}
\label{app:pChi}

\begin{figure} [t]
\includegraphics[width=\linewidth]{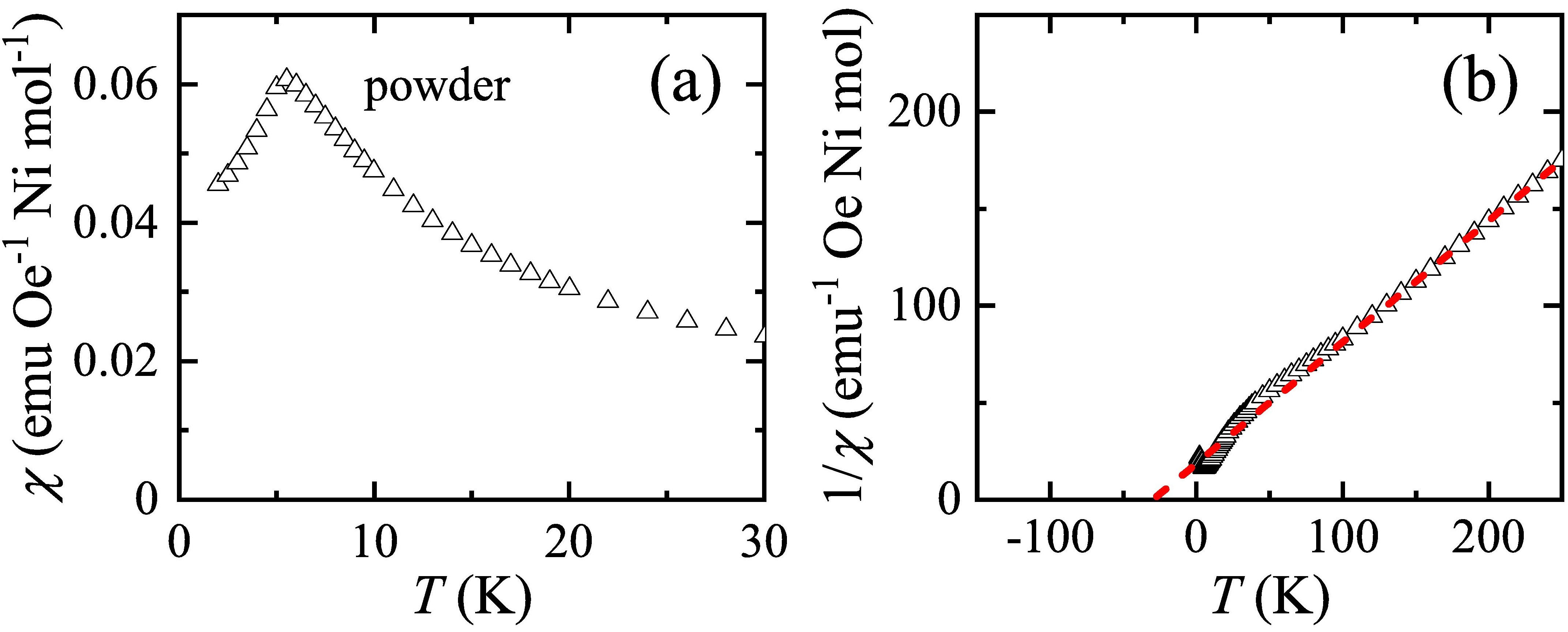}
\caption{(a) Magnetic susceptibility of the polycrystalline sample \NMO used for neutron diffraction ($H=0.1$~T). (b) Inverse magnetic susceptibility. A dashed red line notes the Curie--Weiss fit.
}
\label{Fig:B1}
\end{figure}

Figure~\ref{Fig:B1} shows the temperature-dependent magnetic susceptibility of the polycrystalline sample at $H=0.1$~T. We used this powder for the powder neutron diffraction measurement presented in Sec.~\ref{sec:NPD}. A tentative antiferromagnetic transition from Fig.~\ref{Fig:B1}(a) at approximately 5.5~K is confirmed by powder neutron diffraction, as presented in Fig.~\ref{Fig:C8} and Sec.~\ref{sec:NPD}. From the fits using the Curie--Weiss law, we obtain $\Theta_\textrm{CW}=-29.8(6)$~K and $\mu_\textrm{eff}=3.57$~$\mu_\textrm{B}$, which is somewhat smaller than previously reported values~\cite{Morey2019,Tang2021}. The powder susceptibility can be estimated from the single-crystal data (Fig.~\ref{Fig:2}) as $\chi_\text{powder}$~=~(2$\chi_{ab}$~$+$~$\chi_{c}$)/3, which is located between $\chi_{ab}$ and $\chi_{c}$ (not shown). We note that a larger susceptibility of the powder sample measured could be attributed to the instrumental difference in the PPMS (powder; Fig.~\ref{Fig:B1}) and the MPMS-XL7 [single crystals; Figs.~\ref{Fig:2}(a) and \ref{Fig:2}(b)], different diamagnetic backgrounds, or the sample dependence between the polycrystalline powder and single crystal.

\section{Powder neutron diffraction}
\label{app:pND}
This section provides details for powder neutron diffraction analysis. In the reduction of the data, a pair of bank data (in the left and right direction of the scattering) is summed to increase the signal-to-noise ratios of the peaks; they are labeled as bank 1 to bank 5 detectors in this paper (2$\theta$ = 27.08$^{\circ}$, 58.33$^{\circ}$, 90$^{\circ}$, 121.66$^{\circ}$, and 152.83$^{\circ}$ from bank 1 to bank 5, respectively) with increasing order of the scattering angle.

\begin{figure} [t]
\includegraphics[width=\linewidth]{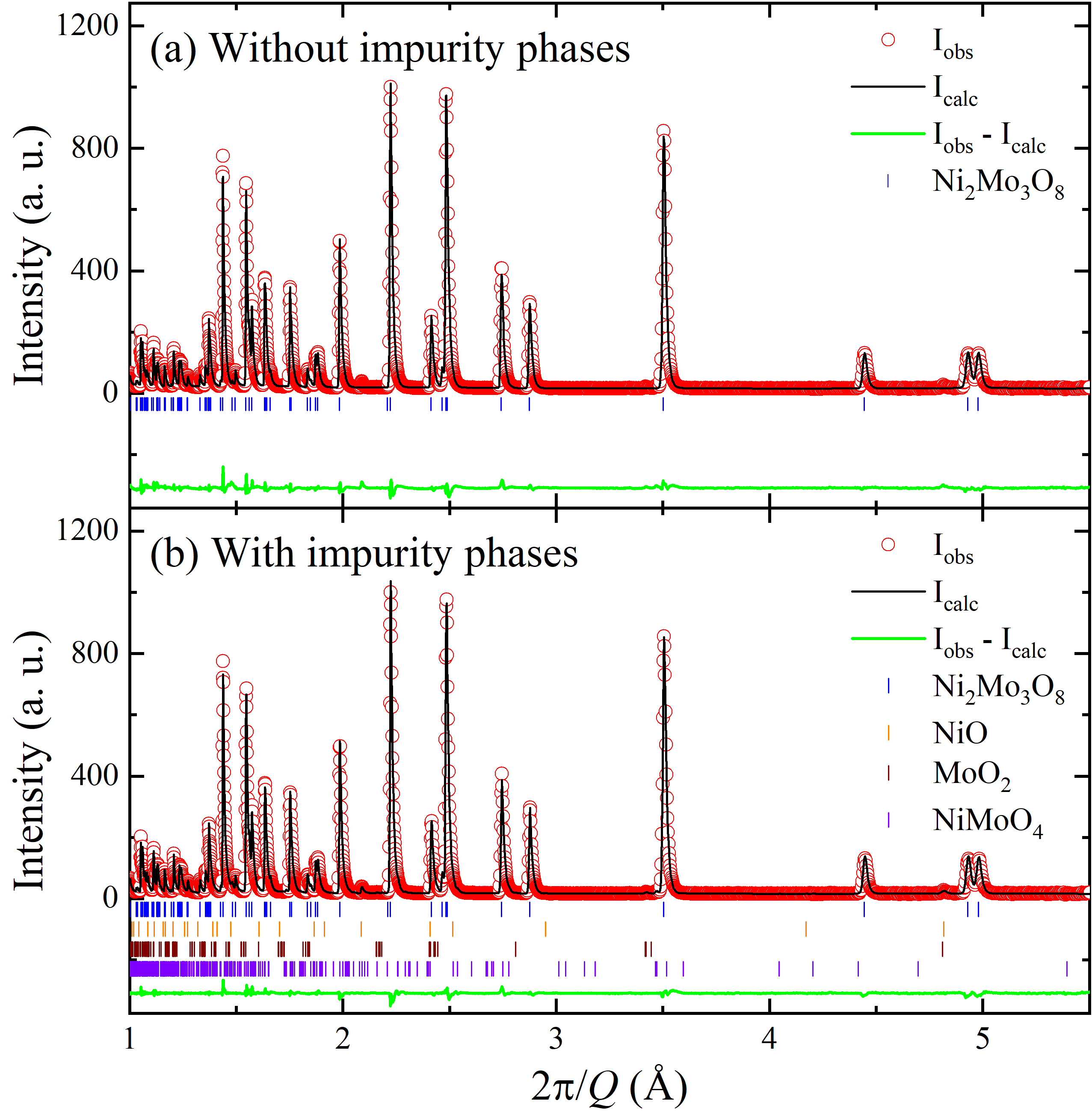}
\caption{Structural refinements (a) without and (b) with three impurity phases with the neutron diffraction data collected at 5.5~K. The reliability factors are GOF = 4.72, Rp = 5.13 \%, and wRp = 6.98 \% for (a) and GOF = 2.50, Rp = 3.00 \%, and wRp = 3.69 \% for (b).
}
\label{Fig:C1}
\end{figure}

In the data, we found tiny impurity peaks. Like for three impurity phases, shown in Fig.~\ref{Fig:4}, we did not mark positions of the nuclear Bragg peaks of impurity phases (NiO, MoO$_{2}$, NiMoO$_{4}$) in this paper for simplicity (except in Fig.~\ref{Fig:C1}) because their mass fractions are tiny. However, we included them in all our refinements with the powder neutron diffraction data shown in this paper to determine more precise nuclear and magnetic structures. For completeness, we explicitly compare the structural refinements with and without these three impurity phases in Fig.~\ref{Fig:C1}; this comparison clearly reveals a tiny difference in the fits. Table~\ref{Table:C1} presents extracted structural parameters from the neutron diffraction analysis at 5.5~K shown in Fig.~\ref{Fig:C1}(b).

\begin{table}[t]
\caption{Refined structural parameters at 5.5~K using powder neutron diffraction data. Lattice parameters are \textit{a} = 5.7488(1)~\AA, \textit{c} = 9.859(1)~\AA. Reliability parameters are GOF = 2.50, Rp = 3.00 \%, and wRp = 3.69 \%. Uiso is the isotropic atomic displacement parameter (\AA$^2$).
}
\vspace{0.2cm}
\label{Table:C1}
\setlength\extrarowheight{4pt}
\begin{tabular}{cccccc}
\hline\hline
\multicolumn{1}{c}{Atoms} ~& Site  ~& x           ~& y            ~& z           ~& Uiso       \\\hline
Ni1                        &2b      & 1/3          & 2/3           & 0.9744(1)    & 0.0288(1)  \\
Ni2                        &2b      & 1/3          & 2/3           & 0.5362(1)    & 0.0229(1)  \\
Mo                         &6c      & 0.1468(1)    & -0.1468(1)    & 0.2740(1)    & 0.0157(1)  \\
O1                         &2a      & 0            & 0             & 0.4164(1)    & 0.0039(1)  \\
O2                         &2b      & 1/3          & 2/3           & 0.1713(1)    & 0.0118(1)  \\
O3                         &6c      & 0.4855(1)    & -0.4855(1)    & 0.3909(1)    & 0.0188(1)  \\
O4                         &6c      & 0.1678(2)    & -0.1678(1)    & 0.6576(1)    & 0.0175(1)  \\
\hline\hline
\end{tabular}
\end{table}

\begin{figure}[]
\includegraphics[width=\linewidth]{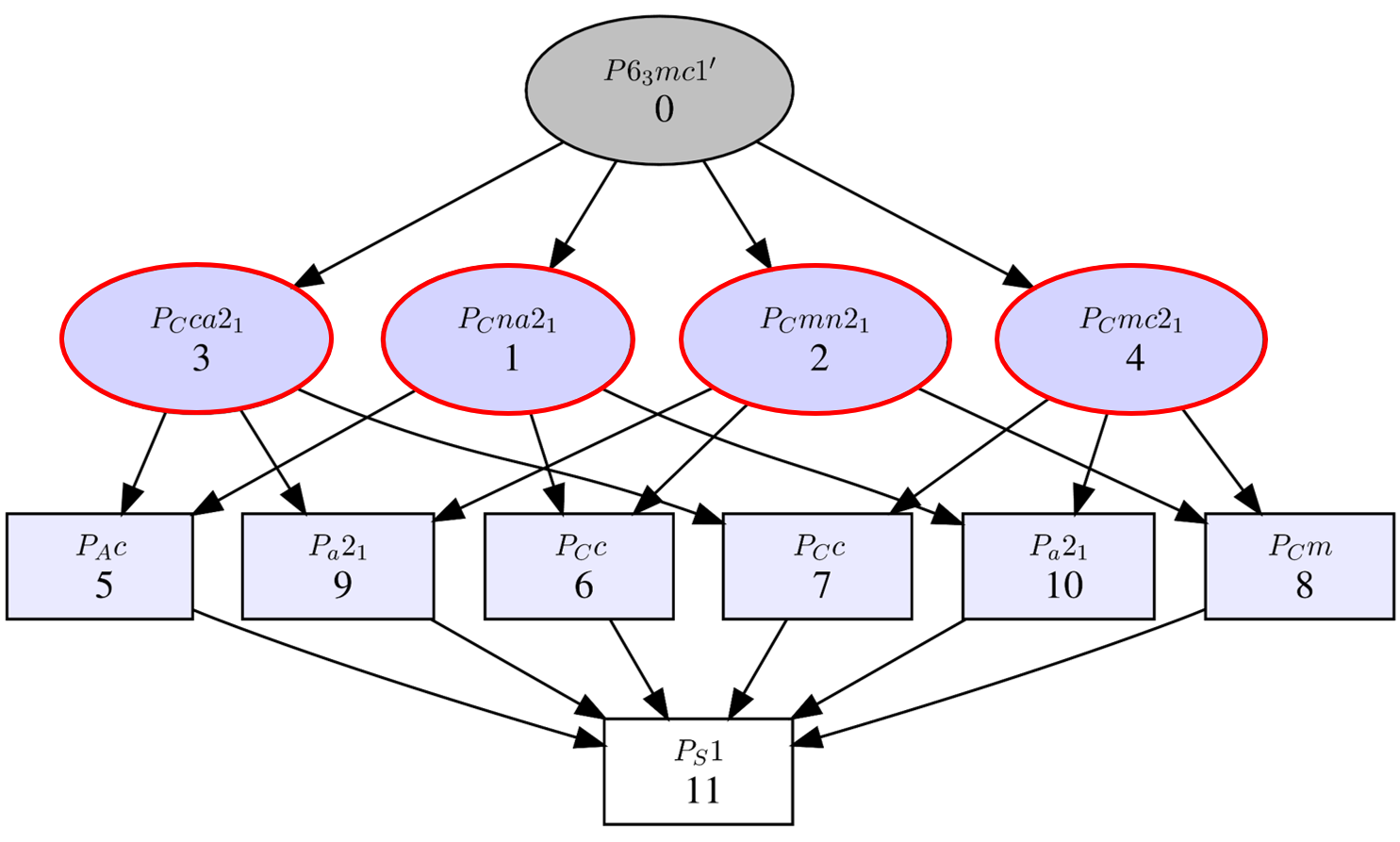}
\caption{Diagram of the possible subgroups of the paramagnetic parent space group \textit{P}6$_3$\textit{mc} with a given $q$. The diagram is generated using the \textit{k}-SUBGROUPSMAG software available on Bilbao Crystallographic Server~\cite{Aroyo2011}. The symmetries of the four maximal subgroups, which are emphasized by red circles, are listed in Table~\ref{Table:C2}.
}
\label{Fig:C2}
\end{figure}

\begin{table}[]
\caption{Symmetry relations of the four maximal magnetic space groups (MSGs) related to \NMO.
}
\vspace{0.2cm}
\label{Table:C2}
\setlength\extrarowheight{4pt}
\setlength{\tabcolsep}{2pt}
\begin{tabular}{>{\centering}m{0.07\textwidth}>{\centering}m{0.11\textwidth}m{0.28\textwidth}}
\hline\hline
M.S.G.                      & S.G. number & (Coordinates $|$ Moments)           \\\hline
\textit{P$_C$na}2$_1$       & 33.154      & (1/6, 2/3, $z$ $|$ 2$m_y$, $m_y$, $m_z$)     \\
                            &             & (1/3,1/3, z+1/2 $|$ 2$m_y$, $m_y$, $-m_z$)   \\
                            &             & (2/3, 2/3, z $|$ $-2m_y$, $-m_y$, $-m_z$)    \\
                            &             & (5/6,1/3, z+1/2 $|$ $-2m_y$, $-m_y$, $m_z$)  \\[0.2cm]\hline
\textit{P$_C$mn}2$_1$       & 31.133      & (1/6,2/3, $z$ $|$ 0, $m_y$,0)                \\
                            &             & (1/3,1/3, $z+1/2$ $|$ 0, $-m_y$,0)           \\
                            &             & (2/3,2/3, $z$ $|$ 0, $-m_y$,0)               \\
                            &             & (5/6,1/3, $z+1/2$ $|$ 0, $m_y$,0)            \\[0.2cm]\hline
\textit{P$_C$ca}2$_1$       & 29.109      & (1/6,2/3, $z$ $|$ $2m_y$, $m_y$, $m_z$)      \\
                            &             & (1/3,1/3, $z+1/2$ $|$ $-2m_y$, $-m_y$, $m_z$)\\
                            &             & (2/3,2/3, $z$ $|$ $-2m_y$, $-m_y$, $-m_z$)   \\
                            &             & (5/6,1/3, $z+1/2$ $|$ $2m_y$, $m_y$, $-m_z$) \\[0.2cm]\hline
\textit{P$_C$mc}2$_1$       & 26.76       & (1/6,2/3, $z$ $|$ 0, $m_y$,0)                \\
                            &             & (1/3,1/3, $z+1/2$ $|$ 0, $m_y$,0)            \\
                            &             & (2/3,2/3, $z$ $|$ 0, $-m_y$,0)               \\
                            &             & (5/6,1/3, $z+1/2$ $|$ 0, $-m_y$,0)           \\
\hline\hline
\end{tabular}
\end{table}

\begin{table*}[]
\caption{Representative magnetic structural models tested in \textit{P$_C$ca}2$_1$ with obtained reliability parameters.
}
\label{Table:C3}
\setlength\extrarowheight{4pt}
\setlength{\tabcolsep}{5pt}
\begin{tabular}{cccccccccc}
\hline\hline
Condition                     &                   & Atom     &$M_{a}$     &$M_{b}$     &$M_{c}$     &M          & GOF           & Rp(All)     & wRp(All)  \\\hline
$M_{T}$(Ni1) $>$ $M_{O}$(Ni2) & A (stripy/stripy) & Ni1      & -0.417(58) & -0.209(15) & 1.499(11)  & 1.542(61) & 11.67         & 7.15          & 7.63        \\\vspace{0.3cm}
                              &                   & Ni2      & -0.538(54) & -0.269(14) & -0.894(11) & 1.008(57) &               &               &             \\
                              & B (zigzag/stripy) & Ni1      & 0.988(14)  & 0.494(3)   & 0.897(12)  & 1.240(19) & 11.72         & 7.16          & 7.67        \\\vspace{0.3cm}
                              &                   & Ni2      & -0.105(13) & -0.052(3)  & -1.466(11) & 1.469(17) &               &               &             \\\hline
$M_{T}$(Ni1) $<$ $M_{O}$(Ni2) & C (stripy/stripy) & Ni1      & -0.570(43) & -0.285(11) & 0.892(12)  & 1.020(46) & 11.68         & 7.16          & 7.64        \\\vspace{0.3cm}
                              &                   & Ni2      & -0.375(47) & -0.187(12) & -1.492(11) & 1.527(49) &               &               &             \\
                              & D (zigzag/stripy) & Ni1      & 0.053(12)  & 0.027(3)   & -0.876(11) & 0.878(17) & 11.75         & 7.18          & 7.68        \\\vspace{0.3cm}
                              &                   & Ni2      & -0.950(14) & -0.475(4)  & 1.476(11)  & 1.689(18) &               &               &             \\
\hline\hline
\end{tabular}
\end{table*}

\begin{figure}  [t]
	\includegraphics[width=\linewidth]{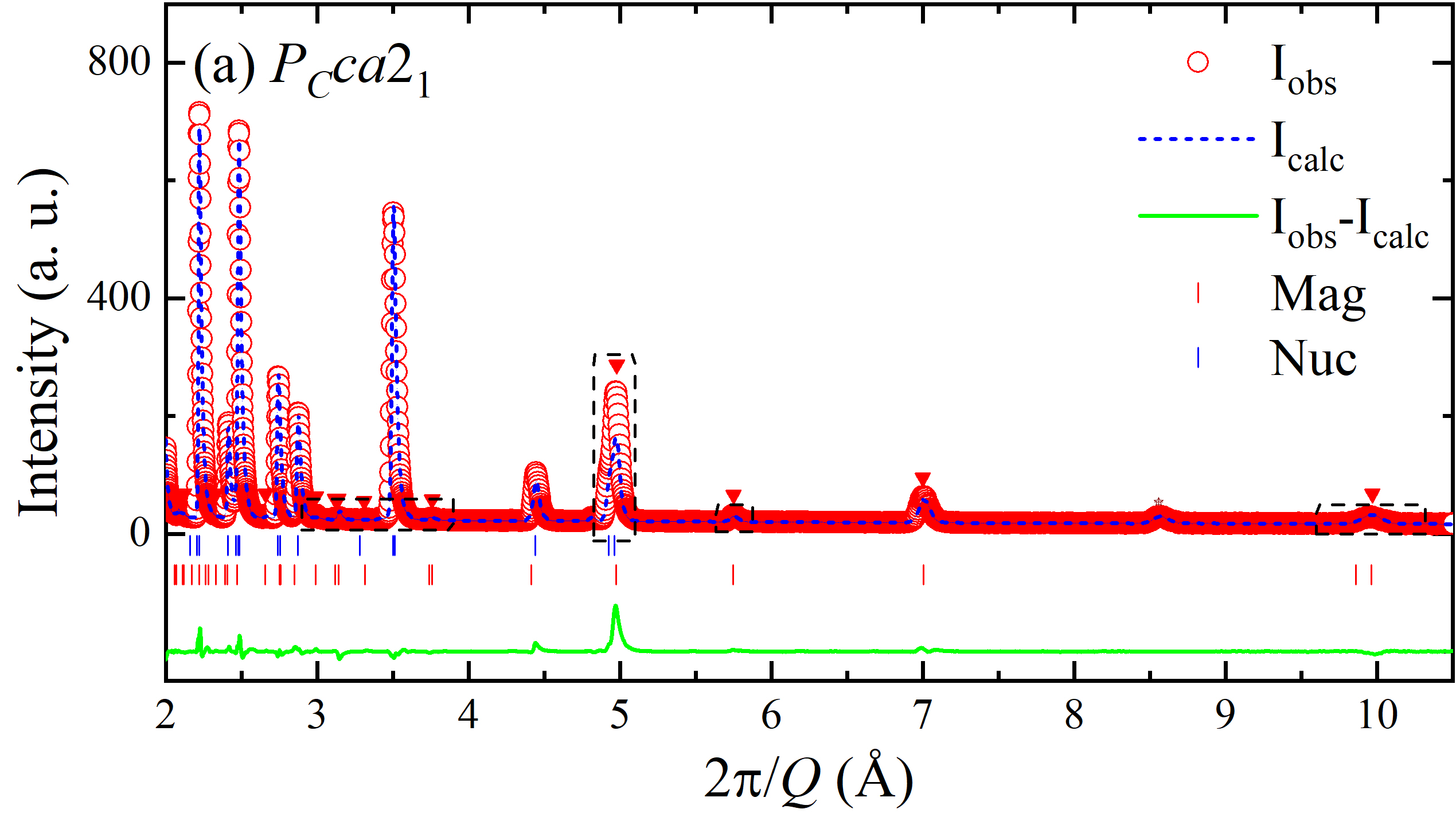}
	\vspace{0.2cm}
	\includegraphics[width=\linewidth]{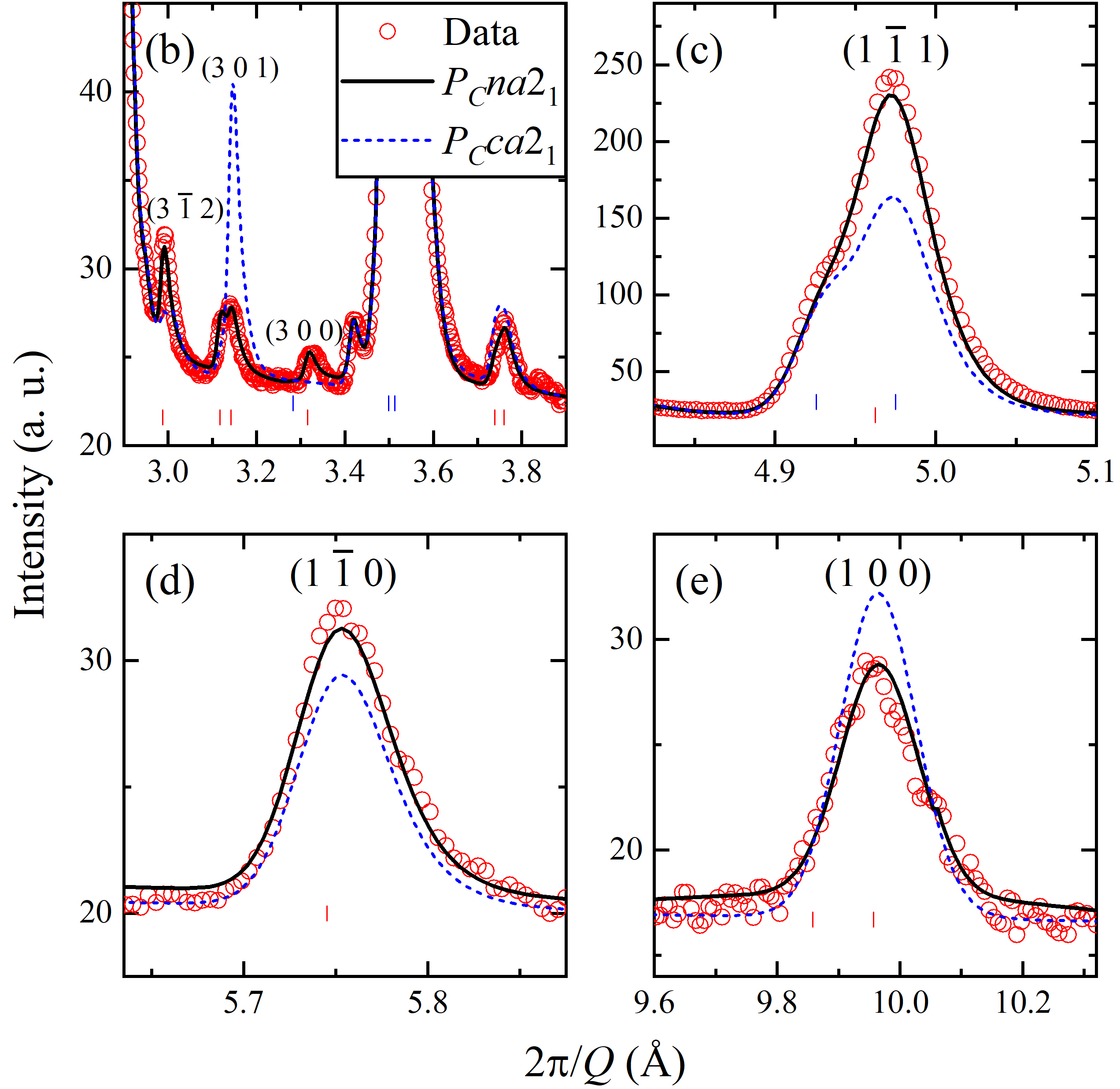}
	\caption{(a) Results from magnetic structural refinements in \textit{P$_C$ca}2$_1$ (model A of Table~\ref{Table:C3}) using the 1.5~K data. (b) - (e) Enlarged views for selective magnetic Bragg peaks in (a) reveal a strong disagreement in the fit when using \textit{P$_C$ca}2$_1$, compared to $I_{\rm calc}$ from \textit{P$_C$na}2$_1$ (Fig.~\ref{Fig:6}). Model A (stripy/stripy) of Table~\ref{Table:C3} was used as an example.
	}
	\label{Fig:C4}
\end{figure}

Among maximal subgroups for the magnetic structural determination (Fig.~\ref{Fig:C2}), we describe how we rule out \textit{P$_C$ca}2$_1$ in detail. As discussed in the main text, we performed magnetic refinements in two magnetic space groups (\textit{P$_C$ca}2$_1$ and \textit{P$_C$na}2$_1$) because they can support magnetic moments along the $a$ or $c$ axis, unlike the other two. Indeed, in Fig.~\ref{Fig:6}, the magnetic structure in \textit{P$_C$na}2$_1$ explains the 1.5~K data very well. However, the magnetic refinements using \textit{P$_C$ca}2$_1$ do not work (see Fig.~\ref{Fig:C4}).

\begin{figure} [t]
\includegraphics[width=\linewidth]{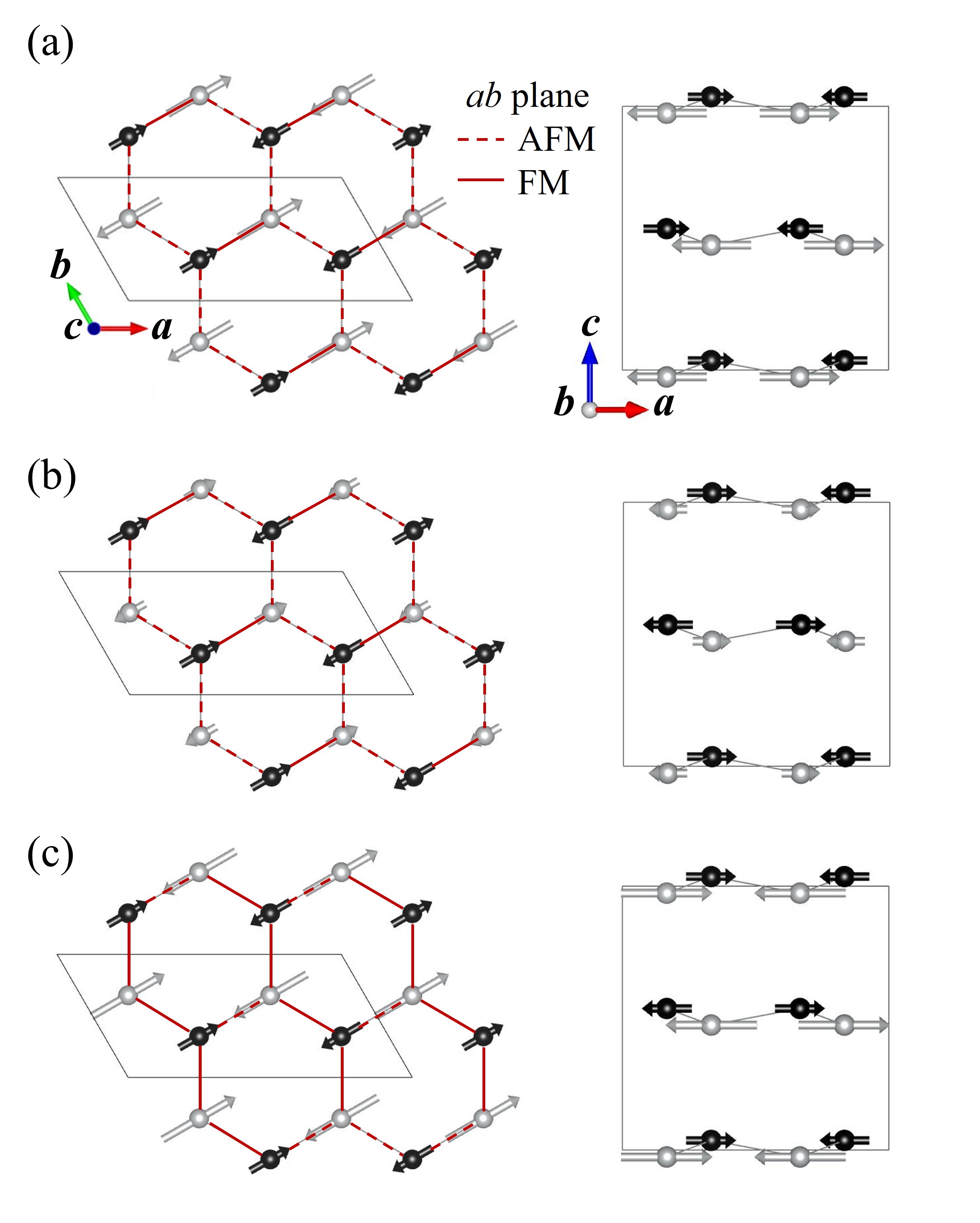}
\caption{Refined magnetic structures after setting $M_{c}$~=~0 to compare the intralayer (within the $ab$ plane) and interlayer (along the $c$ axis) coupling in a simpler way. (a) Stripy order in the plane with an antiferromagnetic interlayer coupling (simplified from model 1 in Table~\ref{Table:I} in \textit{P$_C$na}2$_1$). (b) Zigzag order in the plane with a ferromagnetic interlayer coupling (model A of Table~\ref{Table:C3} in \textit{P$_C$ca}2$_1$). (c) Zigzag order in the plane with an antiferromagnetic interlayer coupling (model B from Table~\ref{Table:C3} in \textit{P$_C$ca}2$_1$). The moment values of (b) and (c) are scaled at both Ni sites for a visualization purposes. Red lines represent the connection of the $ab$-plane spin component.
}
\label{Fig:C3}
\end{figure}

Also, for the more intuitive comparison, we set $M_{c}$~=~0 in the refined magnetic structures, by illustrating a qualitative difference in the fitted magnetic structures in \textit{P$_C$na}2$_1$ and \textit{P$_C$ca}2$_1$ (Fig.~\ref{Fig:C3}). The simplified magnetic structure in \textit{P$_C$na}2$_1$ provides a stripy magnetic structure both in the $ab$ plane and along $c$ axis; the interlayer coupling is antiferromagnetic in the fitted structure [Fig.~\ref{Fig:C3}(a)]. However, \textit{P$_C$ca}2$_1$ cannot support it by symmetry (Table~\ref{Table:C2}); instead, it allows for either a stripy order in the $ab$ plane with ferromagnetic interlayer coupling [Fig.~\ref{Fig:C3}(b)] or a zigzag order in the $ab$ plane with antiferromagnetic interlayer coupling [Fig.~\ref{Fig:C3}(c)]. This qualitative distinction clearly explains why the magnetic refinement did not work completely with magnetic structures in \textit{P$_C$ca}2$_1$, as explicitly illustrated in Fig.~\ref{Fig:C4}. A representative list of refined magnetic structures in \textit{P$_C$ca}2$_1$ is presented in Table~\ref{Table:C3} to compare the reliability factors with those from \textit{P$_C$na}2$_1$ (Table~\ref{Table:I}).

\begin{figure}  [t]
	\includegraphics[width=\linewidth]{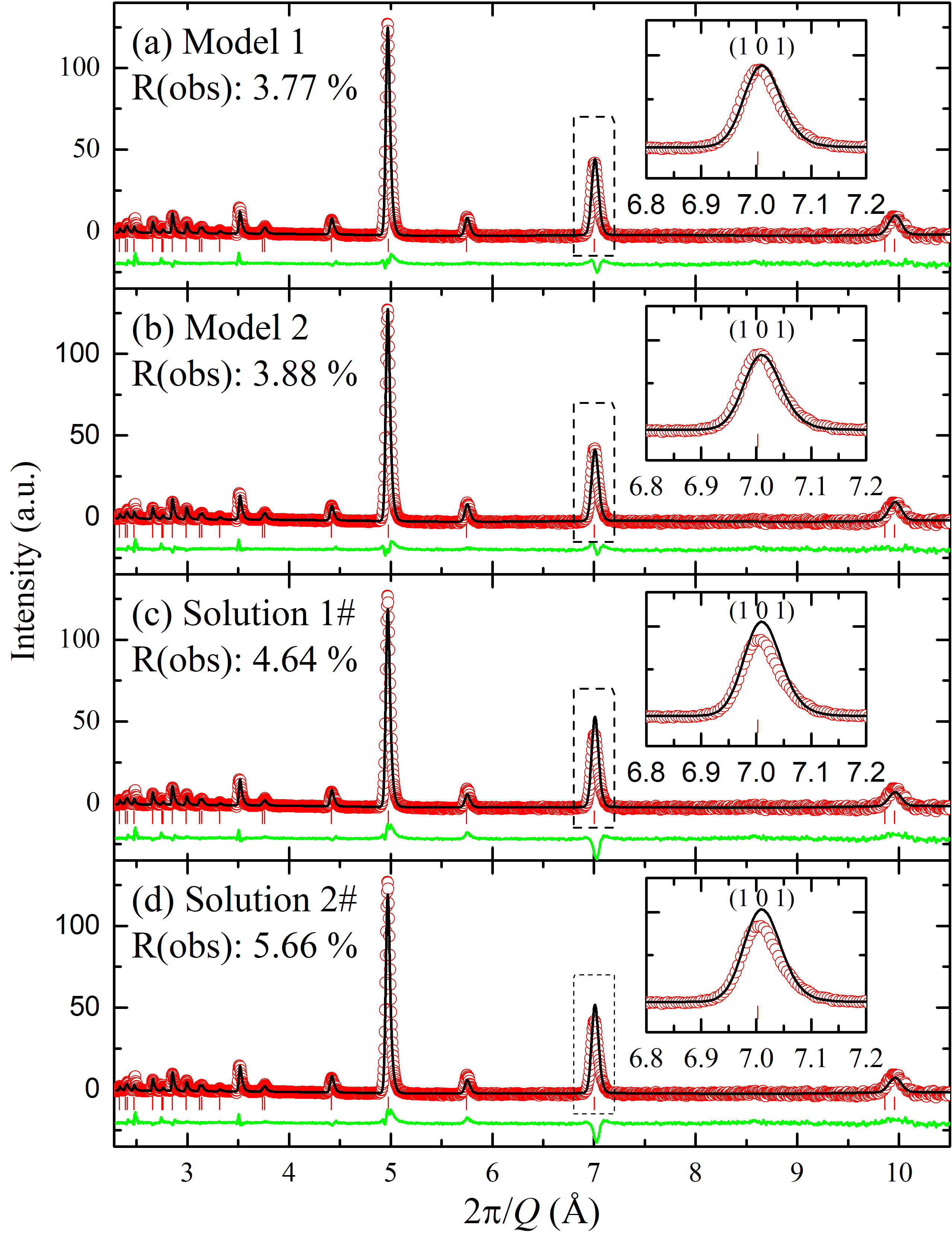}
	\caption{Comparison of results from magnetic structural refinements using the powder data in models 1 and 2 and solutions 1\# and 2\#. Only magnetic signal at 1.5~K (after the 10~K data are subtracted) is used. (a) Model 1. (b) Model 2. (c) Solution 1\#. (d) Solution 2\#. Insets enlarging the magnetic (100) peak clearly confirm the better fit with models 1 and 2 compared to solutions 1\# and 2\#.
	}
	\label{Fig:C6}
\end{figure}

We now provide full details about magnetic refinements using models 1 and 2 and solutions 1 and 2. In the main text, we showed that the magnetic structure of \NMO is stripy/stripy in the plane with the ferromagnetic interlayer coupling, which differs from the previously reported stripy/zigzag structure in the plane with the antiferromagnetic interlayer coupling~\cite{Morey2019}. In Ref.\ \cite{Morey2019}, two statistically identical stripy/zigzag magnetic structures were found. To test their solutions with our neutron diffraction data, we first tried to fit our data starting with the stripy/zigzag structures, but the fit always converged to the stripy/stripy model after running a few refinement cycles. This means that the stripy/stripy model is more stable than the stripy/zigzag solution. Thus, we simulated the reported solutions 1 and 2 with constraints in the fit, and slightly adjusted the moment values to obtain the best possible results, labeled as solutions 1\# and 2\#. The agreement factors for all four cases are listed in Table~\ref{Table:I} for direct comparison (also see Fig.~\ref{Fig:8}). A comparison of the fitting result and agreement factors indicates that the stripy/stripy models yield slightly better fitting results, compared to the stripy/zigzag solutions. Moreover, the same conclusion was reached when we repeated magnetic refinements using the pure magnetic signal at 1.5~K (see Fig.~\ref{Fig:C6})). Furthermore, a consistent result was obtained from the single-crystal neutron diffraction analysis (Table~\ref{Table:II}).

\begin{figure} [t]
\includegraphics[width=\linewidth]{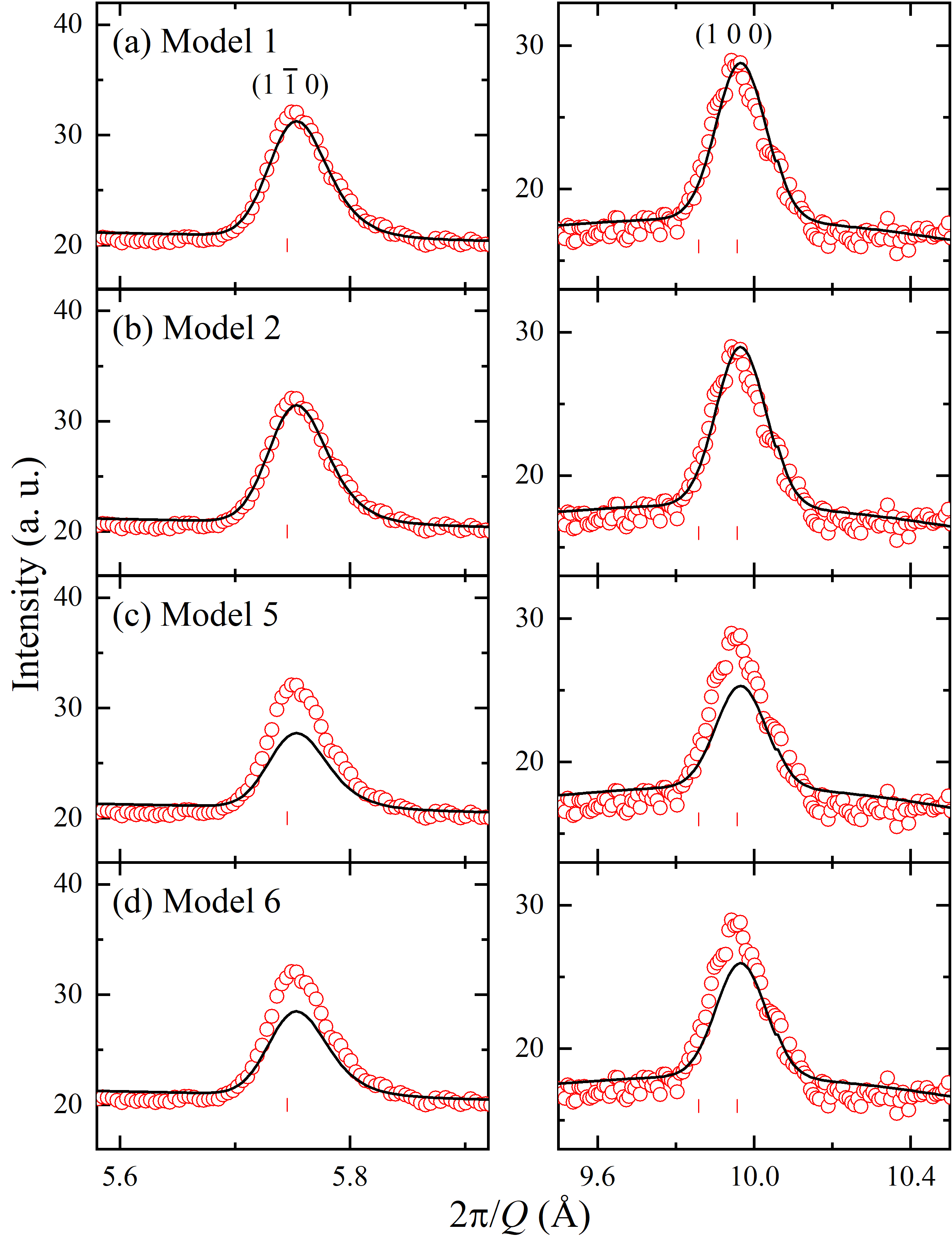}
\caption{Comparison of results from magnetic structural refinements in (a) model 1, (b) model 2, (c) model 5 [$M_{c}$(Ni2) = 0 from model 1], and (d) model 6 [$M_{c}$(Ni1) = 0 from model 1] of Table~\ref{Table:C4}.
}
\label{Fig:C5}
\end{figure}

For completeness, we present magnetic refinement results of additional magnetic structural models tested with powder neutron diffraction data collected at 1.5~K, as summarized in Table~\ref{Table:C4}. In Table~\ref{Table:I}, we show that the magnetic space group \textit{P$_C$na}2$_1$ fits all observed magnetic peaks well in the stripy/stripy model. We note that the $M_{c}$ value of the Ni2 (Ni1) site in model 1 (2) is very small (i.e., 0.18~$\mu_B$ and 0.10~$\mu_B$, respectively). This may imply an absence of $M_{c}$ in the accurate magnetic structure. Therefore, we tested this hypothesis by imposing $M_{c}$~=~0 in models 1 to 4, labeled models 5 to 8 in Table~\ref{Table:C4}. Then, refinements became marginally worse, as explicitly compared in Fig.~\ref{Fig:C5}. Both the (1, 0, 0) and (1, $\overline{1}$, 0) peaks are slightly less fitted in models 5 and 6, and the agreement factors (GOF, wRp, and Rp) increase compared to those in models 1 and 2. This means that such small and finite $M_{c}$ components can be determined in models 1 and 2 within the resolution of our high-quality data.

\begin{table*}[t]
\caption{Results from magnetic structural refinements in \textit{P$_C$na}2$_1$ using the all-bank data at 1.5~K. Reliability parameters for each bank data are provided separately for a more detailed comparison.
}
\vspace{0.2cm}
\label{Table:C4}
\setlength\extrarowheight{4pt}
\setlength{\tabcolsep}{5pt}
\begin{tabular}{ccccccccccccc}
\hline\hline
Model                &  Atoms & $M_{a}$    & $M_{b}$   & $M_{c}$    & M         & Parameter    & Bank1 & Bank2 & Bank3 & Bank4 & Bank5 & All     \\\hline
                                            \multicolumn{13}{c}{M$_T$(Ni1) $>$ M$_O$(Ni2)}                                              \\
1                    & Ni1    & -2.040(8)  & -1.020(2) & -0.708(11) & 1.904(13) & GOF & 2.88  & 5.53  & 6.87  & 7.54  & 7.15  & 6.23    \\
                     & Ni2    & 1.181(10)  & 0.590(2)  & -0.179(9)  & 1.038(14) & Rp(All)  & 5.91  & 3.72  & 2.91  & 2.77  & 2.55  & 4.22    \\\vspace{0.3cm}
                     &        &            &           &            &           & wRp(All) & 2.87  & 3.64  & 3.47  & 3.54  & 3.33  & 3.45    \\
2                    & Ni1    & -2.040(8)  & -1.020(2) & -0.100(9)  & 1.769(12) & GOF & 2.88  & 5.52  & 6.87  & 7.54  & 7.15  & 6.23    \\
                     & Ni2    & 1.181(11)  & 0.591(3)  & -0.793(10) & 1.295(15) & Rp(All)  & 5.90  & 3.72  & 2.90  & 2.77  & 2.55  & 4.22    \\\vspace{0.3cm}
                     &        &            &           &            &           & wRp(All) & 2.87  & 3.64  & 3.46  & 3.54  & 3.33  & 3.45    \\
5                    & Ni1    & -2.036(7)  & -1.018(2) & -0.756(8)  & 1.918(11) & GOF & 3.08  & 5.62  & 6.90  & 7.52  & 7.17  & 6.27    \\
                     & Ni2    & 1.154(10)  & 0.577(2)  & 0.000(0)   & 0.999(10) & Rp(All)  & 6.02  & 3.78  & 2.97  & 2.77  & 2.57  & 4.29    \\\vspace{0.3cm}
                     &        &            &           &            &           & wRp(All) & 3.08  & 3.70  & 3.48  & 3.53  & 3.34  & 3.48    \\
6                    & Ni1    & -2.094(8)  & -1.047(2) & 0.000(0)   & 1.814(8)  & GOF & 3.05  & 5.65  & 6.86  & 7.49  & 7.12  & 6.25    \\
                     & Ni2    & 1.099(9)   & 0.549(2)  & -0.712(9)  & 1.188(13) & Rp(All)  & 6.13  & 3.80  & 2.93  & 2.75  & 2.54  & 4.33    \\\vspace{0.3cm}
                     &        &            &           &            &           & wRp(All) & 3.05  & 3.72  & 3.46  & 3.51  & 3.31  & 3.46    \\\hline
                                            \multicolumn{13}{c}{M$_T$(Ni1) $<$ M$_O$(Ni2)}                                              \\
3                    & Ni1    & -1.150(10) & -0.575(2) & -0.174(9)  & 1.011(14) & GOF & 2.91  & 5.55  & 6.87  & 7.52  & 7.15  & 6.23    \\
                     & Ni2    & 2.047(7)   & 1.024(2)  & -0.708(11) & 1.909(13) & Rp(All)  & 5.55  & 3.73  & 2.91  & 2.76  & 2.55  & 4.24    \\\vspace{0.3cm}
                     &        &            &           &            &           & wRp(All) & 2.91  & 3.66  & 3.47  & 3.53  & 3.33  & 3.45    \\
4                    & Ni1    & -1.158(10) & -0.579(3) & -0.783(10) & 1.272(15) & GOF & 2.91  & 5.56  & 6.87  & 7.51  & 7.13  & 6.23    \\
                     & Ni2    & 2.043(8)   & 1.021(2)  & -0.106(10) & 1.772(12) & Rp(All)  & 5.95  & 3.74  & 2.90  & 2.75  & 2.53  & 4.24    \\\vspace{0.3cm}
                     &        &            &           &            &            & wRp(All) & 2.90  & 3.66  & 3.47  & 3.53  & 3.32  & 3.45   \\
7                    & Ni1    & -1.120(10) & -0.560(2) & 0.000(0)   & 0.970(10) & GOF & 3.18  & 5.69  & 6.88  & 7.48  & 7.14  & 6.27    \\
                     & Ni2    & 2.041(8)   & 1.021(2)  & -0.718(9)  & 1.908(12) & Rp(All)  & 6.21  & 3.82  & 2.96  & 2.75  & 2.56  & 4.36    \\\vspace{0.3cm}
                     &        &            &           &            &           & wRp(All) & 3.17  & 3.75  & 3.47  & 3.51  & 3.32  & 3.48    \\
8                    & Ni1    & -1.099(9)  & -0.549(2) & -0.723(9)  & 1.195(13) & GOF & 3.06  & 5.66  & 6.86  & 7.48  & 7.11  & 6.25    \\
                     & Ni2    & 2.084(8)   & 1.042(2)  & 0.000(0)   & 1.805(8)  & Rp(All)  & 6.16  & 3.81  & 2.93  & 2.74  & 2.54  & 4.34    \\
                     &        &            &           &            &           & Rwp & 3.05  & 3.73  & 3.46  & 3.51  & 3.31  & 3.46    \\
\hline\hline
\end{tabular}
\end{table*}

In addition, as cross-checks, we repeated the magnetic refinements for models 1 and 2 by using pure magnetic signals at 1.5~K, as shown in Fig.~\ref{Fig:C6}; the insets confirm that fits work better in models 1 and 2 than solutions 1\# and 2\#, consistent with all other analysis.

\begin{figure} [t]
\includegraphics[width=\linewidth]{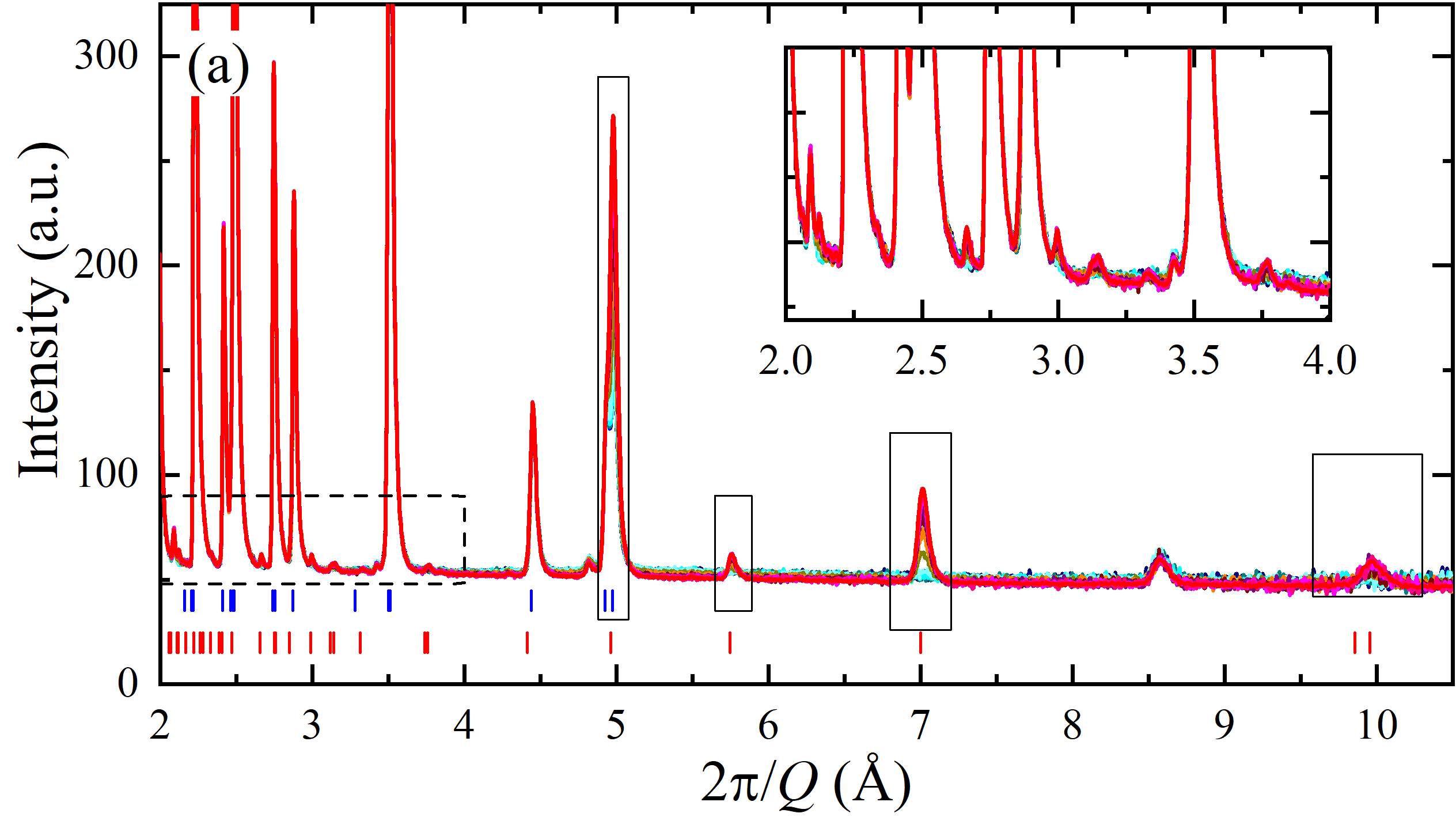}
\includegraphics[width=\linewidth]{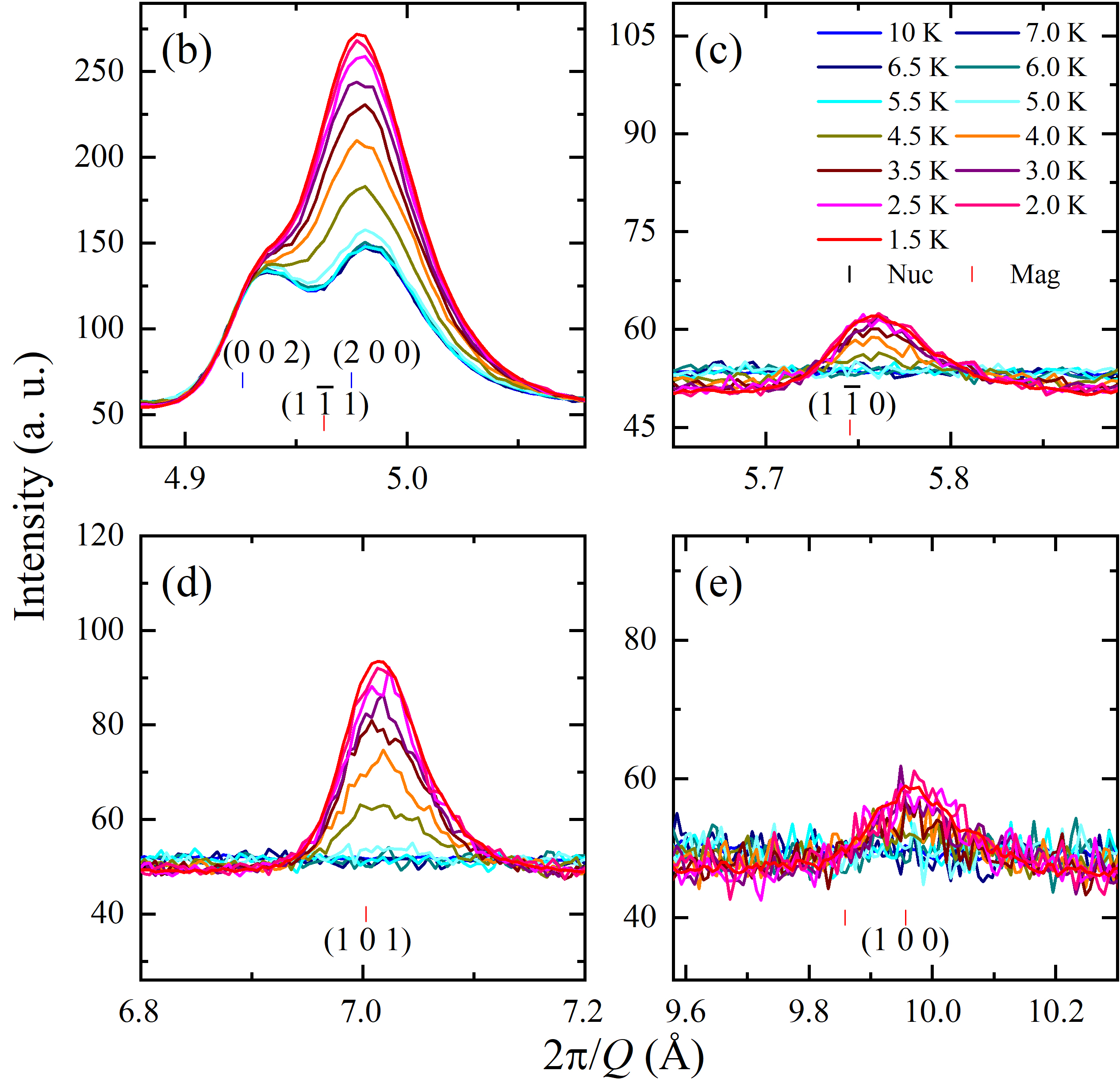}
\caption{Powder neutron diffraction data of \NMO collected from 1.5 to 10~K. Positions of Bragg reflections are obtained from fitting the 1.5~K data.
}
\label{Fig:C7}
\end{figure}

Also, we describe how our high-quality powder neutron diffraction data resolve the previous confusion over the indexing of the magnetic Bragg peak. As shown in Fig.~\ref{Fig:C7}(b) (see also Fig.~\ref{Fig:6}), we can clearly separate nuclear and magnetic Bragg peaks located nearby in the narrow range of the $d$ spacing. At 1.5~K, the intensity near the (200) nuclear peak (\textit{d} $\sim$ 4.96~\AA), which was originally indexed as (100) in the paramagnetic phase, increases because of the appearance of new magnetic peaks of (1$\overline{1}$1) and (1$\overline{1}$$\overline{1}$) reflections, which our high-resolution data can resolve in magnetic refinements. In a previous report~\cite{Morey2019}, this magnetic peak was indexed as (004), and it was taken as evidence of a significant perpendicular magnetic moment $M_{c}$, and the irreducible representations $\Gamma$1 and $\Gamma$3 were disregarded because they cannot have the $M_{c}$ magnetic moment component. However, we cannot discard these irreducible representations based on their arguments because the (004) reflection is not a magnetic peak. Also, even if (004) was a magnetic reflection, we could not confirm the presence of $M_{c}$ because neutrons can see the magnetic moment only perpendicular to the wave vector. Therefore, the presence of ``magnetic" (00$L$) Bragg peak in the neutron diffraction data would mean a possible magnetic moment within the $ab$ plane. Note that we did not observe a magnetic (001) Bragg peak (see Table~\ref{Table:C6}).

\begin{figure} [t]
\includegraphics[width=\linewidth]{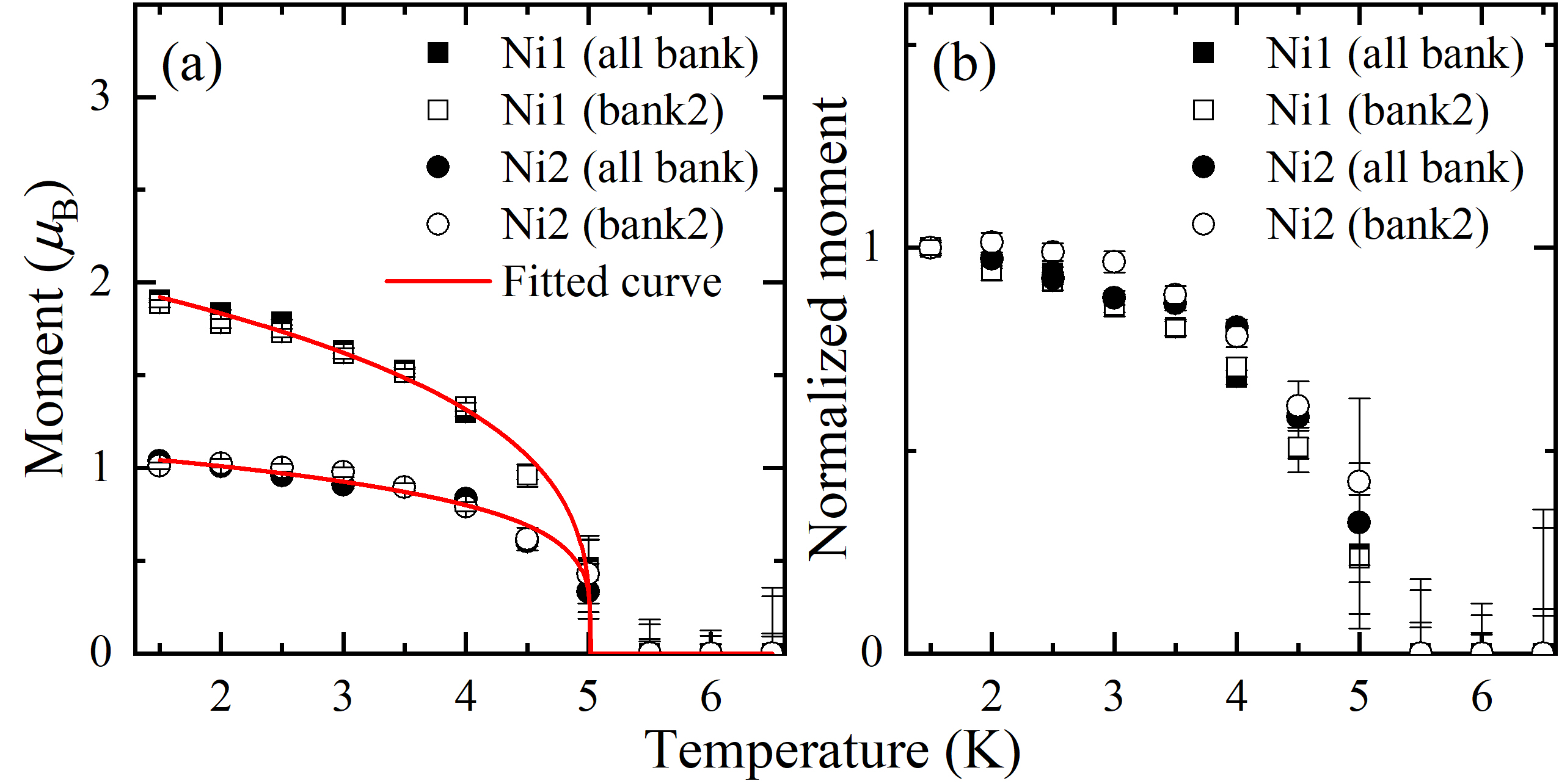}
\caption{Evolution of fitted magnetic moments with temperature using the all-bank and bank 2 data separately. (a) Magnetic moments at the Ni1 and Ni2 sites. (b) Normalized magnetic moments. Solid red lines in (a) are from the fits (see texts). Model 1 of Table~\ref{Table:I} is used in the refinement as a representative structure.
}
\label{Fig:C8}
\end{figure}

Further, we collected neutron diffraction data with increasing temperature to study the evolution of magnetic Bragg peaks towards the paramagnetic phase. Figure~\ref{Fig:C7} shows the temperature-dependent data between 1.5~K and 10~K. It shows the data for the selective range of the \textit{d} spacing for clarity, where we can see that the magnetic Bragg peaks are gradually suppressed with heating and become absent at around 5.5~K. For a more accurate determination of magnetic moments with temperature (using model 1 in Table~\ref{Table:I}), the all-bank and bank 2 data are used separately in magnetic refinements. The transition temperature extracted from the neutron diffraction data via the fit is between 5~K (Fig.~\ref{Fig:C8}) and 5.5~K (not shown), and it is consistent with $T_\textrm{N}$ found from magnetic, thermodynamic, and electric measurements (Fig.~\ref{Fig:2}). We also attempted to fit the data above 5.5~K, although it does not reveal magnetic Bragg peaks, to estimate the error bars of fitted moments. Then, we set the zero magnetic moment values for the data above 5.5 K in refinements because no magnetic peaks are observed above 5 K. In Table~\ref{Table:C5}, we present the results of the magnetic refinements, which are used for the critical analysis, shown in Fig.~\ref{Fig:C8}.

Figure~\ref{Fig:C8} presents the change in the fitted magnetic moments upon a change in temperature, which clearly demonstrates the reliably fitted moments in both types of the data used (see Table~\ref{Table:C5} for refinement results). Figure~\ref{Fig:C8}(a) shows that the fitted magnetic moments of Ni$^{2+}$ at the tetrahedral (Ni1) site are almost twice those of the octahedral (Ni2) site. However, the normalized magnetic moments at both sites show a comparable temperature dependence, as illustrated in Fig.~\ref{Fig:C8}(b). The solid lines are from the critical analysis using $M(T)=M_0(1-T/Tc)^\beta$~\cite{Stanley}. The temperature dependence of the Ni1 and Ni2 moments exhibits a power-law behavior with critical exponents of $\beta$ = 0.31(1) and 0.21(2), respectively. It might be understood by a conventional interpretation based on the criticality theory; the $\beta$ value at the Ni1 site is close to the value of the three-dimensional (3D) Heisenberg ($\beta=0.36$) or 3D XY ($\beta=0.35$) interactions, whereas its value at the Ni2 site is close to that of the two-dimensional XY interaction ($\beta=0.23$)~\cite{Bramwell1993, Lee2019}. However, we should recall that this could be misleading because there could be a non-negligible coupling between two Ni sites.

\begin{table}[]
	\caption{List of magnetic Bragg peaks observed in the powder neutron diffraction data at 1.5~K (bank 2) based on magnetic refinements using the pure magnetic signal. $I_{\rm obs}$ represents the observed intensity.
	}
	\vspace{0.2cm}
	\label{Table:C6}
	\setlength\extrarowheight{4pt}
	\setlength{\tabcolsep}{10pt}
	\begin{tabular}{|ccc|c|c|}
\hline
		H & K  & L  & \textit{d} spacing (\AA) & $I_{\rm obs}$ (a.u.) ~\\\hline
		1 & 0  & 0  & 9.95677            & 186.762     \\
		0 & 0  & -1 & 9.85832            & 1.78543~$\times$~10$^{-10}$ \\
		0 & 0  & 1  & 9.85832            & 1.78543~$\times$~10$^{-10}$ \\
		1 & 0  & 1  & 7.00551            & 208.426     \\
		1 & 0  & -1 & 7.00551            & 208.426     \\
		1 & -1 & 0  & 5.74843            & 75.4373     \\
		1 & -1 & -1 & 4.96602            & 366.353     \\
		1 & -1 & 1  & 4.96602            & 366.353     \\
		1 & 0  & 2  & 4.41762            & 24.9786     \\
		1 & 0  & -2 & 4.41762            & 24.9786     \\
		3 & -1 & 0  & 3.76316            & 12.3329     \\
		1 & -1 & 2  & 3.74176            & 4.47674     \\
		1 & -1 & -2 & 3.74176            & 4.47674     \\
		3 & -1 & 1  & 3.51583            & 27.6295     \\
		3 & -1 & -1 & 3.51583            & 27.6295     \\
		3 & 0  & 0  & 3.31892            & 6.34412     \\
		3 & 0  & -1 & 3.14532            & 4.54668     \\
		3 & 0  & 1  & 3.14532            & 4.54668     \\
		1 & 0  & 3  & 3.12059            & 6.20378     \\
		1 & 0  & -3 & 3.12059            & 6.20379     \\
		3 & -1 & 2  & 2.99122            & 10.9114     \\
		3 & -1 & -2 & 2.99122            & 10.9114     \\
		1 & -1 & 3  & 2.85281            & 20.5409     \\
		1 & -1 & -3 & 2.85281            & 20.5409     \\
		3 & -2 & 0  & 2.76149            & 4.42909     \\
		3 & 0  & 2  & 2.75293            & 1.62474     \\
		3 & 0  & -2 & 2.75293            & 1.62474     \\
		2 & -1 & -3 & 2.74246            & 8.93724~$\times$~10$^{-13}$ \\
		2 & 0  & 3  & 2.74246            & 1.28346~$\times$~10$^{-12}$ \\
		3 & -2 & 1  & 2.65923            & 10.5654     \\
		3 & -2 & -1 & 2.65923            & 10.5654     \\
		3 & -1 & 3  & 2.47516            & 5.3427       \\
		3 & -1 & -3 & 2.47516            & 5.34269     \\
		3 & -2 & 2  & 2.40905            & 3.95495     \\
		3 & -2 & -2 & 2.40905            & 3.95495     \\
		1 & 0  & 4  & 2.3924              & 4.03382     \\
		1 & 0  & -4 & 2.3924              & 4.03383     \\
		3 & 0  & 3  & 2.33533            & 3.57791     \\
		3 & 0  & -3 & 2.33533            & 3.57791     \\
\hline
	\end{tabular}
\end{table}

\begin{table*} [t]
\caption{Results from magnetic structural refinements using the powder neutron diffraction data collected from 1.5 to 5~K.
}
\vspace{0.2cm}
\label{Table:C5}
\setlength\extrarowheight{4pt}
\setlength{\tabcolsep}{4pt}
\begin{tabular}{ccccccccccccc}
\hline\hline
T (K) & Atoms & $M_{a}$    & $M_{b}$    & $M_{c}$     & M                  & Parameter     & Bank1 & Bank2 & Bank3 & Bank4 & Bank5 & Overall \\\hline
1.5    & Ni1   & -2.040(8)  & -1.020(2)  & -0.708(11)  & 1.904(13)        & GOF  & 2.88  & 5.53  & 6.87  & 7.54  & 7.15  & 6.23       \\
         & Ni2   & 1.181(10)  & 0.590(2)   & -0.179(9)   & 1.038(14)        & Rp(All)   & 5.91  & 3.72  & 2.91  & 2.77  & 2.55  & 4.22   \\\vspace{0.3cm}
         &         &                   &                 &                  &                         & wRp(All)  & 2.87  & 3.64  & 3.47  & 3.54  & 3.33  & 3.45 \\
2.0    & Ni1   & -1.983(8)  & -0.992(2)  & -0.652(13)  & 1.837(15)        & GOF  & 1.11  & 2.01  & 2.32  & 2.50  & 2.38  & 2.14      \\
         & Ni2   & 1.143(11)  & 0.572(3)   & -0.194(11)  & 1.009(16)       & Rp(All)   & 5.81  & 4.91  & 3.41  & 3.06  & 2.87  & 4.93   \\\vspace{0.3cm}
         &         &                  &                  &                   &                       & wRp(All)  & 3.37  & 4.20  & 3.71  & 3.71  & 3.48  & 3.72  \\
2.5    & Ni1   & -1.927(8)  & -0.964(2)  & -0.636(12)  & 1.786(15)       & GOF  & 1.12  & 1.95  & 2.34  & 2.47  & 2.41  & 2.13       \\
         & Ni2   & 1.089(10)  & 0.545(3)   & -0.176(11)  & 0.960(15)      & Rp(All)   & 5.98  & 4.83  & 3.39  & 3.04  & 2.88  & 4.90    \\\vspace{0.3cm}
         &         &                  &                  &                    &                      & wRp(All)  & 3.42  & 4.07  & 3.74  & 3.65  & 3.52  & 3.70   \\
3.0    & Ni1   & -1.757(10) & -0.879(3)  & -0.592(14)  & 1.633(17)      & GOF      & 1.11  & 2.00  & 2.55  & 2.70  & 2.58  & 2.28   \\
         & Ni2   & 1.033(13)  & 0.517(3)   & -0.168(13)  & 0.910(18)      & Rp(All)    & 5.74  & 4.87  & 3.56  & 3.28  & 3.21  & 4.88   \\\vspace{0.3cm}
         &         &                  &                  &                   &                       & wRp(All)  & 3.41  & 4.21  & 4.06  & 4.02  & 3.83  & 3.99  \\
3.5    & Ni1   & -1.638(11) & -0.819(3)  & -0.563(14)  & 1.526(18)      & GOF       & 1.15  & 1.93  & 2.36  & 2.43  & 2.32  & 2.10  \\
         & Ni2   & 1.017(14)  & 0.508(3)   & -0.167(13)  & 0.896(19)      & Rp(All)    & 5.86  & 4.82  & 3.41  & 2.91  & 2.84  & 4.85  \\\vspace{0.3cm}
         &         &                  &                  &                   &                       & wRp(All) & 3.51 & 4.05  & 3.77  & 3.61  & 3.44  & 3.68   \\
4.0    & Ni1   & -1.375(14) & -0.688(4)  & -0.511(15)  & 1.296(21)      & GOF      & 1.07  & 1.96  & 2.28  & 2.43  & 2.32  & 2.09  \\
         & Ni2   & 0.952(17)  & 0.476(4)   & -0.123(14)  & 0.834(22)      & Rp(All)   & 5.09  & 4.91  & 3.27  & 2.94  & 2.81  & 4.66   \\\vspace{0.3cm}
         &         &                  &                  &                   &                       & wRp(All) & 3.22 & 4.13  & 3.64  & 3.62  & 3.45  & 3.65   \\
4.5    & Ni1   & -1.048(18) & -0.524(4)  & -0.320(26)  & 0.962(32)      & GOF  & 1.11  & 1.90  & 2.23  & 2.44  & 2.35  & 2.07      \\
         & Ni2   & 0.683(21)  & 0.342(5)   & -0.125(24)  & 0.605(32)      & Rp(All)   & 5.75  & 4.80  & 3.20  & 2.93  & 2.89  & 4.79   \\\vspace{0.3cm}
         &         &                  &                  &                   &                      & wRp(All) & 3.42 & 4.00  & 3.56  & 3.64  & 3.48  & 3.63    \\
5.0    & Ni1   & -0.534(41) & -0.267(10) & -0.053(141) & 0.465(147) & GOF  & 1.05  & 1.89  & 2.24  & 2.44  & 2.37  & 2.07       \\
         & Ni2   & 0.371(48)  & 0.186(12)  & -0.093(139) & 0.335(147) & Rp(All)   & 5.21  & 4.70  & 3.21  & 2.90  & 2.88  & 4.58    \\\vspace{0.3cm}
         &         &                  &                  &                  &                       & wRp(All)  & 3.18  & 3.97  & 3.57  & 3.62  & 3.51  & 3.62  \\
\hline\hline
\end{tabular}
\end{table*}

\section{Single-crystal neutron diffraction}
\label{app:sND}
\begin{figure} []
\includegraphics[width=\linewidth]{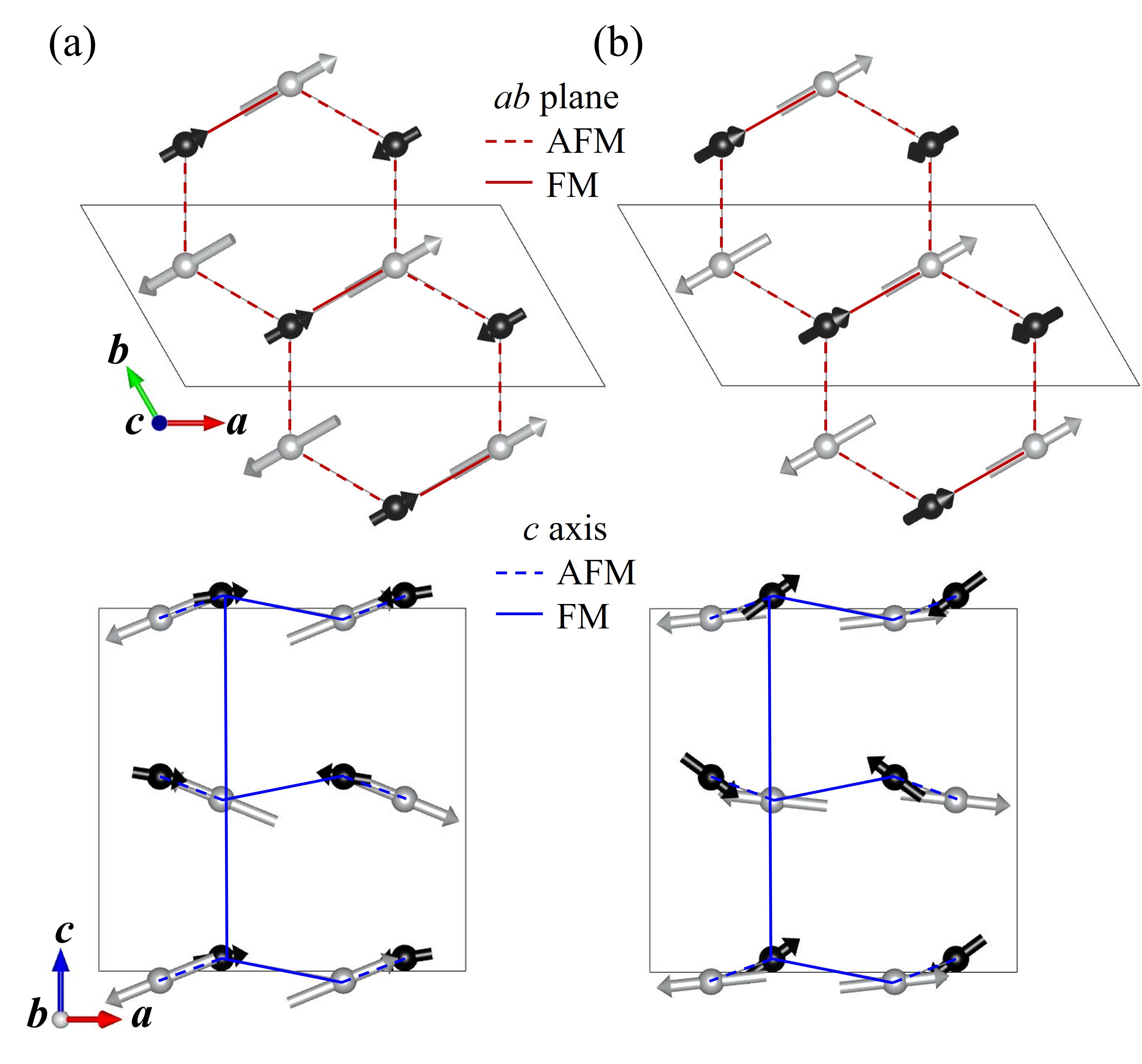}
\includegraphics[width=\linewidth]{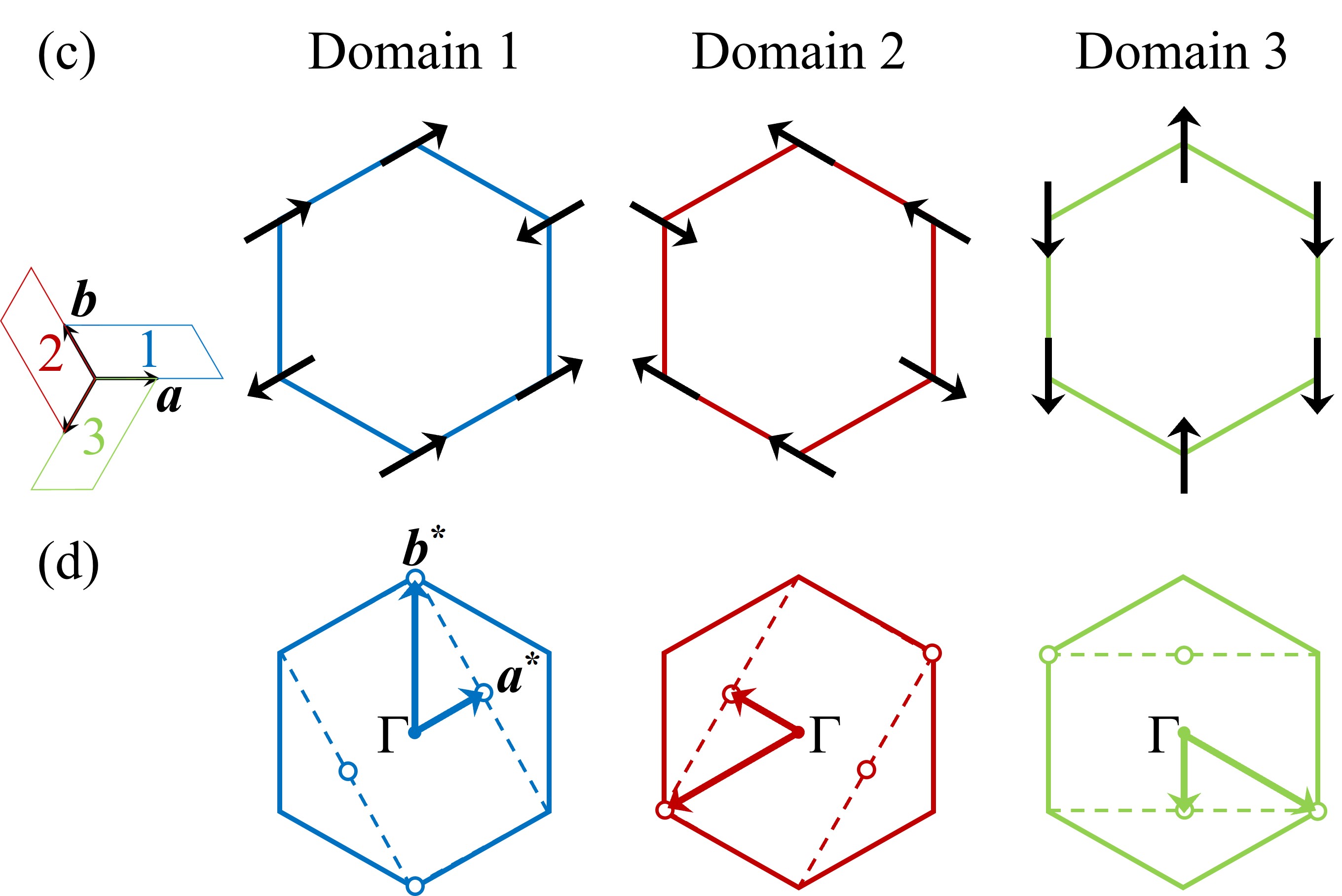}
\caption{(a) and (b) Refined magnetic structures from Figs.~\ref{Fig:9}(c) and \ref{Fig:9}(d), respectively. The parallelogram-shaped black solid line denotes the magnetic unit cell. The crystallographic $a$ and $b$ axes based on the extended unit cell are presented. Black (gray) ions are octahedral (tetrahedral) Ni sites. Red and blue lines represent the connection of the $ab$-plane and $c$-axis spin components, respectively. (c) Three equivalent magnetic domains in a collinear stripy order with $M_{c}$~=~0 for simplicity. (d) Reciprocal space of (c). The blue, red, and green dashed lines show the reciprocal lattice, corresponding to the magnetic unit cells of domains 1, 2, and 3 in real space, respectively.
}
\label{Fig:D1}
\end{figure}

This Appendix provides details about single-crystal neutron diffraction analysis. First, the structural parameters extracted from refinements using the neutron diffraction data at 10~K are listed in Table~\ref{Table:D1}. Also, we present refined magnetic structures in Figs.~\ref{Fig:D1}(a) and \ref{Fig:D1}(b) which are consistent with those from the powder result. Our analysis found three nearly equally populated magnetic domains, as shown in real space in Fig.~\ref{Fig:D1}(c), and the corresponding reciprocal lattices are shown in Fig.~\ref{Fig:D1}(d). We observed all three types of magnetic Bragg peaks in the single-crystal neutron diffraction data measured at 1.5~K from three magnetic domains. We tested a single magnetic domain in magnetic refinements, but the test did not work because of the finite intensity of the magnetic Bragg peaks belonging to other domains.

\begin{table}[t]
\caption{Obtained structural parameters from the single-crystal neutron diffraction analysis (10 K). Lattice parameters are \textit{a} = 5.7595~\AA~and \textit{c} = 9.8699~\AA. Reliability parameters are GOF = 2.50, Rp = 4.39 \%, and wRp = 4.97 \%.
}
\vspace{0.2cm}
\label{Table:D1}
\setlength\extrarowheight{4pt}
\begin{tabular}{cccccc}
\hline\hline
\multicolumn{1}{c}{Atoms} ~& Site      ~& x           ~& y            ~& z            ~& Uiso      \\\hline
Ni1                       ~&2b         ~& 1/3         ~& 2/3          ~& 0.9738(1)    ~& 0.0021(2) \\
Ni2                       ~&2b         ~& 1/3         ~& 2/3          ~& 0.5362(1)    ~& 0.0017(2) \\
Mo                        ~&6c         ~& 0.1460(1)   ~& -0.1460(1)   ~& 0.2742(1)    ~& 0.0011(1) \\
O1                        ~&2a         ~& 0           ~& 0            ~& 0.4164(1)    ~& 0.0027(2) \\
O2                        ~&2b         ~& 1/3         ~& 2/3          ~& 0.1715(1)    ~& 0.0027(2) \\
O3                        ~&6c         ~& 0.4879(1)    &-0.4879(1)	  ~& 0.3918(1)    ~& 0.0029(2) \\
O4                        ~&6c         ~& 0.1688(1)    &-0.1688(1)    ~& 0.6580(1)    ~& 0.0029(2) \\
\hline\hline
\end{tabular}
\end{table}

\begin{figure}[t]
\includegraphics[width=\linewidth]{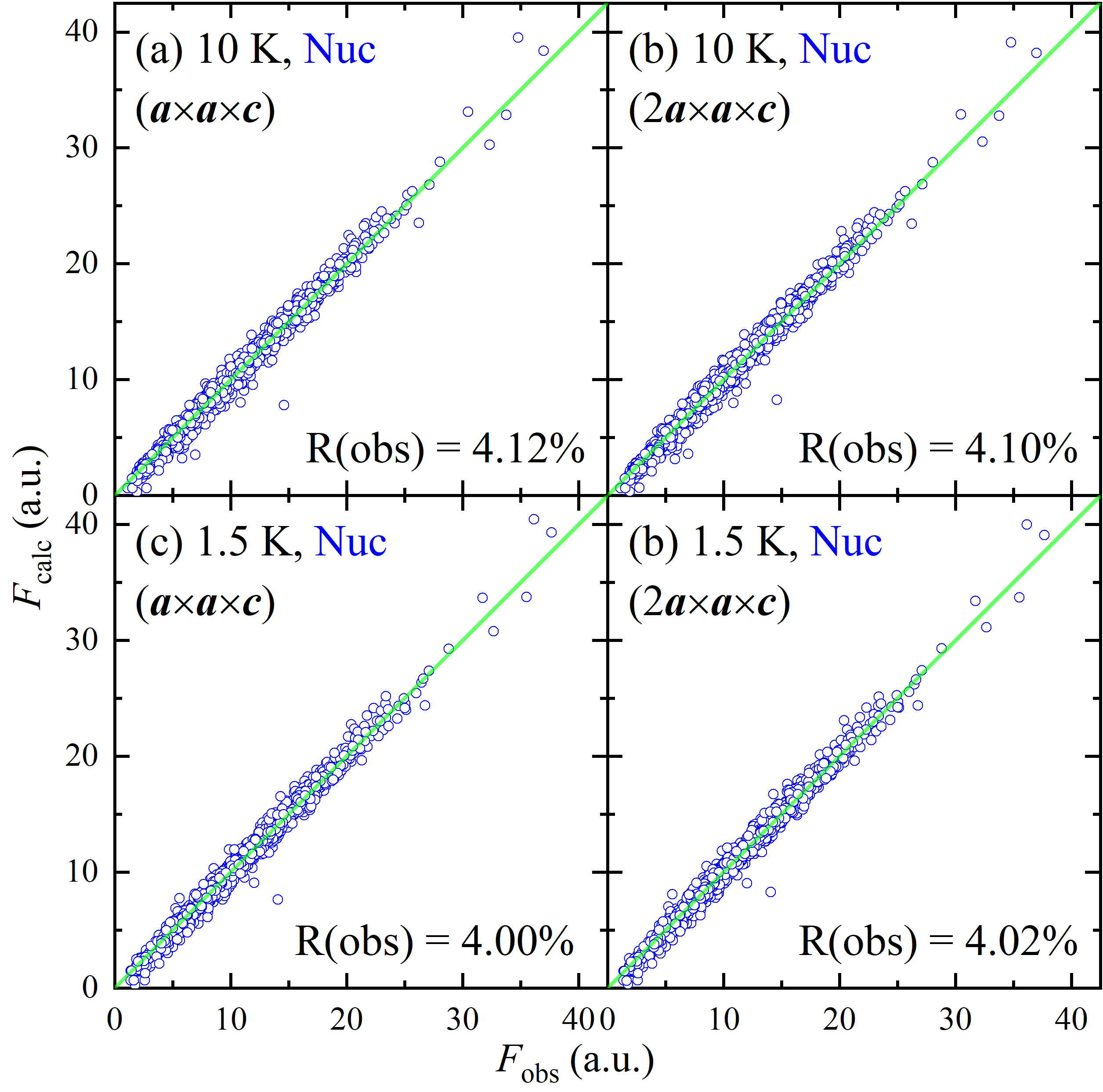}
\caption{Comparison of structural refinements at 10 and 1.5~K using the single-crystal data. (a) 10 K in the \textit{P}6$_3$\textit{mc} space group. (b) 10 K with the extended unit cell (2$a$$\times$$a$$\times$$c$). (c) 1.5 K in the \textit{P}6$_3$\textit{mc} space group. (d) 1.5 K with the extended unit cell (2$a$$\times$$a$$\times$$c$). The two-rotation data are used.
}
\label{Fig:D2}
\end{figure}

In the magnetic refinements, shown in Figs.~\ref{Fig:9}(b)-\ref{Fig:9}(d) in the main text, we use the extended unit cell, 2$a$$\times$$a$$\times$$c$. For completeness, we present a repetitive structural refinement using the 10~K data with the extended unit cell (Fig.~\ref{Fig:D2}) to obtain accurate structural parameters such as extinction and scale parameters, which are fixed for more reliable magnetic refinements with the 1.5~K data in our analysis [Figs.~\ref{Fig:9}(b)-\ref{Fig:9}(d)].

\begin{figure} [ht]
\includegraphics[width=\linewidth]{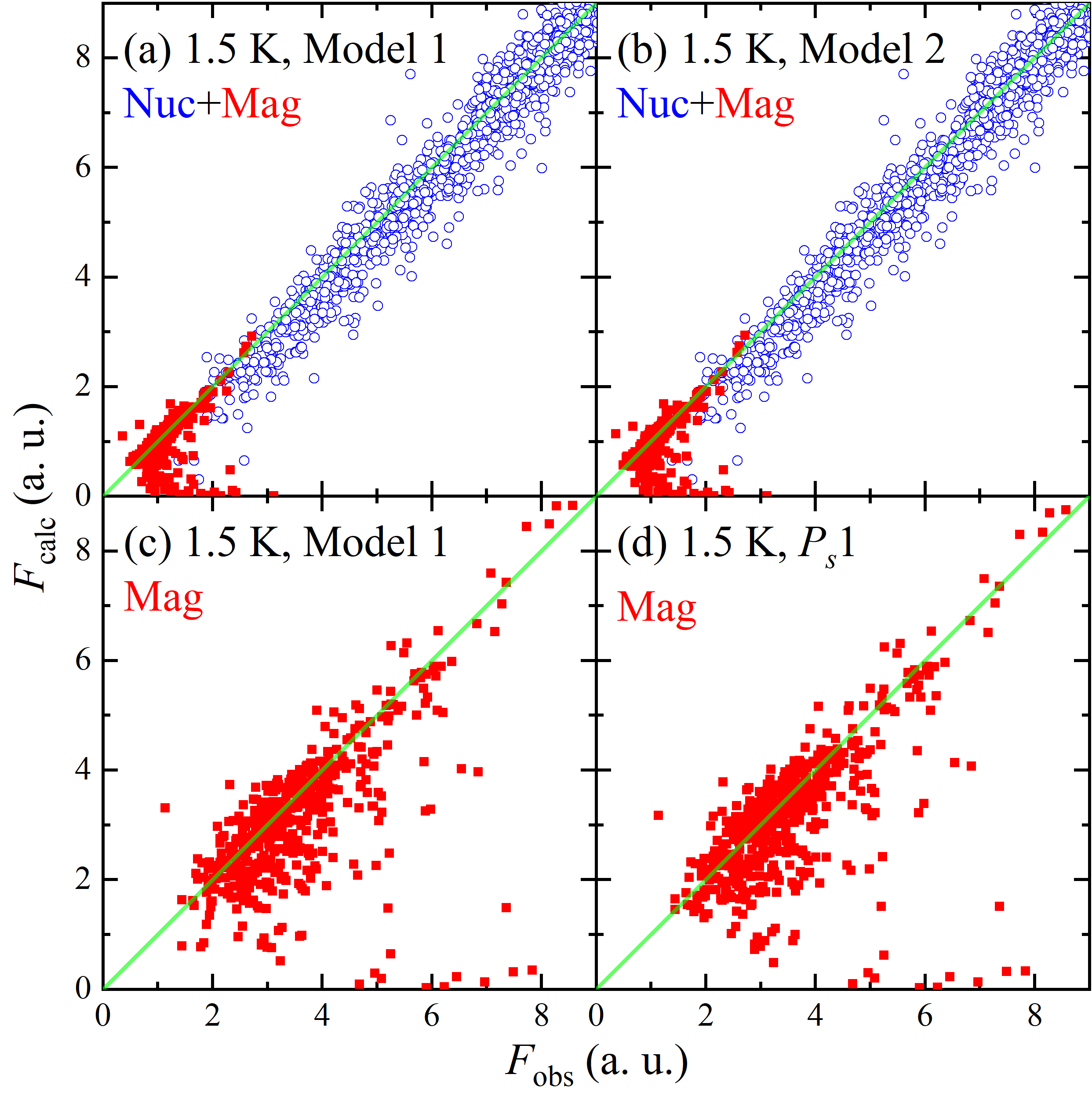}
\caption{Comparison of magnetic refinements using the 1.5~K data for the single-crystal neutron diffraction, focusing on weaker peaks. (a) Model 1 and (b) model 2 with the two-rotation data. (c) Model 1 (\textit{P$_C$na}2$_1$) and (d) the least symmetric magnetic space group (\textit{P$_S$}1) obtained using only forbidden reflections (537 peaks) in the five-rotation data for the comparison.
}
\label{Fig:D3}
\end{figure}

We present the structural refinements results at 1.5~K using only the allowed Bragg peaks. Our aims are twofold: (i) to confirm that the structural parameters extracted from the 10~K data can be reliably used for magnetic refinements with 1.5~K data and (ii) to see whether our data are sensitive to any possible structural distortion from a hexagonal to an orthorhombic structure via a magnetic transition. We used the only identical nuclear Bragg peaks (2341 nuclear reflections) collected from two identical rotation angles at 1.5 and 10~K while fixing all parameters except the atomic position and thermal parameters in the refinements to determine whether we can obtain any hint of a structural transition. We use nuclear space group \textit{P}6$_3$\textit{mc} and magnetic space group \textit{P$_C$na}2$_1$ in the extended unit cell (2$a$$\times$$a$$\times$$c$), as shown in Figs.~\ref{Fig:D2}(c) and \ref{Fig:D2}(d), respectively. The obtained nuclear structures are comparable. Thus, we did not observe any evidence of the structural distortion in the nuclear structure from 10 to 1.5~K within the resolution of our single-crystal neutron diffraction data.

We discuss some excluded forbidden peaks in Figs.~\ref{Fig:9}(c)-\ref{Fig:9}(e) in the main text because they are not fitted even with the lowest magnetic space group. Figures~\ref{Fig:D3}(a) and \ref{Fig:D3}(b) present the magnetic refinement results after including the additional peaks, where one can see weaker structure factors $F_{\rm calc}$ in both models 1 and 2. If they are magnetic peaks, their existence may imply the symmetry of a true magnetic space group is lower than \textit{P$_C$na}2$_1$. We fitted the data to the lower-symmetric magnetic space groups to test this possibility. Lowering the symmetry of all four orthorhombic maximal subgroups resulted in six possible monoclinic candidates [\textit{P$_A$c} (5), \textit{P$_a$}2$_1$ (9), \textit{P$_C$c} (6), \textit{P$_C$c} (7), \textit{P$_a$}2$_1$ (10), and \textit{P$_C$m} (8)] and one triclinic candidate [\textit{P$_S$}1 (11); Fig.~\ref{Fig:C2}]. The number in parentheses denotes the subgroup number in the group analysis, as depicted in Fig.~\ref{Fig:C2}. Using the two-rotation data, we tested all possible subgroups originating from \textit{P$_C$na}2$_1$, e.g., \textit{P$_A$c} (5), \textit{P$_C$c} (6), \textit{P$_a$}2$_1$ (10), and the least symmetric magnetic space group, \textit{P$_S$}1, based on model 1. However, none of them explained the additional peaks. Thus, these additional reflections at 1.5 K are not sensitive to the magnetic order. They could be attributed to extrinsic effects, such as low statistics and/or multiple diffraction~\cite{Chang2004}, which cannot be considered in the simple extinction model in our analysis. We note that the forbidden signals at 10~K were at least already subtracted from the 1.5~K data for identical peaks before any magnetic refinements were done. We obtain the same result that the additional peaks from all five-rotation data are not explained by the lowest-symmetric magnetic space group (Fig.~\ref{Fig:D3}). Therefore, we removed these Bragg peaks from Figs.~\ref{Fig:9}(b)-\ref{Fig:9}(d) in the main text for simplicity. Magnetic refinement results are presented in Tables~\ref{Table:D2} and~\ref{Table:D4} for the two-rotation-angle and five-rotation-angle data, respectively, for comparison.

\begin{table*} [t]
\caption{Magnetic refinements results in lower-symmetry magnetic space groups with the 1.5~K single-crystal neutron diffraction data (the two-rotation data, with 2342 allowed and 229 forbidden peaks). Note that the Becker and Coppens model (a type 1 Lorentzian shape) was used for absorption~\cite{Petricek2014,Becker1974,Becker1975}.
}
\vspace{0.2cm}
\label{Table:D2}
\setlength\extrarowheight{4pt}
\begin{tabular}{ccccccccccc}
\hline\hline
                    &Atom   &(Coordinates $|$ Moments)                    & $M_{a}$    & $M_{b}$    & $M_{c}$    & M         & GOF   & R(All)  & R(Nuc)   & R(Mag) \\\hline
\textit{P$_A$c}     &Ni1\_1 &(1/6,2/3,z $|$ 2$m_y$,$m_y$,$m_z$)           &-2.966(160) &-1.483(40)  &-0.199(200) &2.576(259) &2.99   &4.96     &4.68      &28.42   \\
                    &       &(2/3,2/3,z $|$ -2$m_y$,-$m_y$,-$m_z$)        &            &            &            &           &       &         &          &        \\
                    &Ni1\_2 &(1/3,1/3,z+1/2 $|$ 2$m_y$,$m_y$,$m_z$)       &-1.858(179) &-0.929(45)  & 0.510(225) &1.688(291) &       &         &          &        \\
                    &       &(5/6,1/3,z+1/2 $|$ -2$m_y$,-$m_y$,-$m_z$)    &            &            &            &           &       &         &          &        \\
                    &Ni2\_1 &(1/6,2/3,z $|$ 2$m_y$,$m_y$,$m_z$)           & 1.599(153) & 0.799(38)  &-0.421(212) &1.447(264) &       &         &          &        \\
                    &       &(2/3,2/3,z $|$ -2$m_y$,-$m_y$,-$m_z$)        &            &            &            &           &       &         &          &        \\
                    &Ni2\_2 &(1/3,1/3,z+1/2 $|$ 2$m_y$,$m_y$,$m_z$)       & 0.948(161) & 0.474(40)  & 0.865(214) &1.192(271) &       &         &          &        \\
                    &       &(5/6,1/3,z+1/2 $|$ -$2m_y$,-$m_y$,-$m_z$)    &            &            &            &           &       &         &          &        \\[0.2cm]\hline
\textit{P$_C$c}     &Ni1    &(1/6,2/3,z $|$ $m_x$,$m_y$,$m_z$)            &-2.418(50)  &-0.575(151) &-0.487(161) &2.241(227) &2.91   &4.87     &4.59      &28.05   \\
                    &       &(1/3,1/3,z+1/2 $|$ $m_x$,$m_x$-$m_y$,-$m_z$) &            &            &            &           &       &         &          &        \\
                    &       &(2/3,2/3,z $|$ -$m_x$,-$m_y$,-$m_z$)         &            &            &            &           &       &         &          &        \\
                    &       &(5/6,1/3,z+1/2 $|$ -$m_x$,-$m_x+m_y$,$m_z$)  &            &            &            &           &       &         &          &        \\
                    &Ni2    &(1/6,2/3,z $|$ $m_x$,$m_y$,$m_z$)            & 1.203(63)  & 0.250(159) &-0.497(157) &1.206(232) &       &         &          &        \\
                    &       &(1/3,1/3,z+1/2 $|$ $m_x$,$m_x$-$m_y$,-$m_z$) &            &            &            &           &       &         &          &        \\
                    &       &(2/3,2/3,z $|$ -$m_x$,-$m_y$,-$m_z$)         &            &            &            &           &       &         &          &        \\
                    &       &(5/6,1/3,z+1/2 $|$ -$m_x$,-$m_x+m_y$,$m_z$)  &            &            &            &           &       &         &          &        \\[0.2cm]\hline
\textit{P$_a$}2$_1$ &Ni1    &(1/6,2/3,z $|$ $m_x$,$m_y$,$m_z$)            &-2.465(54)  &-1.219(92)  &-0.916(62)  &2.323(124) &2.99   &4.93     &4.66      &28.06   \\
                    &       &(1/3,1/3,z+1/2 $|$ $m_x$,$m_y$,-$m_z$)       &            &            &            &           &       &         &          &        \\
                    &       &(2/3,2/3,z $|$ -$m_x$,-$m_y$,-$m_z$)         &            &            &            &           &       &         &          &        \\
                    &       &(5/6,1/3,z+1/2 $|$ -$m_x$,-$m_y$,$m_z$)      &            &            &            &           &       &         &          &        \\
                    &Ni2    &(1/6,2/3,z $|$ $m_x$,$m_y$,$m_z$)            & 1.207(68)  & 0.293(101) &-0.133(66)  &1.098(138) &       &         &          &        \\
                    &       &(1/3,1/3,z+1/2 $|$ $m_x$,$m_y$,-$m_z$)       &            &            &            &           &       &         &          &        \\
                    &       &(2/3,2/3,z $|$ -$m_x$,-$m_y$,-$m_z$)         &            &            &            &           &       &         &          &        \\
                    &       &(5/6,1/3,z+1/2 $|$ -$m_x$,-$m_y$,$m_z$)      &            &            &            &           &       &         &          &   \\[0.2cm]\hline
\textit{P$_S$}1     &Ni1\_1 &(1/6,2/3,z $|$ $m_x$,$m_y$,$m_z$)            &-1.910(179) &-1.830(191) &-0.233(159) &1.885(306) &2.99   &4.97     &4.70      &27.80   \\
                    &       &(2/3,2/3,z $|$ -$m_x$,-$m_y$,-$m_z$)         &            &            &            &           &       &         &          &        \\
                    &Ni1\_2 &(1/3,1/3,z+1/2 $|$ $m_x$,$m_y$,$m_z$)        &-2.619(223) &-0.747(227) & 0.789(138) &2.466(347) &       &         &          &        \\
                    &       &(5/6,1/3,z+1/2 $|$ -$m_x$,-$m_y$,-$m_z$)     &            &            &            &           &       &         &          &        \\
                    &Ni2\_1 &(1/6,2/3,z $|$ $m_x$,$m_y$,$m_z$)            & 1.130(140) & 0.469(178) &-0.120(156) &0.991(275) &       &         &          &        \\
                    &       &(2/3,2/3,z $|$ -$m_x$,-$m_y$,-$m_z$)         &            &            &            &           &       &         &          &        \\
                    &Ni2\_2 &(1/3,1/3,z+1/2 $|$ $m_x$,$m_y$,$m_z$)        & 1.684(205) & 0.171(200) & 0.872(134) &1.827(317) &       &         &          &        \\
                    &       &(5/6,1/3,z+1/2 $|$ -$m_x$,-$m_y$,-$m_z$)     &            &            &            &           &       &         &          &        \\
\hline\hline
\end{tabular}
\end{table*}

For completeness, we present results of magnetic structural refinements using all five-rotation data collected only at 1.5~K, which include the two identical rotation data at both temperatures; in detail, we collected additional three-rotation-angle data only at 1.5~K to maximize the signal-to-noise ratio of the magnetic Bragg peaks. The same analysis as described in the main text using two-rotation single-crystal data was repeated with the five-rotation single-crystal data (13126 nuclear and 537 magnetic reflections). We fitted the 13126 nuclear reflections collected at 1.5 K with an extended unit cell, 2$a$$\times$$a$$\times$$c$, to obtain accurate structural parameters, including extinction parameters. The obtained structural parameters were then fixed for subsequent magnetic refinements.

The fitted magnetic moments with agreement parameters after magnetic refinements using various magnetic models are presented in Table~\ref{Table:D3}. Total magnetic moments at the tetrahedral and octahedral sites are comparable with those from the analysis using the two rotation data (compare Tables~\ref{Table:II} and \ref{Table:D3}). However, Table~\ref{Table:D3} indicates that the R(Mag) values are smaller than those obtained from the two-rotation angle data, indicating better statistics for the former data. Note that the difference in the R(Mag) values increases between models 1 and 2, which shows that model 1 could be better than model 2 based on these magnetic refinements. The results of magnetic refinements in \textit{P$_C$na}2$_1$ (model 1 was used as an example) and \textit{P$_S$}1 (the lowest symmetry) are illustrated in Fig.~\ref{Fig:D3} for comparison with results from the two-rotation data analysis.

\begin{table*} [h]
\caption{Magnetic refinements results in \textit{P$_C$na}2$_1$ with the 1.5~K single-crystal neutron diffraction data (the five-rotation data, with 537 forbidden peaks).
}
\vspace{0.2cm}
\label{Table:D4}
\setlength\extrarowheight{4pt}
\setlength{\tabcolsep}{8pt}
\begin{tabular}{ccccccccc}
\hline\hline
Condition                     &Model    & Atom & $M_{a}$    & $M_{b}$    & $M_{c}$    & M         & GOF     & R(Mag)   \\\hline

$M_{T}$(Ni1) $>$ $M_{O}$(Ni2) & Model 1 & Ni1  & -2.399(35) & -1.199(9)  & -0.950(43) & 2.284(56) & 3.23    & 21.46    \\\vspace{0.3cm}
	                          &         & Ni2  & 1.241(45)  & 0.621(11)  & -0.161(44) & 1.087(64) &         &          \\
                              &Model 2  & Ni1  & -2.409(36) & -1.205(9)  & -0.186(52) & 2.095(64) & 3.30    & 21.80    \\\vspace{0.3cm}
 	                          &         & Ni2  & 1.251(48)  & 0.625(12)  & -0.889(46) & 1.401(68) &         &          \\\hline

$M_{T}$(Ni1) $<$ $M_{O}$(Ni2) & Model 3 & Ni1  & 1.166(48)  & 0.583(12)  & -0.044(42) & 1.011(65) & 3.46    & 22.67    \\\vspace{0.3cm}
                              &         & Ni2  & -2.423(36) & -1.211(9)  & -0.983(41) & 2.317(56) &         &          \\
                              &Model 4  & Ni1  & 1.269(55)  & 0.634(14)  & -0.843(54) & 1.385(78) & 3.70    & 23.67    \\
                              &         & Ni2  & -2.408(41) & -1.204(10) & -0.165(59) & 2.092(73) &         &          \\
\hline\hline
\end{tabular}
\end{table*}

\begin{table*} [t]
\caption{Magnetic refinements results in lower-symmetry magnetic space groups with the 1.5~K single-crystal neutron diffraction data (the five-rotation data, with 537 forbidden peaks).
}
\vspace{0.2cm}
\label{Table:D3}
\setlength\extrarowheight{4pt}
\setlength{\tabcolsep}{4.5pt}
\begin{tabular}{cccccccccc}
\hline\hline
                           &  Atoms &Coordinates                      & $M_{a}$   & $M_{b}$    & $M_{c}$    & M           & GOF    & R(Mag)   & wR(Mag)  \\\hline
\textit{P$_A$c}     & Ni1\_1 &(1/6,2/3,z), (2/3,2/3,z)         &-1.610(79) &-0.805(20)  &-0.580(102) & 1.510(131)  &3.19    & 21.26 & 16.84  \\
                    & Ni1\_2 &(1/3,1/3,z+1/2), (5/6,1/3,z+1/2) &-0.888(89) &-0.444(22)  & 0.171(133) & 0.788(161)  &        &       &        \\
                    & Ni2\_1 &(1/6,2/3,z), (2/3,2/3,z)         & 2.962(91) & 1.481(23)  &-0.786(104) & 2.683(140)  &        &       &        \\
                    & Ni2\_2 &(1/3,1/3,z+1/2), (5/6,1/3,z+1/2) & 1.737(93) & 0.869(23)  & 0.611(118) & 1.624(152)  &        &       &        \\[0.2cm]\hline
\textit{P$_C$c}     & Ni1    &(1/6,2/3,z), (1/3,1/3,z+1/2)     &-2.426(36) &-0.972(112) &-0.886(67)  & 2.293(136)  &3.20    & 21.29 & 16.89  \\
                    &        &(2/3,2/3,z), (5/6,1/3,z+1/2)     &           &            &            &             &        &       &        \\
                    & Ni2    &(1/6,2/3,z), (1/3,1/3,z+1/2)     &1.296(47)  & 0.539(122) &-0.212(67)  & 1.147(147)  &        &       &        \\
                    &        &(2/3,2/3,z), (5/6,1/3,z+1/2)     &           &            &            &             &        &       &        \\[0.2cm]\hline
\textit{P$_a$}2$_1$ & Ni1    &(1/6,2/3,z), (1/3,1/3,z+1/2)     &-2.378(35) & -1.100(72) &-0.959(41)  & 2.273(90)   &3.20    & 21.34 & 16.92  \\
                    &        &(2/3,2/3,z), (5/6,1/3,z+1/2)     &           &            &            &             &        &       &        \\
                    & Ni2    &(1/6,2/3,z), (1/3,1/3,z+1/2)     &1.240(45)  &  0.449(79) &-0.142(43)  & 1.097(100)  &        &       &        \\
                    &        &(2/3,2/3,z), (5/6,1/3,z+1/2)     &           &            &            &             &        &       &        \\[0.2cm]\hline
\textit{P$_S$}1     & Ni1\_1 &(1/6,2/3,z), (2/3,2/3,z)         &-1.760(111)& -0.649(143)&-0.608(124) & 1.657(220)  &3.18    & 21.14 & 16.70  \\
                    & Ni1\_2 &(1/3,1/3,z+1/2), (5/6,1/3,z+1/2) &-2.891(106)& -1.476(117)& 0.815(111) & 2.633(193)  &        &       &        \\
                    & Ni2\_1 &(1/6,2/3,z), (2/3,2/3,z)         &0.930(100) &  0.185(137)&-0.145(138) & 0.865(219)  &        &       &        \\
                    & Ni2\_2 &(1/3,1/3,z+1/2), (5/6,1/3,z+1/2) &1.587(91)  &  0.678(117)& 0.561(108) & 1.489(183)  &        &       &        \\
\hline\hline
\end{tabular}
\end{table*}

\begin{table*} [t]
\caption{Comparison of magnetic moments of two Ni sites of \NMO and their ratios in our study and the literature. Those from the electron spin resonance (ESR)~\cite{Morey2019}, powder neutron diffraction (PND) refinements of the reference (solutions 1 and 2)~\cite{Morey2019}, PND refinements of our data, single-crystal neutron diffraction (SND) with the two-rotation data (TR), and with the five-rotation data (FR) are compared. Solutions 1 and 2 are obtained directly from the reference, while the solutions marked with \#  are the slightly modified solutions 1 and 2 that fit our pND data.
}
\vspace{0.2cm}
\label{Table:D5}
\setlength\extrarowheight{4pt}
\setlength{\tabcolsep}{35pt}
\begin{tabular}{ccccc}
\hline\hline
Data                        & Model   & $M_{T}$     & $M_{O}$     & Ratio  \\\hline
ESR (10 K)~\cite{Morey2019} &         & 4.32        & 2.43        & 1.78   \\\hline
                            & 1~\cite{Morey2019}		  & 1.727       & 1.431       & 1.21   \\
PND                     & 2~\cite{Morey2019}		  & 1.997       & 0.891       & 2.24   \\
                            & 1\# 	  & 1.771       & 1.237       & 1.43   \\
                            & 2\# 	  & 1.983       & 1.024       & 1.94   \\\hline
                            & 1       & 1.904       & 1.038       & 1.83   \\
PND (1.5~K)         & 2       & 1.769       & 1.295       & 1.37   \\
                            & 3       & 0.994       & 1.911       & 1.92   \\
                            & 4       & 1.272       & 1.772       & 1.39   \\\hline
                            & 1	      & {2.256}     & {1.052}     & {2.14} \\
SND (1.5~K), TR        & 2       & {2.096}     & {1.315}     & {1.59} \\
                            & 3	      & {1.020}     & {2.304}     & {2.25} \\
                            & 4       & {1.324}     & {2.146}     & {1.62} \\\hline
                            & 1       & {2.284}     & {1.087}     & {2.10} \\
SND (1.5~K), FR     & 2       & {2.095}     & {1.401}     & {1.50} \\
                            & 3       & {1.011}     & {2.317}     & {2.10} \\
                            & 4       & {1.385}     & {2.092}     & {1.51} \\
\hline\hline
\end{tabular}
\end{table*}

\begin{table*}
	\centering
	\caption{Definitions of the reliability factors used in the refinements of this study on the powder~\cite{PetricekManaul} and single-crystal data~\cite{Petricek2000}.
		The labeling, All, Mag, Nuc, inside the parenthesis in the R factor used in this study means all, only magnetic, and only nuclear reflections, respectively.
	}
	\label{DefinitionR}
	\setlength\extrarowheight{14pt}
	\begin{tabular}{M{2.5cm}M{6.1cm}|M{2.5cm}M{6.1cm}}
		\hline\hline
		Parameters                                & Powder data
		& Parameters                            & Single-crystal data                                                                                                                                                        \\ [14pt] \hline
		Profile R factor (\%)                  & Rp=$\cfrac{\sum_{i}|y_i(\text{obs})-y_i(\text{calc})|}{\sum_{i}y_i(\text{obs})}\times100$
		& R factor (\%)                          & $\text{R} =\cfrac{\sum_{hkl}||F(\text{obs})|-|F(\text{calc})||}{\sum_{hkl}|F\text{obs}|}\times100$                           \\ [14pt] \hline
		Weighted profile R factor (\%) & wRp $=\sqrt{\cfrac{\sum_{i}w_i(y_i(\text{obs})-y_i(\text{calc}))^2}{\sum_{i}w_iy_i(\text{obs})^2}}\times100$                                                                                                                                                                                                               & Weighted R factor (\%)        &  $\text{wR} =\sqrt{\cfrac{P}{\sum w|F(\text{obs})|}}\times100$                                                                                  \\ [14pt] \hline
		Weight                                       & $w_i = \cfrac{1}{\sigma^2[y_i(\text{obs})]}$                                                                                                                                                                                                                                                                               & Weight                                   &  $w = \cfrac{1}{\sigma^2(|F(\text{obs})|)+(uF(\text{obs}))^2}$                                                                                    \\ [14pt] \hline
		Goodness of fit                        & GOF$=\cfrac{\text{wRp}}{\text{Rexp}}$                                                                                                                                                                                                                                                                                      & Goodness of fit                    & $\text{GOF}=\sqrt{\cfrac{\sum w(F(\text{obs})-F(\text{calc}))^2}{m-n}}$                                                                     \\ [14pt] \hline
		Experimental R factor (\%)     & Rexp$ =\sqrt{\cfrac{\sum_{i}w_iy_i(\mbox{obs})^2}{n-p}}\times100$                                                                                                                                                                                                                                                          &   Minimized function             & $\text{P}=\sum w(|F(\text{calc})|-|F(\text{obs})|)^2$                                                                                                    \\ [14pt] \hline
		\multicolumn{2}{M{8.6cm}|}{\makecell{- $i$: the point of the profile \\ - $y_i(\text{obs})$: the observed intensity \\
				- $y_i(\text{calc})$: the calculated intensity \\ - $n$: the total number of profile points \\ - $p$: the number of refined parameters \\ - RF and RFw: R factor and weighted R factor, \\ using the structure factor (F)}}
		& \multicolumn{2}{M{8.6cm}}{\makecell{- $F(\text{obs})$: the observed structure factor \\ - $F(\text{calc})$: the calculated structure factor \\
				- $u$: an instability factor (a default value, 0.01, used) \\ - $m$: the number of reflections \\ - $n$: the number of parameter refined}}                                      \\ [14pt] \hline\hline
	\end{tabular}
\end{table*}

\end{document}